\documentclass[a4paper,UKenglish,cleveref, autoref, thm-restate]{lipics-v2021}
\newcommand{\remove}[1]{}

\usepackage{amsmath}

\usepackage{float}
\usepackage{amsthm}
\usepackage{hyperref}
\usepackage{subcaption}
\usepackage{color,soul,enumitem}
\usepackage{scrextend}
\newcommand{\edit}[1]{{\color{black}#1}}
\newtheorem{assumption}{Assumption} %[section]

\newenvironment{changemargin}[2]{%
\begin{list}{}{%
\setlength{\topsep}{0pt}%
\setlength{\leftmargin}{#1}%
\setlength{\rightmargin}{#2}%
\setlength{\listparindent}{\parindent}%
\setlength{\itemindent}{\parindent}%
\setlength{\parsep}{\parskip}%
}%
\item[]}{\end{list}}

\remove{
\usepackage{amsmath}

\usepackage{float}
\usepackage{amsthm}
\usepackage{hyperref}
\usepackage{subcaption}
\usepackage{color,soul,enumitem}

\newcommand{\edit}[1]{{\color{black}#1}}

\theoremstyle{claim}

\newtheorem*{claim*}{Claim}

\theoremstyle{plain}
\newtheorem{definition}{Definition} %[section]
\newtheorem{lemma}{Lemma}
\newtheorem{theorem}{Theorem}
\newtheorem{corollary}{Corollary}%[theorem]

\newtheorem*{definition*}{Definition} %[section]
\newtheorem*{lemma*}{Lemma}
\newtheorem*{theorem*}{Theorem}
\newtheorem*{corollary*}{Corollary}%[theorem]

}

\makeatletter
\newcommand{\setword}[2]{%
  \phantomsection
  #1\def\@currentlabel{\unexpanded{#1}}\label{#2}%
}
\makeatother

\DeclareMathOperator{\Nash}{Nash}
\DeclareMathOperator{\Triadic}{Triadic}

\DeclareMathOperator{\Distortion_once}{Distortion}
\DeclareMathOperator{\SupDistortion}{Sup Distortion}

\DeclareMathOperator{\Prob}{Pr}

\newcommand{\combi}[1]{\mathcal{C}({#1})}
\newcommand{\combiwithr}[1]{\mathcal{C}_r({#1})}
\newcommand{\power}[1]{\mathcal{P}({#1})}
\newenvironment{hproof}{%
  \renewcommand{\proofname}{Proof Sketch}\proof}{\endproof}

\newfont{\bssdoz}{cmssbx10 scaled 1200}
\newfont{\bssten}{cmssbx10}

\makeatletter
\newcommand{\printfnsymbol}[1]{%
  \textsuperscript{\@fnsymbol{#1}}%
}
\makeatother

\title{Low Sample Complexity Participatory Budgeting}
\author{Mohak Goyal$^{1}$}{Stanford University \and \url{https://sites.google.com/view/mohakg}}{mohakg@stanford.edu}{https://orcid.org/0000-0002-1176-5549}{}

\author{Sukolsak Sakshuwong$^{1}$}{Stanford University \and \url{https://sukolsak.com/}}{sukolsak@gmail.com}{}{}

\author{Sahasrajit Sarmasarkar\footnote{In alphabetical order}}{Stanford University \and \url{https://sahasrajit123.github.io/}}{sahasras@stanford.edu}{https://orcid.org/0000-0002-6652-4881}{}

\author{Ashish Goel}{Stanford University \and \url{https://web.stanford.edu/~ashishg/}}{ashishg@stanford.edu}{}{}

\authorrunning{M. Goyal, S. Sakshuwong, S. Sarmasarkar, A. Goel} %TODO mandatory. First: Use abbreviated first/middle names. Second (only in severe cases): Use first author plus 'et al.'

\ccsdesc[500]{Applied computing} %TODO mandatory: Please choose ACM 2012 classifications from https://dl.acm.org/ccs/ccs_flat.cfm 

\keywords{Social Choice, Participatory budgeting, Nash bargaining} %TODO mandatory; please add comma-separated list of keywords

\acknowledgements {We would like to thank Lodewijk Gelauff,  (Stanford University), Geoff Ramseyer(Stanford University) and Kamesh Munagala(Duke University) for valuable insights and discussions on the paper and the introduction.}%optional

\begin{document}

\nolinenumbers

\maketitle

\begin{abstract}
    We study low sample complexity mechanisms in participatory budgeting (PB), where each voter votes for a preferred allocation of funds to various projects, subject to project costs and total spending constraints. We analyse the distortion that PB mechanisms introduce relative to the minimum-social-cost outcome in expectation. The Random Dictator mechanism for this problem obtains a distortion of $2$. In a special case where every voter votes for exactly one project, \cite{fain2017} obtain a distortion of $4/3.$  We show that when PB outcomes are determined as any convex combination of the votes of two voters, the distortion is $2$. When three uniformly randomly sampled votes are used, we give a PB mechanism that obtains a distortion of at most $1.66,$ thus breaking the barrier of $2$ with the smallest possible sample complexity.
    
    We give a randomized Nash bargaining scheme where two uniformly randomly chosen voters bargain with the disagreement point as the vote of a voter chosen uniformly at random. This mechanism has a distortion of at most 1.66. We provide a lower bound of 1.38 for the distortion of this scheme. Further, we show that PB mechanisms that output a median of the votes of three voters chosen uniformly at random, have a distortion of at most 1.80. 
\end{abstract}

%% The next command prints the information defined in the preamble.

%\maketitle 

%%%%%%%%%%%%%%%%%%%%%%%%%%%%%%%%%%%%%%%%%%%%%%%%%%%%%%%%%%%%%%%%%%%%%%%%

\section{Introduction}
%More than 1500 cities around the globe have begun adopting {\it Participatory Budgeting} (PB) \cite{wainwright2003,ganuza2012}, a process through which residents can vote directly on a city government's use of public funds.  residents might, for example, vote directly on how to allocate a budget of reserved funds between projects like street repairs or library renovations.  PB has been shown to promote government transparency, resident engagement, and good governance \cite{wampler2007}. 

More than 1500 cities around the globe have begun adopting {\it Participatory Budgeting} (PB) \cite{wainwright2003,ganuza2012}, 
a process through which residents can vote
directly on a city government's use of public funds.  Residents might, for example,
vote directly on how to allocate a budget of reserved funds between projects like
street repairs or library renovations.  PB has been shown to promote
government transparency, resident engagement, and good governance \cite{wampler2007}.

We study a PB setup similar to \cite{freeman2021truthful} where each vote is an allocation of funds to projects (we call it a ``preferred budget'')  subject to the constraint that the sum of allocations to all projects is equal to one. Projects have a fixed cost, and allocations to any project cannot exceed its cost. However, allocations less than the project's cost are allowed. (\cite{freeman2021truthful} consider all project costs equal to one). In this model, therefore, every vote and the outcome of the PB election can be represented as a point on the unit simplex.

We study the {\it distortion} (Definition~\ref{def:distortion}) that PB mechanisms introduce in expectation relative to the social cost minimizing allocation in the worst case of PB instances, following the lines of \cite{anshelevich2017randomized}. We adopt the $\ell_1$ distance as the cost function where a voter with preferred budget $a$ experiences a cost of $d(a,b) = \|a-b\|_1$ from an outcome budget $b$ (Definition ~\ref{def:cost}). 

Several preference elicitation methods have been studied for PB 
\cite{brams2002voting,arrow2010handbook,goel2019knapsack,benade2021preference}. %, yet no matter the mechanism,
Policymakers must then transform a list of votes into a real-world allocation of funds.
Furthermore, even though there may be an ``optimal'' allocation (under natural notions
of social welfare),
this allocation may be intractable to compute \cite{jain2020,peters2020proportional} or difficult to reliably estimate if turnout is low \cite{ewens2019organizational}. In some situations, policymakers need to obtain a quick estimate of the budgetary region in which preferences may lie. In these cases, and when running a fully-fledged PB election is costly or difficult, low-sample complexity PB mechanisms are an attractive choice.

Low-sample complexity preference elicitation mechanisms have also been of interest recently in computational social choice \cite{fain2017,fain2019,anshelevich2017randomized,fain2020concentration} -- in this work, we give low-sample complexity mechanisms (using the preferred budgets of a small number of sampled voters) for PB, which achieve a distortion of less than 2. Note that $2$ is a natural barrier for the distortion in this problem since the {\it Random Dictator} mechanism
achieves a distortion of 2 in our model of PB. The Random Dictator mechanism chooses the outcome as the preferred budget of a uniformly randomly chosen voter. From Theorem 5 of \cite{anshelevich2017randomized}, its distortion is at most $2$, and from our Lemma~\ref{dist2}, it is $2$.  We further prove that a mechanism that chooses any linear combination of two randomly sampled votes ({\it Random Diarchy}) also attains a distortion of 2 (Lemma \ref{distortion_two_bargaining}). Another low sample-complexity mechanism, {\it Random Referee} \cite{fain2019}, asks a randomly chosen voter (``the referee'') to choose one out of two possible outcomes, which are random samples from the preferred budgets of the voters. This mechanism also attains a distortion of at least 2 in our setup (Lemma~\ref{distortion_random_referee}). 
We give a PB mechanism which samples three voters uniformly at random and attains a distortion of at most 1.66.

\remove{Social choice has been studied in \cite{fain2017,fain2020concentration,fain2019,boutilier2015optimal,anshelevich2017randomized,freeman2021truthful} under various voting schemes and metric spaces with elements denoting all possible alternatives and voters with distance In particular, low sample complexity elicitation mechanisms have been studied in \cite{fain2017,fain2019,fain2020concentration}. In this work, we focus on a low sample complexity bargaining scheme where two uniformly at randomly selected voters bargain with the disagreement point being the preferred budget of another uniformly at random sampled voter. Though a similar setup has been studied in \cite[Appendix B]{fain2017} with $\ell_1$ norm as the notion of distance in the metric space, their result only holds for the case when every voter can only vote for a budget allocating all funds to only a single project. We study a more generalized setup where every voter votes for a fund allocation to various projects subject to a budget constraint (say unity), mathematically each vote is a point in the unit budget simplex and thus the results in \cite[Appendix B]{fain2017} do not directly hold for our case. Another low sample complexity mechanism namely {\it random referee} was studied in \cite{fain2019} where each voter only reports an ordinal preference of various alternatives(budgets). We also show that distortion of the outcomes of {\it random referee} mechanisms proposed in \cite{fain2019} is higher than the outcomes of the bargaining schemes in our setup where every vote is a point in the unit simplex.
}

%In particular, we focus on a low sample complexity mechanism for participatory budgeting. In this mechanism, two randomly selected voters bargain with the disagreement point being the preferred budget of another randomly sampled voter where the preferred budget of a voter could be any point in the unit budget simplex. Observe that in this model, each voter votes for a budget allocating funds for every project subject to the budget constraint. A similar model was considered in \cite[Appendix B]{fain2017} with the $\ell_1$ norm as the measure of disutility but with the additional constraint of every voting for exactly one project which however does not generalize to the case where the vote could be any point in the simplex.

%Note that \cite{fain2017} consider a similar low complexity model and prove a stronger result on distortion without the budget constraint where every voter voting for a particular project, however, this does not generalize directly to the setup considered here.

%More specifically, we study a randomised Nash 

%Note that in most such scenarios, it may be difficult to ask every voter to list out   %budget which has to decide how much funds to allocate between various projects. 

\remove{

PB implementations have used various mechanisms to elicit preferred budgets from residents 
\cite{brams2002voting,arrow2010handbook,goel2019knapsack,benade2021preference}. %, yet no matter the mechanism,
Policymakers must then transform a list of votes on a budget into a real-world allocation of funds.
Furthermore, even though there may be an ``optimal'' allocation (under natural notions
of social welfare),
this allocation may be intractable to compute \cite{jain2020,peters2020proportional} or difficult to reliably estimate if turnout is low \cite{ewens2019organizational}.
Our topic of study in this work is the {\it distortion} that PB mechanisms introduce
between the theoretically ``optimal'' allocation and the output by the mechanism.

}
%Participatory budgeting (PB) is a process in which residents vote on the public budget. For example, residents of a city may vote on how to spend a million dollars of the city's funds on projects such as renovating the library or improving the streets. PB first started in the 1980s in Brazil \cite{wainwright2003} and is gaining popularity, with more than 1,500 cities worldwide already adopting PB \cite{ganuza2012}. %PB has been shown to promote transparency, resident involvement, and good governance \cite{wampler2007}.
%
%Several voting methods have been used in PB, such as %\emph{K-approval voting}, in which voters approve up to $K$ projects from the list of all projects,
%\emph{approval voting} \cite{brams2002voting,arrow2010handbook}, in which voters select any number of projects that they like, and \emph{knapsack voting} \cite{goel2019knapsack}, in which voters select as many projects as they like, subject to the budget constraint. Major challenges in PB include low voter turnaround \cite{ewens2019organizational}, and the design of ballots accounting for substitute and complementary projects \cite{fairstein2021proportional, jain2020, jain2021partition}.

\remove{
In particular, we focus on a process known as {\it sequential deliberation} \cite{fain2017,fishkin2018democracy},
a mechanism for collective decision-making % (in contexts where residents can debate directly)
 that we \remove{adapt}{\color{red}  use as the first pre-processing step} %into a mechanism 
 for PB. {\color{red} Typically in most PB instances, there involves a deliberation step where the projects are discussed and bundled accordingly as described in \cite{deliberation_participatory_budegting,gilman_deliberation,https://doi.org/10.1002/pad.1853}.}
In each round of sequential deliberation,
two randomly-selected voters debate to decide on a compromise, which becomes the ``disagreement point''
for the next round.  If the voters in a round fail to reach a compromise, their choice defaults to the disagreement
point of the previous round; as such,
the disagreement point, in essence, captures an aggregate preference of the deliberators of previous rounds.

}
%
%To extend this process to PB (where a ``deliberation'' must be between static votes, not living voters),
%we propose and analyze several \XXX{several? one? two?} classes of ``bargaining'' mechanisms for algorithmically aggregating two random budget votes
%and a disagreement point into a new disagreement point (and let the disagreement point of the first round be a third, random vote).
%
%On another but related note, public deliberation has been gaining popularity as a democratic process for collective decision-making \cite{fishkin2018democracy}. %It has several good properties, including the potential for coming up with more innovative solutions, and enabling people to understand each others' point of views \cite{fishkin1991democracy}.
%Sequential deliberation has been proposed as a method for implementing deliberation among agents \cite{fain2017} in which randomly chosen agents deliberate in pairs in rounds, and the outcome of the previous round is the disagreement point for the new round. (The disagreement point of a bargaining process is the outcome if no agreement is reached by the parties). The disagreement point, in essence, distills the preferences of all the agents who participated in the previous rounds. In this paper, we propose a sequential deliberation-based mechanism for PB. 
%
\remove{
Our theoretical guarantees are for the outcome of one round of deliberation. This mechanism takes only three samples of the voters (one for the disagreement point and two for deliberation) and is therefore also interesting for the goal of minimizing the sample-complexity of the mechanism.\footnote{See also Fain et al. \cite{fain2019} for another PB mechanism with a sample-complexity of 3.}% An important part of our sequential deliberation scheme is that the agents are chosen at each round without the knowledge of the precise bargaining scheme adopted in each round of deliberation, hence it may be adopted to a variety of settings.}
 }

%Note that we could extend this bargaining scheme to multiple rounds  

\remove{
The bargaining scheme described here can be extended to multiple rounds where the outcome of every round can be used as the disagreement point for the next round, and in each round bargaining voters may be sampled uniformly at random to obtain the \emph{sequential deliberation} mechanism as described in \cite{fain2017}. Though we do not provide explicit theoretical guarantees for this setup, we do perform simulations on real participatory budget (PB) data in Appendix \S\ref{sec:exp} that sequential deliberation can very rapidly converge to a solution that has a limited "distortion". This means that with a few rounds of deliberations, %each only requiring participation from a very small sample,
we can get an estimate of what budgetary region the result of a vote could lie in. This could be particularly important in scenarios where turnout is low, or a PB ballot must be simplified before deployment \cite{aitamurto2015five}.
}

\remove{\color{red} Sequential deliberation in this setup\remove{The use of sequential deliberation in PB} has some properties that are especially interesting to policymakers. For example, we will demonstrate with simulations in Appendix \S\ref{sec:exp} that sequential deliberation can very rapidly converge to a solution that has a limited "distortion". This means that with a few rounds of deliberations, %each only requiring participation from a very small sample,
we can get an estimate of what budgetary region the result of a vote could lie in. This could be particularly important in scenarios where turnout is low, or a PB ballot must be simplified before deployment \cite{aitamurto2015five}.

}
\remove{
There are also some intrinsic advantages to using sequential deliberation to collect budgeting preferences. For example, the bargaining process will encourage voters to exchange ideas and opinions in a short, focused period of time that allows the organizer to collect the arguments in connection with certain disagreements in the budget space efficiently. Any deliberative process where different voters constructively interact with other voters would expose voters to opinions from other voters, which has potential benefits to the democratic process \cite{christ2019intergroup}. 
}

\remove{
\subsection{Project Interactions in PB}
Most theoretical work on PB assumes that different projects can be implemented independently. However, in many practical situations, this assumption is not valid. Several projects can be substitutes for each other. As an example, consider the two projects of building a park each on road A and road B. If the two roads are close to each other, the residents of the area would see the two projects as substitutes for each other and pick one arbitrarily in their vote. We define a group $S$ of projects to be \emph{perfect substitutes} if no voter wants more than one project in $S$ to be funded. The utility that voters can derive from group $S$ is the maximum funding of a project in $S$. %For example, if two companies are paid $0.2$ and $0.5$ to do the same work, only $0.5$ will be used, and $0.2$ is wasted. 
In another scenario, projects can be complements of each other. For example, buying new computer hardware and software may be seen as complementary projects by the voters. We define a group of projects $R$ to be perfect complements if the utility that voters can derive from each project in $R$ is the minimum funding of any project in $R$. For example, if the software and hardware projects are funded $0.2$ and $0.5,$ they can only use $0.2$ each, and the excess funding of $0.3$ for the hardware project is wasted. These are special cases of submodular and supermodular utility functions for substitute and complement projects, respectively. Further, we define a group of projects to be contradictions if it is physically impossible to implement more than one of them. %As noted by Jain, Talmon, and Bulteau in \cite{jain2021partition}, the utility of agents may be given by a submodular function over the set of \emph{substitute} projects, and by a supermodular function over the set of \emph{complementary} projects. 
 %Such project interactions may sometimes be unknown to the PB designers or may be difficult to account for in the ballot.
} 

%Jain, Sornat, and Talmon \cite{jain2020} introduce a model in which the project interactions are specified in the ballot and the voters do approval voting. They provide algorithms for vote aggregation aimed at utility maximization under different functions of  

\subsection{Our Contributions}

When the PB mechanism samples three voters uniformly at random, we show that aggregation schemes that choose a median of their preferred budgets achieve a distortion of at most 1.80.  We refer to such schemes as the {\it median schemes} and denote this class of schemes by $\mathcal{M}$.

%Specifically, we show that deliberation schemes constrained to output the median of the preferred budgets of three sampled voters also called {\it median scheme} achieves a distortion of at most 1.80. Furthermore, we also give a \emph{randomized Nash bargaining scheme} achieving a distortion of at most $1.66$ {\color{red} where two of the sampled voters bargain with the preferred budget of the third voter serving as the disagreement point}.

We then turn to the case where two uniformly randomly chosen voters can come together and ``bargain'' with a third voter's preferred budget (again chosen uniformly at random) as the ``disagreement point.'' We formulate the bargaining rules for the voters via the well-studied Nash bargaining framework \cite{binmore1986nash}. When these bargaining rules can be further specified by a randomized rule (\S\ref{exact_soln_constr}), we show that the distortion of the resulting mechanism is at most $1.66$ (Theorem \ref{distortion_Nash_bargaining_randomised}). We call this mechanism the \emph{randomized Nash bargaining scheme} $\mathfrak{n}_{rand}$.

{A key technical tool we use is the analysis of \emph{pessimistic distortion} (PD) (Definition~\ref{pess_dist_defn}) first proposed by \cite{fain2019}. PD is a form of distortion where the comparison is made with a counterfactual which chooses a separate outcome for every small subset of voters (of a fixed size $\kappa$), thereby attaining a lower social cost than the true `optimal'. In this work, we use $\kappa = 6$. This choice is due to computational constraints. We show that the PD with $\kappa = 6$ is an upper bound on the distortion of our proposed mechanisms with any number of voters $n$. 

We then reduce the problem of computing the PD into a set of linear programs for the median schemes $\mathcal{M}$ and bilinear programs of constant size for the randomized Nash bargaining scheme $\mathfrak{n}_{rand}$. For this, we use a projection of the preferred budgets of voters into a space (we call it the {\it incremental allocation space} (\S\ref{sec:projection_smaller_subset_voters})) that captures the common preferences of a subset of voters relative to other voters. In the median schemes, funds are allocated to projects ensuring that the final outcome is the median of the preferred budgets of three randomly sampled voters.
In the randomized Nash bargaining scheme $\mathfrak{n}_{rand}$, the expected funds allocated to a project satisfy additional proportionality constraints (\S\ref{exact_soln_constr}), resulting in bilinear programs. Since the proportionality constant is not fixed, this results in another variable in the optimization formulation. \remove{for \color{red} $\mathcal{M}$ and a set of bilinear programs for $\mathfrak{n}_{\text{rand}}.$ -- undefined terms.} The problem has a complex combinatorial structure due to the nuances of Nash bargaining. However, we are able to exploit symmetries of the problem, enabling us to solve it efficiently. Since the bilinear programs are of a constant size (depends on $\kappa$, which we set to $6$), we can solve these in fixed time.}%We provide the code for solving these problems on GitHub\footnote{\textcolor{red}{PUT LINK TO CODE}}. 

Same as \emph{Random Dictator} and \emph{Random Referee}, our PB mechanism 
$\mathfrak{n}_{rand}$ also naturally respects project interactions such as complementarity and substitution as long as the voters are aware of these interactions. This is because the bargaining outcome between two voters is guaranteed to be Pareto optimal for them. We describe this point in detail in \S\ref{sec:TD_project_interaction}.
%\edit{This means that for all possible instances of PB, the allocation that results from sequential deliberation has a total \emph{disutility} less than $1.66$ times that of the social welfare-maximizing allocation in expectation, where the expectation is over the randomness in the bargaining process.}

%This result is interesting because any voting mechanism constrained to choose outcomes from the voters' preferred budgets has an expected distortion at least 2.
%\edit{In the sequential deliberation approach, the bound on the distortion assuming there are no project interactions is also valid if there are project interactions.} Interestingly, this is the first voting mechanism that provably has distortion below $2$ for PB with interdependent projects; any voting mechanism constrained to choose outcomes from the voters’ preferred budgets has distortion at least $2$ (Lemma~\ref{dist2}). 

\remove{
\edit{When the project interactions are given explicitly by the PB designer, we study a menu-based ballot design. Such ballot designs are growing in popularity in PB and gaining a theoretical insight into them is important. For example, the 2020 city budget consultation in Long Beach conducted on the Stanford City Budgeting Platform\footnote{\url{https://budget.pbstanford.org/longBeach2020}} had a menu-based ballot with some complex project interactions. We study a simplified model of such a ballot. %In this setting, we relax the assumption that all voters agree on the project interactions. 
%We further relax the assumption that projects in a group of substitutes are perfect substitutes. 
%This model can work with both fractional and non-fractional funding of projects. 
 In our model, }the PB designer divides the projects into groups such that the voters' utilities are additively separable over the different groups. Groups correspond to contradictory, substitute, or independent projects. Independent projects correspond to groups of size $1$.  %We consider a natural extension of the overlap utility model. 
For preference aggregation, the voters do approval voting within a group and distribute money across groups as in knapsack voting. 
%with the constraint that the cost of approved projects in a group is at least as much as the money allocated to the group by the voter. 
This ballot design makes voting cognitively easier than knapsack voting, where voters must allocate money to each project separately. %At the same time, it elicits more information regarding preferences than approval or K-approval voting. 
In many practical situations, some voters only care about broad issues such as healthcare, security, education, etc. In our proposed ballot, the project groups may correspond to these broad issues, with individual projects listed within the groups. Voters would spend more cognitive effort in deciding how to allocate funds to the broad issues; and then approve their favored projects in the groups, usually with less cognitive load.

 \edit{ %If voters think that the projects in a group are substitutes or contradictions, they would allocate one unit of funds, and approve all the acceptable projects in the group. If voters wish to fund $2$ projects from a group but find $5$ options acceptable, they can allocate two unit of funds, and approve the $5$ acceptable projects in the group. 
%It is noteworthy that this is the first ballot design which can handle both project substitutions and contradictions by the same preference elicitation and aggregation method. 
 Our proposed ballot enables the handling of project interactions where projects are either substitutes or contradictions, and is the first ballot design capable of doing so.} For unit cost projects, we show that for social welfare-maximizing preference aggregation, our method is strategy-proof when at most one project from a group can be funded.  %If voters think that the projects are complementary, they would fund the project group for as many units as they think should be funded, and approve the projects they think should be funded together. 
%For the menu-based ballot, we consider both fractional and non-fractional allocations. 
We show that under non-fractional allocations, maximizing social welfare is NP-hard by reduction from the maximum set cover problem. We then provide a dynamic programming-based %fixed parameter tractable (FPT) 
algorithm for maximizing social welfare%, parameterized by the maximum size of a group
.}
%
%%%%%%%%%%%%%%%%%%%%%%%%%%%%%%%%%%%%%%
%
\subsection{Related Work}

The \textit{sequential deliberation} (SD) mechanism for social choice was proposed in \cite{fain2017} where the two uniformly randomly chosen voters deliberate in each round under the rules of Nash bargaining, and the outcome for every round is the disagreement point for the next round. The SD for one round corresponds to the randomized Nash bargaining scheme $\mathfrak{n}_{rand}$. They analyzed the mechanism in median spaces, which include median graphs and trees, and found an upper bound of the distortion of the mechanism to be 1.208. They also analyze the distortion in the \emph{budget space} (or unit simplex) in a special setting where each voter only approved funds for a single project. In this case, they show that the distortion in the equilibrium of SD is $4/3$. This paper extends their work in the case of the unit simplex, such that voters in our model do not have to restrict their vote to one project. %{\color{red} Note that in our model projects do not have a pre-determined cost but agents can choose to allocate an arbitrary amount of funds to any project subject to the budget constraint along the lines of \cite{freeman2021truthful}.}

The authors of \cite{goel2016towards} study a model where voters' opinions evolve via deliberations in small groups over multiple rounds. Opinions in their model correspond to preferred budgets in our model; however, unlike preferred budgets, opinions change as a result of deliberations. They study the distortion in single-winner elections setting and show that it is bounded by $O\left(1+\sqrt{\frac{\log n}{n}}\right)$ when voters deliberate in groups of $3$ ($n$ is the number of voters).

\remove{
 The lower bound of \cite[Theorem 1]{goel2016towards} on the expected distortion could be considered stronger than the distortion result on {\it random diarchy} mechanism in Lemma \ref{distortion_two_bargaining} as the outcome in every round in \cite{goel2016towards} depends only on the opinions of two sampled voters in that round and the outcome from the previous round but, in the proposed {\it random diarchy} scheme the linear combination of the two sampled preferred budgets could be optimized over the entire space of preferred budgets. %Also in \cite{goel2016towards}, the voters lie on the real line whereas in our setup the preferred budgets lie on a unit simplex.  Another mechanism (namely {\it triadic consensus}) that samples three candidates at every round was studied in \cite{goel2012triadic} for a single winner election where the sampled candidates at each round are asked which amongst the other two sampled candidates they prefer. However, they study the problem in the setup where the candidates can be ranked on a real line and also study the strategic behavior of the candidates.
}
%which constrains the outcome to only depend on the opinions of the two sampled voters at every round. However, they study it in a different opinion formation model where the opinions lie on a real line, and the opinion formation progresses over multiple rounds.

%Some recent works have proposed randomized voting schemes. %In Fain et al. \cite{fain2019}, voters report a pairwise comparison between randomly chosen budget proposals, 
The work most closely related to ours is \cite{fain2019}; they study the \textit{random referee} mechanism. %, in which two randomly chosen voters are asked for their preferences, and another voter is randomly chosen as a referee to pick one of the two preferences.
We use their technique of analyzing the PD of 6 voters. However, they apply this technique where the underlying decision space is the Euclidean plane and use the underlying geometric structure to perform a grid search. In contrast, we study the PD with Nash bargaining, which leads to a complex structure of outcomes that we capture in linear or bilinear programs. %for finding an upper bound of the distortion in obtaining our results.

%Bhaskar et al. \cite{bhaskar2018truthful} study mechanisms where voters report a set of projects with utility higher than a randomly set threshold; this is somewhat orthogonal to our work where only a small set of voters need to participate. 

The authors of \cite{fain2020concentration} analyze low sample-complexity randomized mechanisms for PB. They  obtain constant factor guarantees for higher moments of distortion, and the distortion bound they provide is much larger than 2. Several additional results and research directions in PB are described in the survey \cite{aziz2021participatory}.
%Another area of focus has been fairness in PB \cite{fain2018fair,conitzer2017fair,fluschnik2019fair}.
% and finding solutions in the core \cite{fain2016core}. 

\remove{\color{red} 

}
%Similarly, the triadic consensus was also studied in \cite{goel2012triadic}. {\color{red} where voters lie in one-dimensional space. }

%Our proposed sequential deliberation-based method is also a randomized voting mechanism in that it chooses bargaining agents uniformly at randomly.
%
%Benade et al. \cite{benade2018} do a theoretical and empirical study of the trade-offs between different preference elicitation methods for PB. %, including knapsack votes, rankings by value or value for money, and threshold approval votes. They found that threshold approval voting is superior in maximizing the voters' unknown underlying utilities. 

%{\raggedleft\textbf{Future directions}}
\subsection{Future Directions}

 A natural direction for future work is to analyze the distortion for multiple rounds of deliberation in our model, with every round's outcome serving as the next round's disagreement point. Another interesting modelling question is to study the deliberation or bargaining process with more than two agents participating together. Closing the gap of the distortion of $\mathfrak{n}_{rand}$ also remains an interesting open problem. %Another direction of work could be to analyze the pessimistic distortion for larger groups of voters and check if the bounds improve.

%{\raggedleft\textbf{Roadmap}}

\subsection{Roadmap}
%We describe the sequential deliberation approach for PB in Sections~\ref{sec:model_SD}-\ref{sec:exp}. Within this, 

We describe the model and preliminaries in \S\ref{sec:model_and_prelim}, introduce a projection operation and give some technical results in \S\ref{sec:projection_smaller_subset_voters}, characterize the outcome of different schemes in \S\ref{Sec:nash_outcomes}. %{\color{blue} an overview of different bargaining schemes in \S~\ref{overview_barg_section}}, 
We derive the distortion of the class of median schemes in %after one round of deliberation in
\S\ref{sec:distort_Nash_relaxed}. We derive the distortion under $\mathfrak{n}_{rand}$ in \S\ref{sec:triadic_deliberation}. We give empirical results on real-world Participatory Budget (PB) data
in \S\ref{sec:exp}, and discuss project interactions in  \S\ref{sec:TD_project_interaction}.

%Then we discuss the second setting, i.e., the menu-based ballot design in Section~\ref{sec:MBPB}
%%%%%%%%%%%%%%%%%%%%%%%%%%%%%%%%%%%%%%

\section{Model and Preliminaries}
\label{sec:model_and_prelim}
%\subsection{Budget Space and Incremental Allocation}
%\label{sec:utility}
Suppose we have $m$ projects and are required to design a budget. A \emph{budget} denotes the fraction of the total funds that are to be spent on each project.  Projects have a maximum possible allocation or ``project costs.'' All votes respect these \emph{project costs}, and consequently, the outcomes of all our mechanisms also respect the project costs.\footnote{There is no constraint on the minimum allocation to a project other than that it must be non-negative.} For notational simplicity, we drop the project costs from the model henceforth and operate under the assumption that project costs are $1$. All our results trivially follow for general project costs.

\begin{definition}
Let $b_j$ denote the funds allotted to project $j$ in budget $b$. We define the \emph{budget simplex}  as the set of valid budgets i.e.,
$\mathbb{B}$ = $\{ b \in \mathbb{R}^m | \sum_{j=1}^m b_j = 1 \text{ and } b_j \geq 0, \forall j \in [m] \}$. 
\end{definition}
%
%\begin{assumption}
%In our model, we do not adopt a fixed cost for a project. We assume that any amount of money, ranging from 0 to all of the funds (normalized to 1) can be allotted to any project, with the constraint that the total funds allotted to all projects sum to 1.
%\end{assumption}
%
 There are $n$ voters, each with a \emph{preferred budget} $v_i \in \mathbb{B}.$ 
A \emph{vote profile} $P$ denotes the list of preferred budgets of all voters, i.e, $P =(v_1, v_2, \ldots, v_n)$. The funds allotted to project $j$ by voter $i$ is $v_{i,j}.$ A vote profile defines an \emph{instance of PB}. The outcome of an instance of PB is a budget in $\mathbb{B}.$
Voters adopt the $\ell_1$ distance as the cost function. (Not to be confused with the project costs, which is a different concept here.)
\begin{definition} \label{def:cost} For $a,b \in \mathbb{B},$  the cost of an outcome $b$ for a voter with preferred budget $a$  is $d(a,b) = \sum_{j=1}^{m} |a_j - b_j|.$ The sum of cost over all budgets, $\sum_{i\in [n]} d(v_i,b),$ is the social-cost of budget $b$. 
\end{definition}

%Since it is easier to work with a notion of utility instead of a cost, 
We define the \emph{overlap utility} which is closely related to the \emph{cost}. Note that this notion of overlap utility has been studied in knapsack voting \cite{goel2019knapsack,freeman2021truthful}.

%Given two voters with preferred budgets
%$a,b \in \mathbb{B}$, we define their overlap utility as follows.
%
\begin{definition}[Overlap Utility]{\label{def:ou}} $u(a,b) = \sum_{j=1}^{m} \min(a_j,b_j).$
\end{definition}
%
%Note that this definition implies that overlap utility of one budget with itself is $1$ i.e. $u(a,a)=1~~~~\forall a \in \mathbb{B}$. 
\begin{lemma}
\label{disutility}
For budgets $a,b \in \mathbb{B}$, $d(a,b) = 2 - 2u(a,b)$.
\end{lemma}
A proof is in Appendix \ref{disutility_proof}. Lemma~\ref{disutility} implies that for a voter, maximizing overlap utility is the same as minimizing the cost.
Note that overlap utility is \emph{symmetric}, i.e, $u(a,b) = u(b,a).$ %Also note that $u(a,a)=1$ for all  $a \in \mathbb{B}.$

%%%%%%%%%%%%%%%%%%%%%%%%%%%%%%%

 \subsection{Distortion }
\label{sec:distortion_defn}
%Having characterized the total utility of the three voters involved in a bargaining process, we now look at the total utility of \textit{all} voters. We evaluate the effectiveness of triadic deliberation in maximizing total utility through a measure called \textit{distortion}, which was first used in the context of social choice by Boutilier et al. in \cite{boutilier2015optimal}.

Here we define \emph{distortion,} which we use as a metric to quantify how good a outcome %our bargaining solution 
is in comparison to the optimal solution for minimizing social-cost.
%
% FIXME: "voting mechanism" or "aggregation method"??
%\begin{definition}
%The \textbf{triadic deliberation mechanism} $\Triadic(P)$ is a voting mechanism that takes a vote profile $P$, performs one round of sequential deliberation, and outputs the outcome of the deliberation.
%\end{definition}
%
We define distortion through the \emph{cost} $d(\cdot, \cdot)$. %[Recall Definition \ref{def:cost}],
%\footnote{Often termed as `cost' in literature; called `social cost' when summed over all agents.}
 %of two budgets. The cost is equal to twice of 1 minus the overlap utility [Lemma~\ref{disutility}]. %The proof of Lemma~\ref{disutility} is in Appendix~\ref{disutility_proof}.
\iffalse
\begin{definition}
For a voter with preferred budget $a \in \mathbb{B}$, the \textbf{cost} from budget $b \in \mathbb{B}$ is $d(a, b) = \sum\limits_{i=1}^{m} |a_i - b_i|$.
\end{definition}
%\edit{Note that the disutility $d(a,b)$ is symmetric in $a$ and $b.$ }
\begin{lemma}
\label{disutility}
For any budgets $a$ and $b$, $d(a,b) = 2 - 2u(a,b)$.
\end{lemma}
Lemma~\ref{disutility} implies that the budget maximizing overlap utility is same as the one minimizing the cost.
\fi
%This shows the relationship between disutilities and overlap utilities. It measures how far two budgets are from each other. The proof is in the Appendix.
%
\begin{definition} 
\label{def:distortion}
The \textbf{distortion} of budget $b$ for vote profile $P$ is
\[ \Distortion_once_P(b) = \frac{\sum_{v \in P} d(v, b)}{\min_{b^* \in \mathbb{B}} \sum_{v \in P} d(v, b^*)}. \]
\end{definition}
Let $h(P)$ be the output of mechanism $h$ for vote profile $P$.
\begin{definition}
The distortion of a class of voting mechanisms $\mathcal{H}$ is:
$ \Distortion_once(\mathcal{H})  = \sup_{n \in \mathbb{Z}^+,~P \in \mathbb{B}^n,~h \in \mathcal{H}} \mathbb{E}[\Distortion_once_P(h(P))].$
\end{definition}

\iffalse
We now define the distortion for classes of voting mechanisms. 
\begin{definition}
The worst case distortion for class $\mathcal{H}$ is
$$\SupDistortion(\mathcal{H}) = \sup_{n \in \mathbb{Z}^+,~P \in \mathbb{B}^n,~h \in \mathcal{H}} \mathbb{E}[\Distortion_once_P(h(P))]$$ 
\end{definition}
\fi

%Further we also define and the average distortion for class  $\mathcal{H}$ is
%$$\ExpDistortion(\mathcal{H}) = \sup\limits_{n \in \mathbb{Z}^+, P \in \mathbb{B}^n} \mathbb{E}_{h \sim \mathcal{H}}[\Distortion_once(h(P))]. $$

% FIXME: What is P? Is it B^|N|? Why use |N|?
% FIXME: What if the denominator is 0?

%\edit{Did we define $h(P)$?}

Note that distortion is defined as a supremum over all instances of PB and all mechanisms in class $\mathcal{H}.$\footnote{We will often study the distortion of a single mechanism, i.e., not a class. In that case, $\Distortion_once(h)$ simply denotes the distortion of the mechanism $h$.}
%
%In $\SupDistortion(\mathcal{H}),$ we consider the worst case distortion over all voting mechanisms in the class $\mathcal{H}.$ 
The expectation is over the randomness of the mechanism, which also includes the randomness in the selection of voters. % for deliberation at each step.

The distortion of a voting mechanism is widely used to evaluate its performance regarding how close its output is to the social cost-minimizing outcome in expectation \cite{anshelevich2017randomized, meir2021representative, aziz2022approximate,  garg2019iterative, fain2019}.
%Naturally, it is desired for voting mechanisms to have a small distortion. 
The  \emph{Random Dictator} \cite{anshelevich2017randomized} voting mechanism has a distortion of $2,$ as shown in Lemma \ref{dist2}. A proof is given in Appendix \ref{proof_dist2}. %Lemma~\ref{dist2}, proved in Appendix~\ref{proof_dist2}, gives a lower bound on the distortion of all voting mechanisms that are constrained to choose an outcome from the vote profile. Lemma~\ref{dist3} also gives a loweer
 %The proof is by an example and is in Appendix~\ref{proof_dist2}.

\begin{lemma}
\label{dist2}
Any aggregation method constrained to choose its outcome as the preferred budget of a uniformly randomly chosen voter has distortion $2$.
\end{lemma}
%

 %\subsection{Distortion of Deliberation with two voters} 

Now, consider a mechanism that chooses the outcome via the deliberation between two voters chosen uniformly at random with preferred budgets $a$ and $b$. Within this class, we consider mechanisms constrained to choose the outcome as a convex combination of budgets $a$ and $b$.

Now, consider a mechanism constrained to choose the outcome as a linear combination of budgets $a$ and $b$ where $a$ and $b$ denote the preferred budgets of randomly sampled voters. That is, $\alpha(P) a + (1-\alpha(P)) b$ for $\alpha(P) \in [0,1].$ Note that $\alpha(P)$ may be optimized over the entire vote profile.%This line segment corresponds to linear combinations of $a$ and $b$ and 
\footnote{All such outcomes maximize the sum of the overlap utilities of the deliberating agents.} 
%\footnote{$z = (a+b)/2$ also maximizes the product  $u(a,z)\times u(b,z)$.}  
We refer to this class of mechanisms as \emph{Random Diarchy} and denote it by $\mathcal{Q}$. Interestingly, the distortion of $\mathcal{Q}$ is 2, the same as that of \emph{Random Dictator}. % Note that any point on this line is also a valid budget in $\mathbb{B}$ and denote all such schemes by $\mathcal{Q}$. 
%
%We now prove the following lemma.
%
\begin{lemma}\label{distortion_two_bargaining} For Random Diarchy %$\mathcal{Q}$,
$\inf_{\mathfrak{q} \in \mathcal{Q}} \Distortion_once(\mathfrak{q}) = 2.$
%$\sup\limits_{n \in \mathbb{Z}^+,~P \in \mathbb{B}^n} \mathbb{E}%_{a,b \sim P}
%[\Distortion_once_P(\mathfrak{q}(P))] \geq 2 \forall \mathfrak{q} \in \mathcal{Q}$
\end{lemma}

%Lemma~\ref{distortion_two_bargaining} states that for any voting mechanism $\mathfrak{q} \in \mathcal{Q}$, there exists a vote profile $P$ for which the distortion is 2. We prove this lemma in Appendix \ref{distortion_two_bargaining_proof_section} with an example.

 A proof is given in Appendix \ref{distortion_two_bargaining_proof_section}.%Lemma~\ref{distortion_two_bargaining} is similar in essence to the lower bound result of \cite[Theorem 1]{goel2016towards}. In \cite{goel2016towards} they study an opinion formation model where the opinions of the voters evolve over multiple rounds. However, at each round the opinion of each voter updates to a function of the current opinion of the sampled voters at that round and its opinion in the previous round with the opinions lying on a real number line. However in {\it Random Diarchy}, the coefficients $\alpha(P)$ may be optimized over the entire vote profile $P$.
%which constrains the outcome to only depend on the preferred alternatives of the two uniformly sampled voters at every round, and the outcome of the previous round. However, they constrain the preferred alternatives of each voter to lie on a real line, unlike the simplex budget space in our PB model.
\remove{$a$ and $b$ (and not have $\alpha$ as a function of the entire vote profile $P$). }
%
%An important point to note is that \cite{goel2016towards} study  Lemmas~\ref{dist2} and~\ref{distortion_two_bargaining} together show the importance of including the consideration of a third budget in the deliberation process towards improving the distortion below the natural barrier of $2$. %In this paper, we do precisely this.
%
%$$ \Distortion_once(h)  = .$$
%
We further show that \emph{Random Referee} scheme described in \cite{fain2017} where one of the two preferred budgets of the bargaining voters is chosen based on the preferred budget of third sampled voter also has a distortion ratio of at least 2 in Lemma~\ref{distortion_random_referee}, proven in Appendix \ref{distortion_random_referee_proof_section}. We denote the class of such mechanisms by $\mathcal{R}$.

\begin{lemma}\label{distortion_random_referee} For Random Referee %$\mathcal{Q}$,
$\inf_{\mathfrak{q} \in \mathcal{R}} \Distortion_once(\mathfrak{q}) \geq 2.$
%$\sup\limits_{n \in \mathbb{Z}^+,~P \in \mathbb{B}^n} \mathbb{E}%_{a,b \sim P}
%[\Distortion_once_P(\mathfrak{q}(P))] \geq 2 \forall \mathfrak{q} \in \mathcal{Q}$
\end{lemma}

 \subsection{Model of preference aggregation\remove{Sequential Deliberation}}
\label{sec:model_SD}
%Let there be $n$ voters and $m$ projects. Define the \emph{budget space} $\mathbb{B}$ as $\{ b \in \mathbb{R}^m | \sum_{i=1}^m b_i = 1 \text{ and } b_i \geq 0, \forall i \in [m] \}$. \edit{%It is the space of budgets with sum of allocations to projects at most $1.$
% In our model,} each voter $i$ has a preferred budget $v_i \in \mathbb{B},$ \edit{and adopts the overlap utility function. Define a \emph{vote profile} $P$ as the set of preferred budgets of all $n$ voters.} 

\remove{
 The \emph{sequential deliberation} mechanism runs a pre-defined number $T$ rounds of bargaining between randomly chosen pairs of voters with the disagreement point given by the outcome of the previous round. Formally the mechanism is:
%A \emph{budget} $b$ is a vector of $m$ elements such that $\sum_{i=1}^m b_i = 1$ and $b_i \geq 0$ for all $i \in [m]$.
%Assume that each voter $i$ has a \textbf{preferred budget} $v_i \in B$. %(FIXME: $b_i$ can also mean the $i$ element of $b$.)

%Formally the mechanism is as follows:
%The \emph{sequential deliberation} mechanism is as follows:
\begin{enumerate}
    \item Pick a voter $i$ uniformly at random. Set the disagreement point for the deliberation $c$ to their preferred budget $v_i.$
    \item Repeat the following process $T$ times, %\edit{where $T$ is the number of rounds of deliberations, set by the PB organizer} :
    \begin{enumerate}[leftmargin = 0.1 cm]
        \item Pick two voters \edit{independently and} uniformly at random with replacement. They bargain with $c$ as the disagreement point.
        \item Set the disagreement point $c$ to the outcome of the bargaining.
    \end{enumerate}
    \item The outcome of the process is $c$.
\end{enumerate}
}

Let us define the mechanism formally in steps and we call it $\textbf{Triadic scheme}$.

\begin{enumerate}%[leftmargin = 0.1 cm]
    \item Pick a voter $i$ uniformly at random and set the disagreement point $c$ as the preferred budget of voter $i$.

    \item Now choose two voters $a$ and $b$ uniformly at random with replacement and they bargain with $c$ as the disagreement point. 
\end{enumerate}

%We will see later from the empirical results that $T$ can be reasonably small, and we will prove that even when $T=1$, the outcome has a small distortion, even for the worst case of PB instances. This motivates us to define the triadic scheme.

\remove{

\begin{definition}{\label{def:triadic_deliberation}}
\textbf{Triadic scheme}, $\Triadic_{\mathcal{A}}(P),$ is a voting mechanism that takes a vote profile $P$, and outputs the result under the bargaining scheme $\mathcal{A}$.\remove{of one round of sequential deliberation under bargaining scheme $\mathcal{A}$.}
\end{definition}
}

All our theoretical results in this paper are for the outcome of the triadic scheme. However, as discussed in \cite{fain2017}, we can extend this bargaining scheme to multiple rounds by setting the outcome of the previous round as the disagreement point for the next round and sampling the two bargaining voters uniformly at random without replacement. We provide empirical results for this setup for multiple rounds (upto 10 rounds) in \S~\ref{sec:exp}.

Our bound on pessimistic distortion assumes that the voters are chosen with replacement as done in \cite{fain2019}. This directly gives us a bound on the distortion when voters are sampled without replacement. It is easy to see that the difference in these bounds is of $O(\frac{1}{n}).$ The case where two or more identical voters are sampled out of three defaults to the {\it Random Dictator} mechanism, which has constant distortion -- the probability of this event is of $O(\frac{1}{n})$.

%his scheme is identical to the 

%However, we can show that the probability that two voters sampled with replacement are identical is very small $O\left(\frac{1}{N}\right)$ where $N$ denotes the total number of voters. However, when two of three voters are identical, the deliberation scheme is the same as that of a {\it random dictator} whose distortion is exactly 2. Thus, the bound on distortion under this setup where voters are sampled without replacement would differ from 1.66 only by $O(\frac{1}{N})$. 

%thus another alternate strategy of sampling the voters without replacement would have a minimal change in the distortion guarantee when the number of voters is reasonably large.

%Note that this sampling strategy is in line with \cite[Definition 3]{fain2017} where voters are sampled with replacement and this is necessary for our proof as a key strategy in our proof is pessimistic distortion(PD). 

%However, we give empirical results for multiple rounds of sequential deliberation (up to 10 rounds) in Appendix \S~\ref{sec:exp}.

We consider bargaining schemes satisfying one or more of the following constraints, namely a) Pareto efficiency, b) Invariance to Affine Transformation, c) Symmetry, and d) Independence of Irrelevant Alternatives. Bargaining schemes that satisfy all of these constraints are the class of Nash bargaining schemes denoted by $\mathcal{N}$  \cite{binmore1986nash}.
%and e) {\color{red} maximizing utility of the disagreement point budget ($u(z,c)$) amongst bargaining solutions with same utilities of two bargaining budgets $a$ and $b$ ($u(z,a)$ and $u(z,b)$) }

%We assume that voters bargain under the assumptions of Pareto efficiency, invariance to affine transformation, symmetry, and {\color{red} maximizing utility of the disagreement point budget in case of a tie}. Let us denote all such bargaining schemes by $\mathcal{A}$. We also consider another bargaining model where we add the extra constraint of independence of irrelevant alternatives which would precisely correspond to Nash bargaining solution \cite{binmore1986nash} and denote such set of bargaining solutions by $\mathcal{N}$.

\remove{
\begin{definition}{\label{relaxed_bargaining}}
    We denote $\mathcal{A}$ as the set of all bargaining schemes satisfying Pareto efficiency, invariance to affine transformation, symmetry, and {\color{red} maximizing utility of the disagreement point budget in case of a tie}.
\end{definition}
}

%We also add an extra constraint of independence of irrelevant alternatives to the set of conditions mentioned in Definition \ref{relaxed_bargaining} above and observe that it precisely corresponds to Nash Bargaining solution \cite{binmore1986nash}, also denoted by $\mathcal{N}$.

%Let $\mathcal{A}$ denote the set of all bargaining schemes satisfying Pareto efficiency, invariance to affine transformation, symmetry, and {\color{red} maximizing utility of the disagreement point budget in case of a tie}. 

%Therefore, the outcome from the bargaining corresponds to the Nash bargaining solution \edit{ \cite{binmore1986nash}.}

%Therefore, the outcome from the bargaining corresponds to the Nash bargaining solution \edit{ \cite{binmore1986nash}.} %\edit{ Recall that the disagreement point is the outcome if no agreement is reached.} 
\begin{definition} \label{def:nb}
An outcome of $\mathcal{N}(a, b, c)$, the \textbf{Nash bargaining} between two voters with preferred budgets $a$ and $b$ and the disagreement point $c$, is a budget $z$ which maximizes the Nash product $(u(a, z) - u(a, c)) \times (u(b, z) - u(b, c))$, subject to individual rationality $u(a, z) \geq u(a, c)$ and $u(b, z) \geq u(b, c)$, and in case of a tie between possible outcomes, maximizes $u(c, z)$.
\end{definition}
 
 The fact that $\mathcal{N}$ breaks ties in favor of the disagreement point is crucial for the distortion of triadic scheme with bargaining schemes in $\mathcal{N}$  to be smaller than $2$. It is also crucial for the membership of $\mathcal{N}$ in a class of bargaining schemes that maximize the sum of overlap utilities of the bargaining agents and the disagreement point. We now define this class of bargaining schemes. 

\begin{definition} \label{def:mb} $ \mathcal{M}$ is the class of \textbf{median schemes} if
any outcome $z \in \mathcal{M}(a, b, c)$ maximises the sum of utilities with budgets $a,b$ and $c$ i.e. $u(z,a)+u(z,b)+u(z,c)$.
\end{definition}

%Note that we do not specify $T$ (the number of rounds). The choice of $T$ is up to the PB organizer to decide when to stop the process. 
%

%We show that every mechanism in $\mathcal{N}$ is a median voting rule.
%
The following important result is proved in Appendix \ref{max_utility_Nash_proof_section}. 
\begin{theorem}{\label{max_utility_Nash}}
Every scheme in $\mathcal{N}$ is also a median scheme i.e. $\mathcal{N} \subseteq \mathcal{M}$
\end{theorem}

%That is, for any budgets $a$, $b$, and $c$, any outcome $z$ of $\mathcal{N}(a, b, c)$ maximizes the sum of overlap utilities $u(a, z) + u(b, z) + u(c, z)$.

%We prove this theorem in \S \ref{max_utility_Nash_proof_section}.

%In the subsequent sections, we would consider upper bounds on the distortion defined in \ref{sec:distortion_defn} for any relaxed Nash bargaining scheme in $\mathcal{N}$
%and a randomized bargaining scheme amongst schemes in $\mathcal{N}$. In the next section, we discuss the ability of triadic scheme to handle project interactions without input from the PB designer. 
%%%%%%%%%%%%%%%%%%%%%%%%%%%%%%%%%%%%%%

\section{Incremental allocation space}
\label{sec:projection_smaller_subset_voters}

We now give a function that captures the marginal preferences of a subset of  voters $S$ regarding the allocation to project $j$, relative to the preference of the other voters (i.e., $P\setminus S$). This function will be useful as an analytical tool in the paper. Specifically,

\begin{definition}{\label{proj_budget_defn}}
Given a vote profile $P=(v_1, v_2, \ldots, v_n),$ and project $j,$ the \textbf{ incremental project allocation} $X_{j,P}: 2^{[n]} \rightarrow [0,1]$ maps a subset of budgets $S$ to \newline
 $$X_{j,P}(S) = \max \left(\left(\min\limits_{i \in S} v_{i,j} \right) - \left(\max\limits_{i \in P \setminus S} v_{i,j} \right), 0 \right).$$
\end{definition}
%{\color{red} Comment: Define $v_{i,j}$ appropriately\\}
Here $\max$ and $\min$ over $\emptyset$ are defined as 0 and 1, respectively.
$X_{j, P}(S)$ denotes the amount by which the budgets in $S$ \emph{all} agree on \emph{increasing} the allocation to project $j$ above the \emph{maximum} allocation to $j$ by any budget in $P \backslash S$. Summing this quantity over all projects $j \in [m]$ gives us $X_P(S),$ which is defined in the following. 

%Note that we may informally think of the quantity $g_{j, P}(S)$ as what fraction of project $j$ is only allotted to budgets in $S$ but no budget in $P\backslash S$. We may get a sense of this notion from Corollary~\ref{sum_each_proj} as the summation of this quantity over all sets $S$ is 1.

\begin{definition}{\label{common_budget_defn}}
For a vote profile $P,$ the \textbf{incremental allocation} $X_P: 2^{[n]} \rightarrow \mathbb{R}$ is
 $X_P(S) = \sum_{j=1}^m X_{j, P}(S)$ for all $S \subseteq P.$ 
 %\footnote{Max and min over an empty set is defined as 0 and 1 respectively.}
\end{definition}

 We use $X_{P}(.)$ in \S\ref{sec:distort_Nash_relaxed} since its complexity is  dependent only on the number of voters $n$ and not on the number of projects $m$. This helps us give  results valid for arbitrarily large values of $m$. We illustrate the functions $X_{j,P}(\cdot)$ and $X_P(\cdot)$
 in the following example.
%We also define the following expression which informally captures what fraction of project $j$ is allotted to budgets in $S$ but no budget in $P\backslash S$.

%And also we can note from definition that $g_P(S) = \sum\limits_{j=1}^{m} g_{j,P}(S)$.

%Also we $$

%Also, since $$

%Thus, we have $g_P(S)$%Summing over all such projects gives us $g_P(S)$.}

%(Define $g(P)_\emptyset = 0$.)  % should probably be \infinite ? Doesn't matter as we never use it.

%See that $g_P(S)$ tells us how much allocation budgets in $S$ have in common, over and above the allocation in $P\backslash S$. 
\begin{example}{\label{example_only_budget}}
 Consider an instance of PB with three projects and a vote profile $P$ with three budgets $a = \langle 1, 0, 0 \rangle$, $b = \langle 0, 1, 0 \rangle$, and $c = \langle 0.25, 0.25, 0.5 \rangle$. %Let $X(S) = g_{P}(S)$ for each subset $S$ of $P$.
 \footnote{For brevity, we omit braces and commas in the argument of $X$.} 
 Then, $X_{1,P}(a) = 0.75.$ This is because the budget $a$ has allocation $1$ to project $1,$ out of which only $0.75$ is \emph{incremental} on top of $\max(b_1,c_1)$. Also, $X_{2,P}(a) = X_{3,P}(a) = 0.$  As a result, $X_P(a) = 0.75.$ Similarly $X_{2,P}(b) = X_P(b) = 0.75.$ Also, $X_{3,P}(c) = X_P(c) = 0.5$. Further, $X_{P}(ac) = 0.25$. This is because the subset $\{a,c\}$ has a minimum allocation of $0.25$ to project $1$ among themselves. It is also incremental since $b_1 = 0$. Further, we have $X_{1,P}(abc) = X_{2,P}(abc) = X_{3,P}(abc) = X_{P}(abc) = 0.$ This is because the group of all three budgets has no allocation that is common to all. Finally, $X_P(\emptyset) = X_{3,P}(\emptyset) = 0.5$ because no budget allocated funds more than 0.5 to project 3.
\end{example}

We use $\mathcal{P}(P)$ to denote the power set of $P$. We now give an important corollary regarding the function $X_{j,P}(\cdot)$.% We will give results on incremental budget space in \S~\ref{sec:projection_smaller_subset_voters}. %, i.e., the set of all subsets of $P$. %We now give a corollary for the function $g_P(S)$. 

\begin{corollary}{\label{sum_each_proj}}
$\sum_{S \in \mathcal{P}(P)} X_{j,P} (S) = 1,$  $ \forall j \in [m].$
\end{corollary}
\remove{

\begin{corollary}{\label{monotonicity_proj_budget_space}}
    For every project $j \in [m],$ and two subsets of $P$ satisfying $ S \subset \hat{S},$ 
    %s.t. $\hat{S},S \in \mathcal{P}(P)$, 
    we have $X_{j,P}(\hat{S}) \leq X_{j,P}(S)$.
\end{corollary}

\begin{corollary}{\label{total_allocation_budget_space}}
{\color{red}  $\sum_{S \in \mathcal{P}(P)| S \ni v_i}  X_{j,P}(S) = v_{i,j}~\forall j \in [m]$. Summing over all $j \in [m]$, we get  $\sum_{S \in \mathcal{P}(P)| S \ni v_i}  X_{P}(S) = 1$ since total allocation by any voter is $1$ as $v_i \in \mathbb{B}$. }
\end{corollary}

\iffalse
\begin{corollary}{\label{new_budget_z_ineq}}
    $g_{j,P}(S) \geq X_{j,P\cup z} (S \cup z)$ for all vote profiles $P,$ budgets $z \in \mathbb{B},$ and projects $j \in [m].$ 
    
    This result also implies that $g_{P}(S) \geq g_{P\cup z} (S \cup z)$.
\end{corollary}
\begin{corollary}{\label{sum_budget_z}}
    $\sum\limits_{S \in \mathcal{P}(P)} g_{P\cup \{z\}} (S \cup \{z\})=1$ for all $z \in \mathbb{B}.$
\end{corollary}
\fi

 %Definition \ref{common_budget_defn} has the following implication. 
\remove{Since,\\ $\sum\limits_{S \subseteq P} \max \bigg(\Big(\min\limits_{s \in S} v_{s,j} \Big) - \Big(\max\limits_{\tilde{s} \in P \setminus S} v_{\tilde{s},j} \Big), 0 \bigg) =1$ 
\footnote{Can be proved by sorting $\{v_{i,j}\}_{i=1}^{k}$ in descending order and showing that this quantity is non-zero only when $S$ exactly contains top $p$ elements for some $p\leq n$}, we can alternatively think of this quantity as the sum over all fractions of projects
that is allotted to all budgets in $P$, but no budget is $P\backslash S$.}

% over the subsets of $P$.

}

A proof is given in Appendix \ref{app_sum_each_proj}. \remove{~\ref{app_sum_each_proj}} Corollary \ref{sum_each_proj} says that every incremental allocation to project $j$ by budgets in $S$ adds up to 1 when summed over all subsets $S$ (this includes the empty set; $X_{j,P}(\emptyset) >0$ implies that no voter allocated the full 1 unit budget to project $j$). %This result captures, in the space of incremental project allocations $X_{j,P}(\cdot)$, the fact that the maximum allocation to a project can be 1. 

%Recall the incremental project allocation function $X_{j,P}(.)$ as described in Definition \ref{proj_budget_defn} w.r.t a vote profile $P= (v_1,v_2,\ldots,v_n)$.
\subsection{Projection On Incremental Allocations}
%We now give a method to construct the incremental project allocation function w.r.t. a vote profile $Q \subseteq P$ given $\{X_{j,P}(S)\}_{S \in \mathcal{P}(P)}$. 
We now give a \emph{projection} of $X_{j,P}$ from $P$ to $Q\subseteq P$ to get $X_{j,Q}.$
This operation has two applications in this paper. First, it enables us to study the allocations of an outcome $z$ relative to the vote profile $P$ by making projections from $P\cup \{z\}$ to $P.$ Second, it is used to study the outcomes of bargaining with a subset $Q\subseteq P$ of voters with respect to the entire vote profile $P$ via projections from $P$ to $Q$.

\begin{lemma}{\label{proj_budget_new_space_lemma}}
    For any vote profile $P$ and $Q \subseteq P$, the \textbf{projection from $P$ to $Q$} is
    $X_{j,Q}({S}) = \sum_{\hat{S} \in \mathcal{P}(P \setminus Q)}X_{j,P}(\hat{S} \cup {S}) $ for all ${S} \in \mathcal{P}(Q),$ and all $j \in [m]$.
    Summing over $j \in [m]$, $X_{Q}({S}) = \sum_{\hat{S} \in \mathcal{P}(P \setminus Q)}X_{P}(\hat{S} \cup S).$
\end{lemma}

A proof is given in Appendix \ref{proj_budget_new_space_lemma}. \remove{~\ref{proj_budget_new_space_proof}.}
Lemma~\ref{proj_budget_new_space_lemma} captures an important technical fact. To calculate the incremental project allocation function on $S$ over a vote profile $Q\subseteq P$, i.e., $X_{Q}(S)$, we may sum $X_P(\cdot)$ over all subsets of budgets in $P$ which contain all elements of $S$ but no element of $Q\setminus S$. Note that here $S \subseteq Q \subseteq P.$

We now consider the problem of analyzing an outcome $z$ with the help of the incremental allocation function. Towards this, we define the function $Z_{j,P}(S)$ with respect to an outcome $z$ with the help of the projection operation described in Lemma~\ref{proj_budget_new_space_lemma}. %\footnote{ This function will be useful in the optimization problem in Equation~\eqref{opti_relaxed} in \S~\ref{sec:distort_Nash_relaxed}.

%With this definition in mind, we now add a new outcome budget $z$ to the vote profile $P$ and define $Z_{j,P}(S)$ below which is used to compare with the incremental allocation space $X_{j,P}(S)$}

%We now give an important technical result on the project-specific budget function $g_{j,P}(\cdot)$.

%A generalization of Lemma~\ref{utility_common} on the  incremental project allocation function when we construct the vote profile $Q$ s.t $Q\subseteq P$ is given in the following lemma.

\remove{
\begin{lemma}{\label{proj_budget_new_space_lemma}}
    For any vote profile $P=(v_1,v_2,\ldots,v_n)$ and $Q \subseteq P$, we have 
    $X_{Q}(\mathfrak{S}) = \sum\limits_{S \in \mathcal{P}(P \setminus Q)}X_{P}(S \cup \mathfrak{S}) $ for all $ \mathfrak{S} \in \mathcal{P}(Q).$
\end{lemma}
}

%Note that Lemma~\ref{proj_budget_new_space_lemma} gives a construction of the incremental common allocation function over a new vote profile $Q$ given a incremental common allocation function over a vote profile $P \supseteq Q$. 

\begin{definition}{\label{new_budget_proj_budget_defn}}
    For vote profile $P$ and budget $z$, define $Z_{j,P}: 2^{[n]} \rightarrow [0,1]$ as $Z_{j,P}(S) = X_{j,P\cup \{z\}}(S \cup \{z\})$ $\forall S \subseteq P.$ %[defined in Definition \ref{proj_budget_defn}]
\end{definition}

\remove{\color{red} $Z_{j,P}(S)$ captures the quantity by which budgets in $S$ \emph{and} $z$ agree on increasing the allocation to project $j$ above the maximum allocation to project $j$ by any budget in $P\setminus S$. Summing this quantity over all projects $j \in [m]$ gives us $Z_P(S)$.}

Recall from Definition~\ref{proj_budget_defn} that $X_{j,P}(S)$ denotes the amount by which \emph{all} budgets in $S$ want to increase the allocation to project $j$ over the maximum allocation to $j$ by any budget in $P \setminus S.$ The quantity $Z_{j,P}(S)$ denotes the amount by which the outcome budget $z$ ``accepts" this preference of $S$. Naturally, $Z_{j,P}(S) \leq X_{j,P}(S).$ 

Analogous to summing $X_{j,P}(S)$ over all $j \in [m]$ to get $X_P(S),$ we can sum $Z_{j,P}(S)$ over all $j \in [m]$ to get $Z_P(S)$. %Formally:

\remove{
{\color{red} $Z_{j,P}(S)$ intuitively captures what quantity of excess allocation to project $j$ by budgets in $S$ over budgets in $P\setminus S$ (defined as $X_{j,P}(S)$) is "accepted" by outcome budget $z$. More formally, it denotes the minimum of excess allocation by budget $z$ to project $j$ over maximum allocation to project $j$ by budgets in $P\setminus S$ and incremental allocation $X_{j,P}(S)$.}
}

%Corollaries \ref{new_budget_z_ineq}\remove{, \ref{monotonicity_proj_budget_Z_space}} and \ref{sum_proj_budget_z} give some results comparing $Z_{j,P}(S)$ and incremental budget allocation space $X_{j,P}(S)$.

\begin{definition}{\label{new_budget_common_budget_defn}}
    For vote profile $P$ and budget $z$, define $Z_P(S): 2^{[n]}\rightarrow [0,1]$ as $Z_P(S) = \sum_{j =1}^{m} Z_{j,P}(S)$  $ \forall S \subseteq P.$
\end{definition}

$Z_{P}(S)$ informally denotes the amount by which outcome budget $z$ ``accepts" the preference of $S$ for increasing allocations above the allocations of $P\setminus S$ across all projects. See that $Z_{P}(S) \leq X_{P}(S).$

%We can now state the following results comparing $Z_{j,P}(S)$ with $X_{j,P}(S)$  [Definition~\ref{proj_budget_defn}] which will be useful as constraints in the optimization problem ~\eqref{opti_relaxed} in \S~\ref{sec:distort_Nash_relaxed}. %This effectively tries the characterize the addition of a new budget on the incremental budget space.

\begin{corollary}{\label{new_budget_z_ineq}}
    For any vote profile P and budget $z$, $Z_{j,P}(S) \leq X_{j,P} (S)$ for all $j \in [m].$ Summing over $j\in [m]$, $Z_{P}(S) \leq X_{P} (S)$.
\end{corollary}

\begin{proof}
Follows directly from Lemma~\ref{proj_budget_new_space_lemma} since $X_{j,P}(S) = Z_{j,P}(S) + X_{j,P\cup \{z\}}(S)$ (we are projecting from $P\cup \{z\}$ to $P$).
\end{proof}
%Note that this results says the incremental project allocation common to a new budget $z$ is bounded by the incremental project allocation for every subset of voters $S$.

\begin{corollary}{\label{sum_proj_budget_z}}
$\sum_{S \in \mathcal{P}(P)} Z_{j,P} (S)=z_j$ for all vote profiles $P$ and $z \in \mathbb{B}.$ Summing over all projects $j \in [m]$, we get $\sum_{S \in \mathcal{P}(P)} Z_{P} (S)=1.$
\end{corollary}

\begin{proof}
 We have $z_j=X_{j,\{z\}}(\{z\})$ [Definition~\ref{proj_budget_defn}]. Apply Lemma~\ref{proj_budget_new_space_lemma} by doing a projection from $P \cup \{z\}$ to $\{z\}.$
\end{proof}

This result captures, in the incremental common budget space, the fact that the total funds allocated by a budget $z$ to projects $j \in [m]$ is 1. %Corollary~\ref{new_budget_z_ineq} intuitively says that for every subset of budgets $S$, $Z_{j,P}(S)$ is at most the incremental project function $X_{j,P}(S)$. 
The following example illustrates $Z_{j,P}(S)$.

\begin{example}
Consider vote profile $P=\{a,b,c\}$ with two projects. Let the budgets $a,b,$ and $c$ be $\langle 0.2,0.8 \rangle, \langle 0.5,0.5 \rangle,$ and $\langle 0.8,0.2 \rangle$ respectively. Let the outcome budget $z$ be $\langle 0.4,0.6 \rangle$. In this case, $X_{2,P}(ab)=0.3$ and $Z_{2,P}(ab)=0.3$ since the excess allocation by outcome $z$ to project $2$ over the allocation by budget $c$ (i.e., 0.4) is larger than the least excess allocation to project $2$ by budgets $a$ and $b$ over allocation in budget $c$ (i.e., $X_{2,P}(ab)$ which is 0.3). In other words, the entire incremental allocation to project 2 by budgets $a$ and $b$ is accepted by outcome $z$. However, $X_{2,P}(a)=0.3$ but $Z_{2,P}(a)=0.1$ since the incremental allocation to project $2$ by budget $z$ over budgets $b$ and $c$ is 0.1. Thus only a partial incremental allocation to project $2$ by budget $a$ is "accepted" by budget $z$.
\end{example}

%{\color{red} Now, we can observe that $Z_{2,P}(ab) = 0.3$, $Z_{2,P}(a)=0.1$ with $X_{2,P}(ab)=0.3$ and $X_{2,P}(a)=0.3$.}

%Note that we have $0<Z_{2,P}(a) = < X_{2,P}(a)$ and thus $X_{2,P \cup \{z\}}(ac \cup \{z\})=g_{2,P}(ac)=0$ and $g_{2,P \cup \{z\}}(ab \cup \{z\})=g_{2,P}(ab)=0.3$ and $g_{2,P \cup \{z\}}(ab \cup \{z\})=g_{2,P}(ab)=0.2$ thus getting a sense of Lemma~\ref{monotonicity_proj_budget_space}.

\remove{
\begin{lemma}{\label{utility_one_voter_proj_budget_Z_space}}
The overlap utility of budget $a \in P$ from budget $z$, is $ u(a,z) = \sum_{S \in \mathcal{P}(P)|S \ni a} Z_P(S).$
\end{lemma}

\begin{proof}
Follows from  Lemma~\ref{utility_common} by defining $a=a$ and $z$ in place of $b$ and using $P\cup\{z\}$ in place of $P$.% we have $u(a,z) = \sum_{S \in \mathcal{P}(P)\vert S \ni a}Z_P(S)$ %which equals $\sum\limits_{S \in \mathcal{P}(P)} |S| \cdot Z_P(S)$ 
\end{proof}

}

\iffalse
We now state a lemma regarding the monotonicity of $X_{j,P}(S)- Z_{j,P}(S)$ w.r.t the subset of voters $S$.

\begin{lemma}{\label{monotonicity_proj_budget_Z_space}}
    For every project $j \in [m],$ and subset $S \subseteq P$, we have $0<Z_{j,P}(S) < X_{j,P}(S)$ implies $Z_{j,P}(\hat{S}) = X_{j,P}(\hat{S})$ for all $\hat{S}$ such that $\hat{S} \supset S$ and $\hat{S} \subseteq P$.
\end{lemma}
\fi

\remove{
\begin{proof}
The result follows from \textcolor{red}{the projection} in Lemma~\ref{proj_budget_new_space_lemma}, since\\ $X_{j,P}(S)=Z_{j,P}(S) + X_{j,P \cup \{z\}}(S \cup \{z\}). $

Now applying Lemma~\ref{monotonicity_proj_budget_space}, we get the desired result.\qedhere  
\end{proof}
}

%The proof of Lemma~\ref{monotonicity_proj_budget_Z_space} is technical and is in Appendix~\ref{sec:monotonicity_proj_budget_space_Z_proof}. 
%Lemma~\ref{monotonicity_proj_budget_space} gives a monotonicity relation between the functions $Z_{j,P}(S)$. 

\iffalse
\begin{lemma}{\label{utility_proj_budget_Z_space}}
The sum of overlap utilities of all budgets in vote profile $P$ with a budget $z$ is $\sum\limits_{v \in P} u(v,z) = \sum\limits_{S \in \mathcal{P}(P)} |S| \cdot Z_P(S).$
\end{lemma}

\begin{proof}
By Lemma~\ref{utility_common}, we have $\sum\limits_{v \in P} u(z,v) = \sum\limits_{v \in P} \sum\limits_{S \in \mathcal{P}(P)\vert S \ni v}Z_P(S)$ which equals $\sum\limits_{S \in \mathcal{P}(P)} |S| \cdot Z_P(S)$.
\end{proof}
\fi

%then every part of project $j$ common to budgets only in a superset of $S$ is definitely present in the budget $z$. This puts a further constraint on $g_{j,P\cup \{z\}} (S \cup \{z\})$ relative to $g_{j,P}(S)$ apart from the constraint in Corollary~\ref{new_budget_z_ineq} and \ref{sum_budget_z}. We give an example of the same below.

%\input{model.tex}
%\input{distortion_measure.tex}
\section{Overview of Median and Nash bargaining schemes} {\label{Sec:nash_outcomes}}
%\input{project_substitution.tex}
%\subsection{Total Utility of Three Voters}
%\subsection{Bargaining under a triplet of budgets}

%Recall bargaining scheme $\mathcal{A}$ given in Definition~\ref{relaxed_bargaining} which satisfies the four properties of bargaining. 
Recall the triadic mechanism from \S~\ref{sec:model_SD} and we characterize its outcome. Let the disagreement point be $c$ and the preferred budgets of the agents chosen randomly for the mechanism be $a$ and $b$. 
%
%In this section we characterize the outcomes of Nash bargaining in terms of incremental allocations $X_{\{a,b,c\}}(\cdot).$ 
For simplicity of notation, we denote $X_{\{a,b,c\}}(S)$ by $\mathfrak{X}(S)$ for $S$ being any subset of $\{a,b,c\}$\footnote{Note that we do not consider $X_P(\cdot)$ in this section where $P$ is the set of the preferred budgets of all the voters, even those not involved in the bargaining.}.
We also denote the outcome budget of the bargaining by $z$ and  $Z_{(a,b,c)} (S)= X_{\{a,b,c,z\}}(S \cup \{z\})$ by  $\mathcal{Z}(S)$ for $S\subseteq \{a,b,c\}$. \remove{ \textcolor{red}{Note that $\mathcal{Z}(S)$ captures the total incremental fraction of projects common to budgets in $S$ and the bargaining solution $z$.}} %Recall that any budget $z \in \mathbb{B}$ satisfies $\sum_{S \subseteq \{a,b,c\}} \mathcal{Z}(S)=1$ [from Lemma \ref{utility_common} and $u(z,z) = 1$] and $\mathcal{Z}(S) \leq \mathfrak{X}(S)$ [from Corollary~\ref{new_budget_z_ineq}]. 

%
% TODO: Explain more about \mathcal{Z}_S = g(a, b, c, z)_{\{z\} \cup S} here.
%
%Recall that $\mathfrak{X}=g(a, b, c)$. Define $\mathcal{Z}_S = g(a, b, c, z)_{\{z\} \cup S}$ for any $S \in \power{\{a, b, c\}}$. Then we have the following result for $\mathcal{Z}.$
%
%
%\begin{theorem}
%For any budgets $a$, $b$, and $c$, and any outcome $z$ of $\Nash(a, b, c)$, $\langle u(a, z), u(b, z), u(c, z) \rangle$ is unique.
%\end{theorem}
%
%\begin{proof}
%This follows from Lemma~\ref{utility_common} and Lemma~\ref{nash_outcome}.
%\end{proof}
%
%
%
%\subsection{Total Utility Maximization}
%
%comment
% mention the knapsack voting paper ... idea of proof from that
%
%\edit{Further,} we show that outcomes of the Nash bargaining maximize the total utility among the three voters involved, \edit{two who are deliberating and one whose preferred budget is the disagreement point}. %The idea of the proof is that if we can show that the Nash bargaining allocates the budget to the portions that voters agree with most, then the total utility is maximized.
\begin{figure}[h!]
\centering
%\begin{subfigure}{.5\textwidth}
  %\centering
  \includegraphics[scale =0.2]{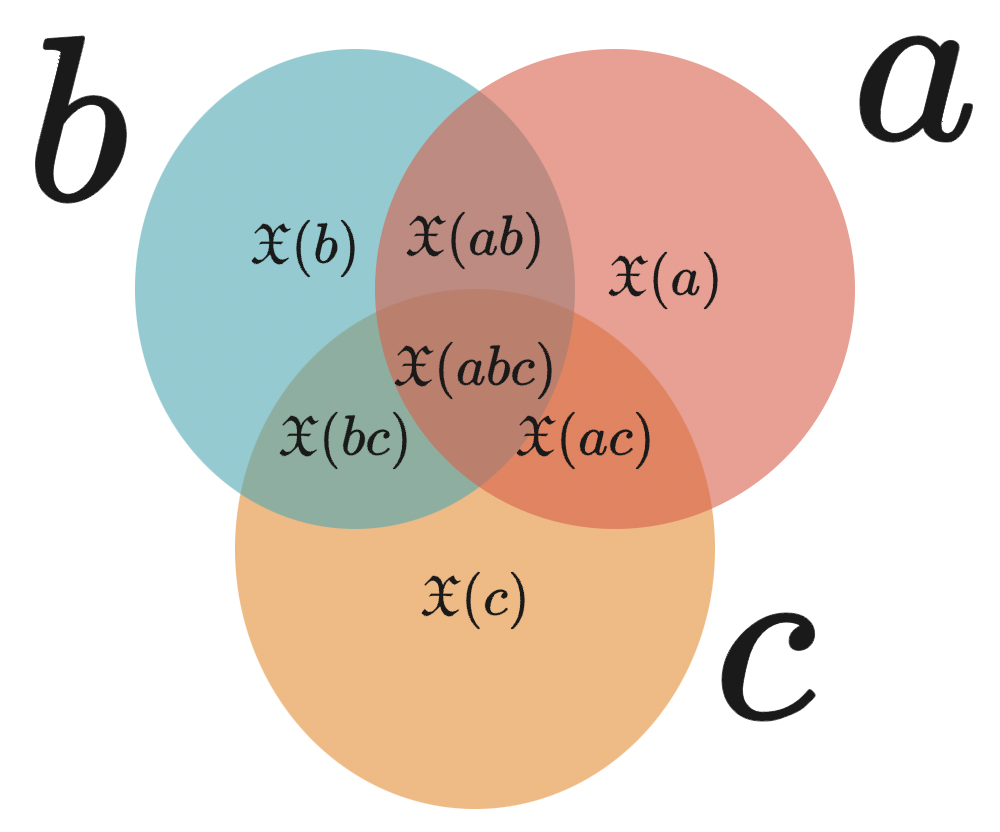}
    \vspace{-0.3cm}
  \caption{Incremental allocations with two preferred budgets of bargaining agents $a$ and $b$,  and disagreement point $c$.}
  \vspace{-0.5cm}
 \label{three_voter_bargaining}
%\end{subfigure}%
\end{figure}
\subsection{Overview of class of schemes $\mathcal{M}$ and $\mathcal{N}$}\label{overview_barg_section}

In Figure ~\ref{three_voter_bargaining}, we illustrate the incremental allocations $\{\mathfrak{X}(S)\}_{S \subseteq \{a,b,c\}}$ with  budgets $a,b,$ and $c$ on a Venn diagram. Recall from Definition \ref{new_budget_common_budget_defn} that $\mathcal{Z}(S)$ denotes what incremental allocation from $\mathfrak{X}(S)$ is "accepted" by outcome $z$. For the construction of $\mathcal{Z}(\cdot)$, the bargaining agents first select all the  allocations "agreed" to by at least two of the three budgets. % into the incremental allocation space $\mathcal{Z}(S).$ 
In Figure ~\ref{three_voter_bargaining}, this corresponds to the area of the overlaps. %as we discuss in \ref{Word:step1} and \ref{Word:step2} below. 
Now, we have two cases, i.e. the total allocation to $\mathcal{Z}$ is less than 1 or exceeds 1. We denote the difference between 1 and the total allocation to $\mathcal{Z}$ by $\textsc{Excess}$.

Consider the case when the total allocation to $\mathcal{Z}(S)$ is less than 1. %which we discuss in \ref{Word:case1} later in this section. 
Here, the agents need to make further allocations worth $\textsc{Excess}.$ Under the class of median schemes $\mathcal{M}$ [described in \S\ref{relax_Nash_defn}], they may select project allocations from $\mathfrak{X}(a),\mathfrak{X}(b),$ and $\mathfrak{X}(c)$ arbitrarily into the outcome $z$  and thus into $\mathcal{Z}(a),\mathcal{Z}(b)$ and $\mathcal{Z}(c)$. In Figure ~\ref{three_voter_bargaining}, this corresponds to the area covered by exactly one of the budgets.
However, under Nash bargaining schemes $\mathcal{N}$, they select allocations worth $\frac{\textsc{excess}}{2}$ from each of $\mathfrak{X}(a)$ and $\mathfrak{X}(b).$ %and select into the outcome $z$.
%And finally under the randomized bargaining scheme $\mathfrak{n}_{\text{rand}}$, we "select" projects with allocations worth $\frac{\textsc{excess}}{2}$ from each of $\mathfrak{X}(a)$ and $\mathfrak{X}(b)$ uniformly at random and put into the bargaining solution $z$. Thus, we have a proportionate "selection" from each of the incremental allocation spaces $\mathfrak{X}(a)$ and $\mathfrak{X}(b)$ into budget outcome $z$ in expectation under scheme $\mathfrak{n}_{\text{rand}}$.

Now, consider the case when the total allocation to $\mathcal{Z}(S)$ is more than 1. %which we discuss in \ref{Word:case2}. 
In this case, under median schemes, $\mathcal{M}$, the participating agents select total project allocations worth $\textsc{Excess}$ arbitrarily from $\mathfrak{X}(ab),\mathfrak{X}(bc)$ and $\mathfrak{X}(ca)$ and remove allocations to these projects. % from the outcome $z$.
In Figure ~\ref{three_voter_bargaining}, this corresponds to the area of the overlap of exactly two budgets.
However, under Nash bargaining schemes $\mathcal{N}$, they select  allocations worth $\frac{|\textsc{Excess}|}{2}$ from each of $\mathfrak{X}(ac)$ and $\mathfrak{X}(bc)$ and remove allocations to these projects from the outcome $z$. %And finally, under the randomized bargaining scheme $\mathfrak{n}_{\text{rand}}$, we "select" projects with allocations worth $\frac{\textsc{excess}}{2}$ from each of $\mathfrak{X}(ac)$ and $\mathfrak{X}(bc)$ uniformly at random and remove allocations to these projects from the bargaining solution $z$. Thus, we have a proportionate "removal" from each of the incremental allocation spaces $\mathfrak{X}(ac)$ and $\mathfrak{X}(bc)$ from budget outcome $z$ in expectation under scheme $\mathfrak{n}_{\text{rand}}$.

%And finally, we formulate our optimization problem
The following lemma characterizes the overlap of the outcome $z \in \mathcal{N}(a, b, c)$ with the budgets $a,b$, and $c$, in terms of the incremental allocation functions $\mathfrak{X}(\cdot)$ and $\mathcal{Z}(\cdot)$.

\begin{lemma}
\label{Nash_barg_outcome}
For any preferred budgets of bargaining agents $a$ and $b$, disagreement point $c$, and outcome $z$ of $\mathcal{N}(a, b, c)$, % we have,
% if $\mathfrak{X}=g(a, b, c)$ and $\mathcal{Z}_S = g(a, b, c, z)_{\{z\} \cup S}$, then %BLAH
\begin{align*}
\mathcal{Z}(abc) &= \mathfrak{X}(abc), \ \ \ \mathcal{Z}(ab) = \mathfrak{X}(ab), \\
\mathcal{Z}(ac) &= \mathfrak{X}(ac) + \min(\textsc{Excess}/2, 0), \\
\mathcal{Z}(bc) &= \mathfrak{X}(bc) + \min(\textsc{Excess}/2, 0), \\
\mathcal{Z}(a) &= \mathcal{Z}(b) = \max(0, \textsc{Excess}/2), \\
\mathcal{Z}(c) &= \mathcal{Z}(\emptyset) = 0.
\end{align*}
Where, $\textsc{Excess} = (1 - \mathfrak{X}(abc) - \mathfrak{X}(ab) - \mathfrak{X}(ac) - \mathfrak{X}(bc)).$
\end{lemma}
\begin{proof}[Proof Sketch]
In Nash bargaining, no part of $z$ is such that it is not preferred by both $a$ and $b$. %that overlap with neither $a$ nor $b$ must be $0$ in an outcome of Nash bargaining. 
That is, $\mathcal{Z}(c) = \mathcal{Z}(\emptyset) = 0$. 
Otherwise,
%$u(a, z)$ and $u(b, z)$ must be less than 1, and
we could construct a new outcome $z'$ that reallocated the funds from $\mathcal{Z}(c)$ or $\mathcal{Z}(\emptyset)$ to $\mathcal{Z}(a)$ and $\mathcal{Z}(b)$. This would increase $u(a, z)$ and $u(b, z)$ and thus $z$ would not be Pareto optimal. %, hence a contradiction for Nash bargaining.
The parts of $z$ that benefit both $a$ and $b$ must be maximized. That is, $\mathcal{Z}(abc) = \mathfrak{X}(abc)$ and $\mathcal{Z}(ab) = \mathfrak{X}(ab)$. Otherwise, we could construct a new outcome $z'$ that reallocates funds from any other project to the project that benefits both $a$ and $b$, thus showing that $z$ is not Pareto optimal. The remaining part of the proof is technical and is in Appendix \ref{Z_nash_proof}.\remove{ ~\ref{Z_nash_proof}.} \qedhere
%, which could not be greater than 1.
%This would increase the Nash product, a contradiction.

%However, when $\mathfrak{X}(ab) + \mathfrak{X}(bc)+\mathfrak{X}(ab)+\mathfrak{X}(abc) \geq 1$, we must have $\mathcal{Z}(b)=\mathcal{Z}(c)=0$. Suppose $\mathcal{Z}(b)>0$ or $\mathcal{Z}(c)>0$, we can move the portions of $\mathcal{Z}$ benefitting only $b$ to those benefiting both $b$ and $c$ keeping $u(z,b)$ and $u(z,a)$ constant while strictly increasing $u(z,c)$ as $\mathfrak{X}(abc) + \mathfrak{X}(ab)+\mathfrak{X}(bc) + \mathfrak{X}(abc) \geq 1$, we can increase $\mathcal{Z}(S)$ all the way to $\mathfrak{X}(S)$ as long as $\sum\limits_{S \subseteq (a,b,c)} \mathcal{Z}(S) =1$. A similar argument can be made for the case when $\mathfrak{X}(ab)+\mathfrak{X}(bc)+\mathfrak{X}(ac)+\mathfrak{X}(abc) \leq 1$.
\end{proof}
\newpage
 We give an explanation of the construction of the Nash bargaining solution $z$ (and correspondingly $\mathcal{Z}$) in three steps.\footnote{The steps are only for illustration purposes. There is no chronology or structure required in bargaining processes. We can only characterize the outcome.}
 
\setword{\textbf{Step 1}}{Word:step1}: The voters with preferred budgets $a$ and $b$  mutually decide to allocate funds to projects that benefit both of them. This means, for all projects $j \in [m],$ $z_j = \min(a_j, b_j).$ In terms of $\mathfrak{X}(\cdot)$ and $\mathcal{Z}(\cdot),$ this corresponds to $\mathcal{Z}(abc) = \mathfrak{X}(abc)$ and $\mathcal{Z}(ab) = \mathfrak{X}(ab).$ At this point, $\mathcal{Z}(\cdot)$ is zero for all other subsets of $\{a,b,c\}.$ 
 %(denoted by $\mathfrak{X}(abc)+\mathfrak{X}(ab)$) to the bargaining solution $z$.
 
  \setword{\textbf{Step 2}}{Word:step2}: At this point, the total allocation to projects in the bargaining outcome $z$ may be less than 1. The bargaining agents now allocate more funds to the projects $j \in [m]$  for which $z_j < \max(a_j,b_j)$ \emph{and} $z_j < c_j.$ %that are currently assigned less funds $z_j$ than one of $a_j$ and $b_j$ \emph{and} also less than $c_j.$
  %That is $\{j ~\vert ~(z_j < a_j \text{~or~} z_j < b_j) \text{~and~} z_j < c_j \}.$ 
  %This corresponds to setting $z_j = \max(\min(a_j, b_j),\min(b_j,c_j),\min(a_j,c_j)).$ 
  %In other words, 
  Now $z_j$ is set to the ``median" of $(a_j,b_j,c_j)$ %\emph{second} value in the \emph{ordered} list of $(a_j, b_j, c_j)$, 
  for \emph{all} projects $j \in [m].$  In terms of $\mathfrak{X}(\cdot)$ and $\mathcal{Z}(\cdot),$ this corresponds to setting $\mathcal{Z}(ac) =  \mathfrak{X}(ac)$ and $\mathcal{Z}(bc) =  \mathfrak{X}(bc).$

 \setword{\textbf{Step 3}}{Word:step3}: Now, two possibilities arise for the total amount of funds allocated in $z$ so far, i.e., the bargaining agents have either over-spent or under-spent the total funds. These cases are central to the analysis in the paper and will be revisited several times.

\setword{\textbf{Case 1}}{Word:case1}: The total funds currently allocated in $z$ is at most 1, i.e., $\mathcal{Z}(ab) + \mathcal{Z}(bc) + \mathcal{Z}(ac) + \mathcal{Z}(abc) \leq 1.$ This is same as:
\begin{align}
  \mathfrak{X}(ab) + \mathfrak{X}(bc) + \mathfrak{X}(ac) + \mathfrak{X}(abc) &\leq 1. \label{case1}
\end{align}
Recall the definition of $\textsc{Excess}$ in Lemma~\ref{Nash_barg_outcome}. In this case, since there is a positive $\textsc{Excess},$ the bargaining agents now allocate more funds to projects with $z_j< \max(a_j,b_j)$. % than in $a$ or $b$.
%That is, $\{j \in [m] ~\vert~  z_j < a_j\}$ or $\{j \in [m] ~\vert~ z_j < b_j\}$.\footnote{Note that it is not possible for any project $j$ that $z_j < a_j$ and $z_j <b_j$ since the previous allocation is at least $\min(a_j, b_j).$}  
Since in Nash bargaining we assume equal importance of the overlap utilities of both the bargaining agents, they divide the $\textsc{Excess}$ equally. They incrementally fund projects with %a positive $(a_j - z_j)$ 
$z_j < a_j$ and the projects with $z_j < b_j$ %a positive $(b_j - z_j)$ 
with $\textsc{Excess}/2$ amount each. They ensure that $z_j \leq \max(a_j,b_j).$ %Denote $S_{a,j} = \max(a_j - z_j, 0)$ and $S_{}$ 
The precise manner of doing so is not important to satisfy the axioms of Nash bargaining. In terms of $\mathcal{Z}(\cdot),$ this corresponds to setting $\mathcal{Z}(a) = \mathcal{Z}(b) = \textsc{Excess}/2$. 

\setword{\textbf{Case 2}}{Word:case2}: The total funds currently allocated in $z$ exceeds 1, i.e., $\mathcal{Z}(ab) + \mathcal{Z}(bc) + \mathcal{Z}(ac) + \mathcal{Z}(abc) \geq 1.$ This is same as:
\begin{align}
  \mathfrak{X}(ab) + \mathfrak{X}(bc) + \mathfrak{X}(ac) + \mathfrak{X}(abc) &\geq 1. \label{case2}
\end{align}
If we are in this case, then the bargaining agents have overspent the funds and $\textsc{Excess}$ is negative. They need to remove $-\textsc{Excess}$ amount of allocations from $z$. %They do so from the allocations that only one of them benefit from. 
Recall that at this point, $z_j$ is set to the %\emph{second} value in the \emph{ordered} list
median of $(a_j, b_j, c_j)$ for \emph{all} projects $j \in [m].$ %They consider projects $j \in [m]$ for which $z_j > a_j$ or $z_j > b_j.$ 
They remove funds from projects with $(z_j > a_j)$ and the projects with $(z_j > b_j)$ with $\textsc{Excess}/2$ amount each. They ensure that $z_j \geq \min(a_j,b_j).$ The precise manner of doing so is not important to satisfy the axioms of Nash bargaining. 
In terms of $\mathcal{Z}(\cdot),$ this corresponds to setting $\mathcal{Z}(ac) = \mathfrak{X}(ac) + \textsc{Excess}/2,$ and $
\mathcal{Z}(bc) = \mathfrak{X}(bc) + \textsc{Excess}/2.$%\footnote{It is important to remember that $\textsc{Excess}$ may be positive or negative depending on which case we are in. It is positive in \ref{Word:case1}~(eq.~\ref{case1}) and negative in \ref{Word:case2}~(eq.~\ref{case2}).}

We now give a randomized way of allocating the $\textsc{Excess}$ funds in \ref{Word:step3} while satisfying the axioms of Nash bargaining.
\subsection{Randomised Nash bargaining solution $\mathfrak{n}_{\text{rand}}$}{\label{exact_soln_constr}}

%We now propose a randomized Nash bargaining scheme $\mathfrak{n}_{\text{rand}}.$ This scheme is in the class $\mathcal{N}$ and is a natural way of assigning the $\textsc{Excess}$ funds to projects in \ref{Word:step3}. The bargaining agents first do \ref{Word:step1} and  \ref{Word:step2} and note that there is no randomness in the process yet. For \ref{Word:step3} we now encounter the same two cases, \ref{Word:case1}~(eq.~\ref{case1}) and \ref{Word:case2}~(eq.~\ref{case2}).

\ref{Word:case1}: %In this case, the total funds allocated in $z$ is less than 1.
 %Consider all projects $j$ with $z_j < a_j.$ 
 Denote $s^a_j = \max\{a_j - z_j, 0\}$ for all projects $j$.\footnote{ This precisely corresponds to $X_{j,Q}(a)$ in the incremental allocation space.} To projects with $s^a_j > 0$, allocate incremental funds $r^a_j$ at random such that $\mathbb{E}[r^a_j]$ is proportional to $s^a_j.$ The sum of $r^a_j$ over all $j \in [m]$ is $\textsc{Excess}/2$ and no incremental allocation $r^a_j$  is more than $s^a_j$.%Denote this incremental allocation to project $j$ by $r^a_j$. 
\footnote{The randomness of this process is the same as the hypergeometric distribution with (discretized) $s^a_j$ balls corresponding to each project $j\in [m]$ in an urn, and we pick (discretized) $\textsc{Excess}/2$ balls without replacement to provide incremental allocations.}
 A similar process is followed for projects $j$ with $z_j < b_j$ by defining $s^b_j = \max\{b_j - z_j, 0\}$ and making incremental allocations $r^b_j$ summing to $\textsc{Excess}/2$, $\mathbb{E}[r^b_j]$ proportional to $s^b_j,$ and with $r^b_j \leq s^b_j.$ %Denote this incremental allocation to project $j$ by $r^b_j$.\footnote{No project can have both $z_j < a_j$ and $z_j < b_j$ after \ref{Word:step2}.}

\ref{Word:case2}:
    %In this case, the total funds allocated in $z$ is more than 1. Consider projects $j \in [m]$ for which $z_j > a_j.$ 
    Denote $t^a_j = \max\{z_j - a_j,0\}$ for all projects $j \in [m]$.\footnote{ This precisely corresponds to $X_{j,Q}(bc)$ in the incremental allocation space.}
    From projects with $t^a_j>0$, remove $r^a_j$ amount of previously allocated funds at random such that $\mathbb{E}[r^a_j]$ is proportional to $t^a_j.$ The sum of $r^a_j$ over all $j \in [m]$ is $-\textsc{Excess}/2$ and with  $r^a_j \leq t^a_j$. %Denote this removed allocation from project $j$ by $r^a_j$.
     A similar process is followed for projects with $z_j > b_j$ by defining $t^b_j = \max\{z_j - b_j,0\}$ and removing allocations $r^b_j$ from project $j$ summing to $\textsc{Excess}/2$, $\mathbb{E}[r^b_j]$ proportional to $t^b_j,$ and with $r^b_j \leq t^b_j.$ % Denote this removed allocation from project $j$ by $r^b_j$.\footnote{No project can have both $z_j > a_j$ and $z_j > b_j$ after \ref{Word:step2}.}

%Note that $\mathfrak{n}_{\text{rand}}$ is in $\mathcal{N},$ i.e, the outcome of $\mathfrak{n}_{\text{rand}}$ satisfies the axioms of Nash bargaining. This is in contrast to the class $\mathcal{M}$ of bargaining schemes. 

%With this setup of Nash bargaining schemes, 
We now give a characterization of median schemes $\mathcal{M}$ in terms of $\mathcal{Z}$
[recall that $\mathcal{N} \subseteq \mathcal{M}$ from Theorem~\ref{max_utility_Nash} in \S\ref{sec:model_SD}].

\subsection{Median schemes $\mathcal{M}$}
\label{relax_Nash_defn}

%We now present a class of \emph{median schemes} $\mathcal{M}$ and present the following theorem for any outcome $z \in \mathcal{M}(a,b,c)$. %by relaxing some of the conditions in Lemma~\ref{Nash_barg_outcome}. 
%Specifically, for any outcome $z \in \mathcal{M}$

\begin{theorem}{\label{median_outcome}}
    For any budgets $a,b,c \in \mathbb{B},$ a budget $z \in \mathbb{B}$ is in  $\mathcal{M}(a,b,c)$ if and only if it satisfies the following conditions.
    \begin{enumerate}[leftmargin=0.2in, start = 0]
    \item $\mathcal{Z}(abc)= \mathfrak{X}(abc)$ and $ \mathcal{Z}(\emptyset) = 0$.
    \item 
    In \ref{Word:case1}: 
     $~~~\mathcal{Z}(ab)=\mathfrak{X}(ab),$ $~~\mathcal{Z}(bc)=\mathfrak{X}(bc),$ $~~\mathcal{Z}(ca)=\mathfrak{X}(ca).$
     \item 
    In \ref{Word:case2}: $~~~\mathcal{Z}(a)=\mathcal{Z}(b)=\mathcal{Z}(c)=0.$
    
    %\item $\mathcal{Z}(ab)=\mathfrak{X}(ab)$; $\mathcal{Z}(bc)=\mathfrak{X}(bc)$; $\mathcal{Z}(ca)=\mathfrak{X}(ca)$ \text{ otherwise }
\end{enumerate}
\end{theorem}

The proof of this theorem is technical and is given in Appendix \ref{median_outcome_proof_section}.\remove{\S \ref{median_outcome_proof_section}.}

\remove{
Specifically, we have that for any budget $a$, $b$ and $c$, an outcome $z$ of $\mathcal{M}(a,b,c)$ satisfies \emph{only} the following conditions.
\begin{enumerate}[leftmargin=0.2in, start = 0]
    \item $\mathcal{Z}(abc)= \mathfrak{X}(abc)$ and $ \mathcal{Z}(\emptyset) = 0$ are always satisfied.
    \item 
    If in \ref{Word:case1}~(eq.~\ref{case1}) then,\\
     $\mathcal{Z}(ab)=\mathfrak{X}(ab),$ $~~\mathcal{Z}(bc)=\mathfrak{X}(bc),$ and $~~\mathcal{Z}(ca)=\mathfrak{X}(ca).$
     \item 
     If in \ref{Word:case1}~(eq.~\ref{case2}) then,\\ $\mathcal{Z}(a)=\mathcal{Z}(b)=\mathcal{Z}(c)=0.$
    
    %\item $\mathcal{Z}(ab)=\mathfrak{X}(ab)$; $\mathcal{Z}(bc)=\mathfrak{X}(bc)$; $\mathcal{Z}(ca)=\mathfrak{X}(ca)$ \text{ otherwise }
\end{enumerate}

}
{Note that all the conditions on the outcomes of the bargaining schemes in $\mathcal{M}$ are \setword{\textit{symmetric}}{Word:symmetric} in all three of $\{a,b,c\}$. However, outcomes in $\mathcal{N}$ also satisfy some additional conditions which may not be symmetric in all three of $\{a,b,c\}$.

%However, this may not be true for schemes in $\mathcal{N}$ in general where some additional constraints which may not be symmetric in all three of $\{a,b,c\}$ hold true.
}
\remove{
Observe that the outcomes of the bargaining schemes in $\mathcal{M}$ are \setword{\textit{symmetric}}{Word:symmetric} in all three of $\{a,b,c\}$. This implies that they give equal importance to the disagreement point as to the preferred budgets of the bargaining agents. {\color{red}This is not true for schemes in $\mathcal{N}$ in general. This is messed up.}
}
%In bargaining schemes in $\mathcal{M}$, the projects in all three budgets are definitely selected. All projects that are in at least two budgets go to the bargaining solution when the total cost of projects in at least two budgets is at most 1. In this case, the surplus is filled by projects which are in exactly one budget. In this condition is not true, no project present in only one budget goes to the bargaining solution. This bargaining solution relaxes the assumption of giving equal importance to budgets $a$ and $b$ as was the case in Lemma~\ref{Nash_barg_outcome}.

We now give a lower bound on $\Distortion_once({\mathcal{N}}).$ Since $\mathcal{M}$ contains ${\mathcal{N}}$, this bound also applies to $\Distortion_once(\mathcal{M}).$ Moreover, the same bound also holds for the distortion of  $\mathfrak{n}_{\text{rand}}.$

%\begin{theorem}\label{thm:lb-nash}
% $\Distortion_once({\mathcal{N}}) > 1.38$
%\end{theorem}
\begin{theorem} \label{thm:lower_bound}
 $\Distortion_once(\mathcal{M}) \geq \Distortion_once(\mathcal{N})> 1.38.$ 
 
 Also, $\Distortion_once(\mathfrak{n}_{\text{rand}}) > 1.38.$
\end{theorem}
%Interestingly, a single example establishes both  Theorems~\ref{thm:lb-nash} and~\ref{thm:lower_bound}.
%The proof is by an example and is in Appendix~\ref{proof_lower_bound}.
\begin{proof}
The proof is by the following example of a PB instance. Suppose there are $n_A + n_B$ voters
and $n_A + 1$ projects for some $n_A, n_B \geq 1$. Let $o_i$ denote the budget where the $i$-th project receives allocation $1$ and all the other projects get allocation $0$. 
Each voter $i$ in group A $(i \in [n_A])$ prefers budget $o_i$.
Each voter $i$ in group B $(i \in [n_A+n_B]\setminus[n_A]$) prefers budget $o_{n_A+1}$. %with analysis in Appendix~\ref{app:lower_bnd}. 
{The analysis  of this example is in Appendix \ref{app:lower_bnd}.  \remove{~\ref{app:lower_bnd} }where we set $n_A = 2200$; $n_B = 3000.$}
\end{proof}

We now give upper bounds of the distortion of $\mathcal{M}$.% and $\mathfrak{n}_{\text{rand}}$ in the next two sections respectively.

\remove{
\subsection{Outcomes of Nash Bargaining}

%In this section, we look at bargaining processes and total utility of their outcomes among the three voters involved.
%We will now look at how bargaining processes can affect the utility. We define the social welfare of a budget to be the sum of the overlap utilities of voters and the budget. We show that, for three voters, outcomes from the Nash bargaining maximize the social welfare among the three voters.

%\subsection{Common Budget Space}

%With the result of Theorem~\ref{cs_main}, it is clear that Nash bargaining takes the project interactions into consideration and does not output budgets with $ \sum_i f(b)_i <1$. Going ahead we do not worry about project interactions and give results which are valid with and without them. To make it easy to prove some properties of the Nash bargaining, we primarily work in the \emph{common budget space}. 
\remove{

\edit{Recall that we refer to the set of preferred budgets of all voters as the vote profile $P$. In the common budget space, we map each subset of the vote profile to a number in the interval $[0,1]$. This quantity for a subset $S$ captures, the sum over all projects $j$, of how much do budgets in $S$ agree on increasing the allocation to $j$ above the maximum allocation to $j$ by any budget in $P \backslash S$. Informally, this quantity captures a notion of similarity of the budgets in a subset $S$ and their dissimilarity with budgets in $P \backslash S$.} Formally,

\begin{definition}
Given a vote profile $P=(v_1, v_2, \ldots, v_k),$ the \textbf{common budget function} $g_P: 2^{[k]} \rightarrow \mathbb{R}$ maps a subset of voters to a point in the common budget space
%whose coordinates are subsets of $\{v_1, v_2, \ldots, v_k\}$
where, for each subset $S$ of voters,
 $g_P(S) = \sum\limits_{j=1}^m \max \bigg(\Big(\min\limits_{s \in S} v_{s,j} \Big) - \Big(\max\limits_{\tilde{s} \in P \setminus S} v_{\tilde{s},j} \Big), 0 \bigg).$

%(Define $g(P)_\emptyset = 0$.)  % should probably be \infinite ? Doesn't matter as we never use it.
\end{definition}

%See that $g_P(S)$ tells us how much allocation budgets in $S$ have in common, over and above the allocation in $P\backslash S$. 
\edit{\textbf{Example 3.2:}} Consider \edit{an instance of PB with three projects and a vote profile $P$ with} three budgets $a = \langle 1, 0, 0 \rangle$, $b = \langle 0, 1, 0 \rangle$, and $c = \langle 0.25, 0.25, 0.5 \rangle$. Let $\mathfrak{X}(S) = g_{P}(S)$ \edit{for each subset $S$ of $P$}.\footnote{For brevity, we omit braces and commas in the argument of $\mathfrak{X}$.} Then, $\mathfrak{X}(a) = 0.75.$ This is because the budget $a$ has allocation $1$ to project $1,$ out of which $0.25$ is shared with budget $c$ and $0.75$ is \emph{exclusive}. Similarly $\mathfrak{X}(b) = 0.75,$ and $\mathfrak{X}(c) = 0.5$. Further, $\mathfrak{X}(ac) = 0.25$. This is because the subset $\{a,c\}$ has an allocation of $0.25$ to project $1$ common to both $a$ and $c$. It is also exclusive since $b$ has no allocation to project $1$. Further, we have $\mathfrak{X}(abc) = 0$ because the group of all three budgets has no allocation common to all.
 
 \edit{Note that the number of points in the \emph{common budget space} is independent of the number of projects on the ballot. It is only a function of the number of voters. This helps make the analysis of distortion in the next section computationally feasible.}
 
 \edit{We now give a formulation of the overlap utility of two budgets $a$ and $b$ in the common budget space. Recall that $\mathcal{P}(P)$ denotes the power set of $P$, i.e., the set of all subsets of the vote profile $P$. }
\begin{lemma}
\label{utility_common}
For any pair of budgets $a$ and $b$ and any vote profile $P$ that includes $a$ and $b$, we have $u(a, b) = \sum\limits_{ S \in \mathcal{P}(P \backslash \{a,b\}) }  g_P(  S \cup \{a,b\})$.
\end{lemma}
}
% FI\mathfrak{X}ME: Type error? The subscripts refer to the variables, not the indices to the variables.

%\subsection{Outcomes of Nash Bargaining}

Recall Nash bargaining given in Definition~\ref{def:nb}. We now characterize the outcomes of the Nash bargaining \edit{in the common budget space}. Given budgets $a$, $b$, and $c$, there may be multiple outcomes of $\Nash(a, b, c)$. However, as we will see, all these outcomes correspond to a unique point in the common budget space.
%
% TODO: Explain more about \mathcal{Z}_S = g(a, b, c, z)_{\{z\} \cup S} here.
%
%Recall that $\mathfrak{X}=g(a, b, c)$. Define $\mathcal{Z}_S = g(a, b, c, z)_{\{z\} \cup S}$ for any $S \in \power{\{a, b, c\}}$. Then we have the following result for $\mathcal{Z}.$
%
%
%\begin{theorem}
%For any budgets $a$, $b$, and $c$, and any outcome $z$ of $\Nash(a, b, c)$, $\langle u(a, z), u(b, z), u(c, z) \rangle$ is unique.
%\end{theorem}
%
%\begin{proof}
%This follows from Lemma~\ref{utility_common} and Lemma~\ref{nash_outcome}.
%\end{proof}
%
%
%
%\subsection{Total Utility Maximization}
%
%comment
% mention the knapsack voting paper ... idea of proof from that
%
\edit{Further,} we show that outcomes of the Nash bargaining maximize the total utility among the three voters involved, \edit{two who are deliberating and one whose preferred budget is the disagreement point}. %The idea of the proof is that if we can show that the Nash bargaining allocates the budget to the portions that voters agree with most, then the total utility is maximized.

\begin{theorem}
\label{max_utility}
For any budgets $a$, $b$, and $c$, any outcome $z$ of\\ $\mathcal{A}(a, b, c)$ maximizes the sum of utilities $u(a, z) + u(b, z) + u(c, z)$. 
\end{theorem}
\begin{comment}
\begin{proof}
%Let $\mathfrak{X}=g(a, b, c)$ and $\mathcal{Z}_S = g(a, b, c, z)_{\{z\} \cup S}$. %BLAH
From Lemma~\ref{nash_outcome} in Appendix~\ref{Z_proof}, we have: $$k_a - \mathfrak{X}(ac) = k_b - \mathfrak{X}(bc),$$ 
where $k_a = (1 - \mathfrak{X}(abc) - \mathfrak{X}(ab) + \mathfrak{X}(ac) - \mathfrak{X}(bc)) / 2$, \\and $k_b = (1 - \mathfrak{X}(abc) - \mathfrak{X}(ab) + \mathfrak{X}(bc) - \mathfrak{X}(ac)) / 2$.

\textbf{Case 1: }If $\mathcal{Z}(a) = 0$, we have
$\mathcal{Z}(a) = \max(0, k_a - \mathfrak{X}(ac)) = 0$. This implies 
$k_a - \mathfrak{X}(ac) \leq 0$ and 
$k_b - \mathfrak{X}(bc) \leq 0$.
Thus, $\mathcal{Z}(b) = \max(0, k_b - \mathfrak{X}(bc)) = 0$.
%Therefore, by Lemma~\ref{knapsack}, the total utility is maximized.

\textbf{Case 2: }If $\mathcal{Z}(a) > 0$, we have
$\mathcal{Z}(a) = \max(0, k_a - \mathfrak{X}(ac)) > 0$. This implies $k_a - \mathfrak{X}(ac) > 0$ and
$k_b - \mathfrak{X}(bc) > 0$.
Thus, $\mathcal{Z}(ac) = \min(k_a, \mathfrak{X}(ac)) = \mathfrak{X}(ac)$ and $\mathcal{Z}(bc) = \min(k_b, \mathfrak{X}(bc)) = \mathfrak{X}(bc)$.

By Lemma~\ref{knapsack} in Appendix~\ref{proof_knapsack}, the sum of utilities $u(a, z) + u(b, z) + u(c, z)$ is maximized in both cases.% Case 1 and Case 2. 
\end{proof}
\end{comment}
The proof of Theorem~\ref{max_utility} is technical and is in Appendix~\ref{proof_max_utility}, along with supporting lemmas, Lemma~\ref{nash_outcome} in Appendix~\ref{Z_proof} and Lemma~\ref{knapsack} in Appendix~\ref{proof_knapsack}. 
An important implication of this result is that the outcome of Nash bargaining is a \emph{median} of the three budgets $a,b,$ and $c.$ Therefore, we can use tools applicable for median voting rules~\cite{comanor1976median} to our framework.

}
\section{Distortion Of Schemes in $\mathcal{M}$}{\label{sec:distort_Nash_relaxed}}
To find an upper bound of the distortion of triadic scheme with any bargaining scheme, we use a technique introduced in \cite{fain2019}, called \emph{pessimistic distortion} (PD). In this technique, we first analyze the distortion for a small group of voters, call it PD, and then show that the distortion over all voters cannot be more than the PD. Specifically, in this paper, we analyze the PD for a group of $6$ voters. %We show that the distortion for any number of voters cannot be higher than the pessimistic distortion. 
The idea is that we allow the counterfactual solution to choose a separate `optimal' budget for every $6$-tuple of voters, thereby attaining a smaller social cost than a common outcome for all voters. On the other hand, for our mechanism, we consider the expected social cost under one outcome. This is why the distortion calculated is \emph{pessimistic}. Formally:
%
%, and $\permu{S}$ be the set of all permutations of $S$.
\begin{definition}{\label{pess_dist_defn}}
The \textbf{pessimistic distortion (PD)} of the class of mechanisms $\mathcal{M}$ with triadic scheme \edit{with 6 voters} is:

% \[ \max_{P \in \mathbb{B}^5} \frac{ \frac{1}{20} \sum_{R \in \combi{P, 3}} \sum_{v \in P \setminus R}  d(\Nash(R), v) }
%{ \min_{p \in \mathbb{B}} \frac{1}{5} \sum_{v \in P} d(p, v) }. \]
$$ PD(\mathcal{M}) = \sup\limits_{P \in \mathbb{B}^6;~h \in \mathcal{M}} \frac{ \displaystyle \frac{1}{20} \sum\limits_{Q \in \combi{[6], 3}} \frac{1}{3} \sum\limits_{i \in [6] \setminus Q}  d(h(Q), P_i) }
{ \min\limits_{p \in \mathbb{B}} \frac{1}{6} \sum\limits_{i \in [6]} d(p, P_i) }. $$

Here $\combi{S, k}$ denotes the set of all $k$-combinations of set $S$.\footnote{For simplicity of notation, we  use $\mathfrak{n}(Q)$ in place of $\mathfrak{n}(P_{Q_1}, P_{Q_2}, P_{Q_3})$ in PD.}
\end{definition}

 %Further, we denote the set of all $k$-combinations-with-replacement of set $S$ by $\combiwithr{S, k}$. %\edit{For an ordered set $R = (v_i,v_j,v_k), \Nash(R)$ denotes the outcome of the Nash bargaining between $v_i$ and $v_j$ with $v_k$ as the disagreement point.}

\begin{figure}[ht!]
\centering
%\begin{subfigure}{.5\textwidth}
  %\centering
  \includegraphics[scale =0.3]{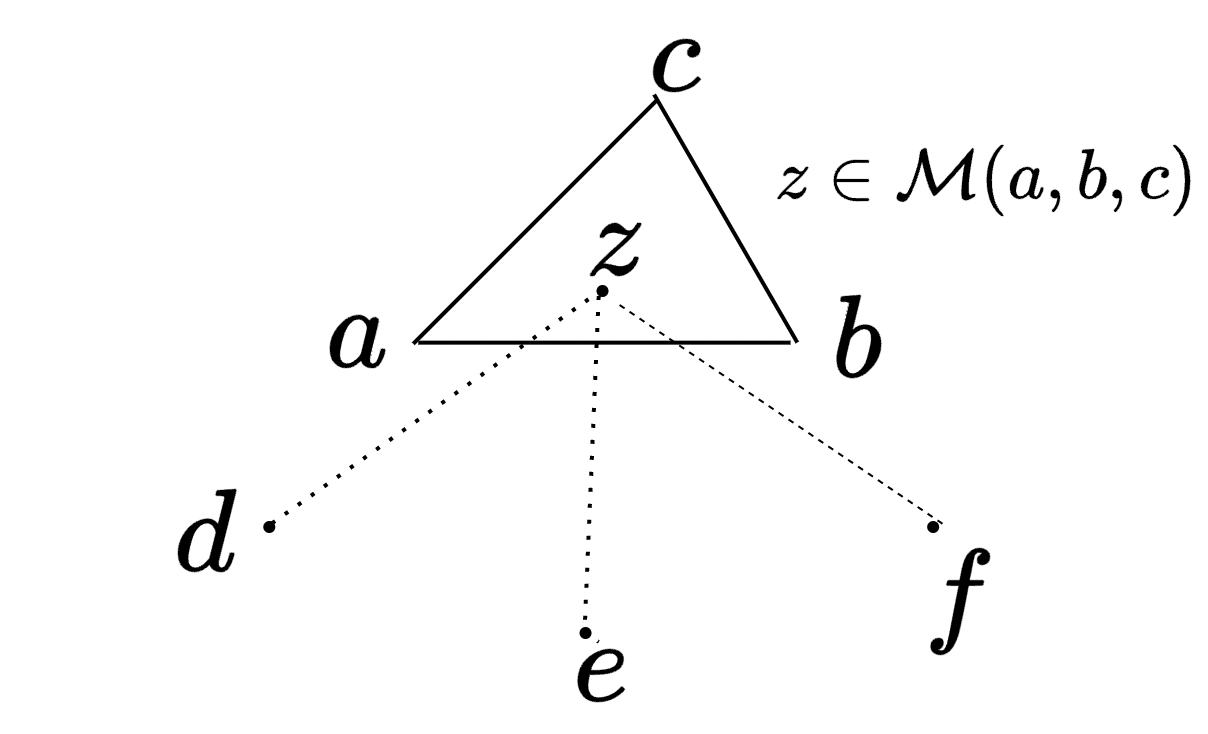}
  \vspace{-0.4cm}
  \caption{Illustration of PD where $a,b,c$ are sampled for the mechanism $\mathcal{M}$, and $\{d,e,f\}$ are the other budgets for which we measure the cost of outcome $z$.}
  \vspace{-0.3cm}
 \label{pd_diagram}
%\end{subfigure}%
\end{figure}

Notice that in the definition of PD, we only consider the cost for the non-bargaining agents (same as in \cite{fain2019}). We illustrate the PD in Figure~\ref{pd_diagram}, where the bargaining is over budgets $\{a,b\},$ the disagreement point is $c,$ and the cost is computed only for $\{d,e,f\},$ the budgets not involved in the bargaining. This definition is more pessimistic than considering all agents' costs. Further, since the outcome of $\mathcal{M}$ is \ref{Word:symmetric} in $\{a,b,c\},$ we can use any \emph{combination} $Q$ of three voters to compute the outcome of bargaining without designating one of the budgets as the disagreement point.  
  The next result, proved in Appendix \ref{proof_distortion} \remove{~\ref{proof_distortion}}, is that the distortion of any bargaining scheme in $\mathcal{M}$ with triadic scheme cannot be more than its PD with triadic scheme with only $6$ voters. %Specifically for $\mathcal{M}$ we have:
\begin{lemma}
\label{distortion_pd}
$\Distortion_once(\mathcal{M}) \leq PD(\mathcal{M})$.
\end{lemma}
We now give a representation of the overlap utilities $u(\cdot,\cdot)$ (equivalently the cost $d(\cdot, \cdot)$), in terms of the incremental allocations $X_{P}(S)$. This representation is of technical importance for proofs. %as defined in \S\ref{common_budget_defn}) useful for characterising the cost $d(\cdot, \cdot)$ between two budgets.

\begin{lemma}
\label{utility_common}
For budgets $\{a,b\},$ and a vote profile $P$ that includes $\{a,b\}$, we have $u(a, b) = X_{(ab)}(ab) \overset{(1)}{=} \sum_{ \hat{S} \in \mathcal{P}(P \backslash \{a,b\}) }  X_P(  \hat{S} \cup \{a,b\})$.
\end{lemma}

\begin{proof}
From Definition~\ref{def:ou}, we have $u(a,b)  = \sum_{j=1}^{m} \min(a_j,b_j) $.  From Definition~\ref{proj_budget_defn} we have $ \sum_{j=1}^{m} \min(a_j,b_j) = \sum_{j=1}^{m} X_{j,(a,b)} (ab) = X_{(ab)}(ab)$.  Now apply Lemma~\ref{proj_budget_new_space_lemma} with $Q=S=\{a,b\}$, to obtain equality $(1)$.
\end{proof}

Lemma~\ref{utility_common} shows that the overlap utility between two budgets $a,b$ is the same as the sum of what $a,b$, and all subsets of the other budgets in $P$ have in common via the incremental allocation function $X_P(S)$.
For example, if $ P = (a,b,c,d),$ then $u(a, b) = X_P(ab) + X_P(abc) + X_P(abd) + X_P(abcd).$ 

Lemma~\ref{utility_common} is useful for the proof of the following important result, which is an upper bound for $PD(\mathcal{M}).$

\remove{The proof of Lemma~\ref{utility_common} is in Appendix \ref{utility_common_proof}. } %Also note that $(1)$ follows from Lemma~\ref{proj_budget_new_space_lemma} on selecting $Q=(a,b)$ and $S=Q$. 
%We now prove some technical results on the incremental budget space $\{X_{j,P}(S)\}_{j \in [m]}$ which will be useful for our analysis later in the paper.%for all sets $S\in \mathcal{P}(P)$ and containing budget $P_i$. 

%The next result is  an upper bound for $PD(\mathcal{M}).$ %and therefore for $\Distortion_once(\mathcal{M})$ as well. 
%The main idea of its proof is as follows. We first consider the cases described in Section~\ref{relax_Nash_defn} and construct an L.P corresponding to each case and solve it for every 3 sized subset of the six voters as defined in Definition ~\ref{pess_dist_defn}.

%The main idea is as follows: we introduce a class of vote aggregation rules called `median voting' rules, which select three voters \edit{independently and} uniformly at random and output a budget that maximizes the total utility of those three voters. According to Theorem~\ref{max_utility}, triadic scheme is in the class of median voting rules. Therefore, it cannot have distortion more than the most adversarial median voting rule. By `adversarial', we refer to the median voting rule that, when there are multiple budgets that maximize the utility of the three voters, chooses the one which minimizes the total utility of all the \textit{remaining} voters. We then analyze the distortion of median voting through the common budget space, making it easier to calculate overlap utilities. We rewrite the problem as linear programs (LP), using case analysis to remove non-linear constraints, and then use a computer to solve the factor-revealing LPs.

\begin{lemma}
\label{pd_triadic_nash_relaxed}
$PD(\mathcal{M}) \leq 1.80$.
\end{lemma}

We give a sketch of the proof here. The detailed proof is in Appendix \ref{pd_triadic_nash_relaxed_proof}. \remove{~\ref{pd_triadic_nash_relaxed_proof}.}

\begin{hproof}
    Let $p^Q$ denote a budget obtained on bargaining with budgets in set $Q$ using a bargaining scheme in $\mathcal{M}$. Note that mechanisms in $\mathcal{M}$ are \ref{Word:symmetric} in  $Q$ therefore, we do not need to designate a disagreement point in $Q$ for analysis.
%
%maximize the total utility of three voters in set $R$.
%That is, $$\Median(R) = \operatorname*{argmax}_{p \in \mathbb{B}} \sum_{v \in R} u(p, v).$$
%
%By Theorem~\ref{max_utility}, $\Nash(R) \subseteq \Median(R)$. Therefore,
\begin{align*}
PD(\mathcal{M}) = & \sup_{P \in \mathbb{B}^6;~h \in \mathcal{M}} \frac{ \frac{1}{60} \sum_{Q \in \combi{[6], 3}} \sum_{i \in [6] \setminus Q}  d(h(Q), P_i) }
{   \frac{1}{6} \min_{v \in \mathbb{B}} \sum_{i \in [6]} d(v, P_i) }, \\
&\leq \sup_{P \in \mathbb{B}^6} \frac
{ \frac{1}{60} \sum_{Q \in \combi{[6], 3}} \sup_{p^Q \in \mathcal{M}(Q)} \sum_{i \in [6] \setminus Q} d(p^Q, P_i) }
{ \frac{1}{6} \min_{v \in \mathbb{B}}  \sum_{i \in [6]} d(v, P_i) }.
\end{align*}

Suppose that $PD(\mathcal{M}) > 1.80$.
Then the following optimization problem has an optimal objective value strictly greater than $0$. 
%
%maximize $\frac{1}{20} \sum_{R \in \combi{P, 3}} \sum_{v \in P \setminus R} d(p^R, v)
%-
%1.93 \cdot \frac{1}{5} \sum_{v \in P} d(p_{\text{all}}, v)$ \\
%\-\ \ \ \ s.t.
%\begin{align*}
%P &\in \mathbb{B}^5 \\
%p^R &\in \Median(R) \ \ \ \ \forall R \in \combi{P, 3} \\
%p_{\text{all}} &\in \mathbb{B}
%\end{align*}
\begin{align}
\text{maximize } ~~~&\frac{1}{60} \sum_{Q \in \combi{[6], 3}} \sum_{i \in [6] \setminus Q} d(p^Q, P_i) - 1.80 \cdot \frac{1}{6} \sum_{i \in [6]} d(v, P_i), \nonumber\\
\text{subject to } ~~~&P \in \mathbb{B}^6, \nonumber\\
~~~&p^Q \in \mathcal{M}(Q) \ \ \ \ \ \ \ \ \ \ \ \ \ \ \ \ \ \ \ \ \ \ \ \ \ \ \ \ \ \ \ \ \ \ \ \ \ \ \ \ \ \ \ \forall Q \in \combi{[6], 3}, \nonumber \\
~~~&v \in \mathbb{B}.  \label{prob:opt}
\end{align}
To convert this problem into a linear program, we map it to the incremental allocation space of the set of 6 budgets $P=\{P_1,P_2,\ldots,P_6\}$. %\footnote{Note the intentional notation overload here, for the purpose of the PD, the set of 6 voters is effectively the entire vote profile, therefore we use the variable $P$ for it.}
% FIXME: Maybe define these later
 Denote $X_{P}(\cdot)$ by $X(\cdot)$ for simplicity of notation in the optimization  programs. Similar to Definition~\ref{new_budget_common_budget_defn}, we define 
$V(S) = X_{(P \cup \{v\})}( S \cup \{v\})$ via the `optimal' budget $v$ and
$Z^Q(S) =~X_{(P~\cup~\{p^Q\})}(S~\cup~\{p^Q\})$ using the outcome of our mechanism $p^Q$, for each $Q \in \combi{[6], 3}$.

By Lemma~\ref{disutility}, we write the cost in terms of the overlap utility $d(p^Q, P_i)= 
2 - 2u(p^Q, P_i),$ which, by Lemma~\ref{utility_common} and the definition of $Z^Q(S)$, 
equals $2 - 2\sum_{ S \in \power{P\backslash P_i } } 
Z^Q(S \cup P_i)$. Similarly, we have $d(v, P_i) = 2 - 2\sum_{ S \in \power{P\backslash P_i } } 
V(S \cup P_i).$ To make the $p^Q \in \mathcal{M}(Q)$ constraints linear, we use case analysis.

% FIXME: Confusing
% FIXME: Confusing

% to make it manageable ?
Consider a given $Q=\{ q_1, q_2, q_3 \} \in  C([6],3)$ and a budget $p^Q \in \mathbb{B}$.
Let $\mathfrak{X}(S) = X_{Q}(S)$ and $\mathcal{Z}(S) = X_{(Q \cup \{p^Q\})}( S \cup \{p^Q\})$. % BLAH
Theorem~\ref{median_outcome} implies that $p^Q \in \mathcal{M}(Q)$ if and only if the following holds: %$Z(123) = Y(123)$, $Z(\emptyset) = 0$, and
\begin{itemize}[leftmargin = 0.3cm]
%\item (Case 0) If $X_{123} \geq 1$, then $p \in \Median(R)$ if and only if $Z_{12} = Z_{13} = Z_{23} = Z_1 = Z_2 = Z_3 = Z_\emptyset = 0$. %$Z_{123}=1$ and other $Z$'s are 0.
% \item (Case 1) If $X_{123} \leq 1$ and $X_{123} + X_{12} + X_{13} + X_{23} \geq 1$, then $Z_1 = Z_2 = Z_3 = 0$.
\item \ref{Word:case1}: If $\mathfrak{X}(q_1q_2q_3) + \mathfrak{X}(q_1q_2) + \mathfrak{X}(q_1q_3) + \mathfrak{X}(q_2q_3) \geq 1$, $\mathcal{Z}(q_1) = \mathcal{Z}(q_2) = \mathcal{Z}(q_3) = 0$.
\item \ref{Word:case2}: If $\mathfrak{X}(q_1q_2q_3) + \mathfrak{X}(q_1q_2) + \mathfrak{X}(q_1q_3) + \mathfrak{X}(q_2q_3) \leq 1$,\\ $\mathcal{Z}(q_1q_2)=\mathfrak{X}(q_1q_2), \mathcal{Z}(q_1q_3)=\mathfrak{X}(q_1q_3),  \mathcal{Z}(q_2q_3)=\mathfrak{X}(q_2q_3)$.
\end{itemize}
%
%Now we apply Lemma~\ref{proj_budget_new_space_lemma} to argue that $\mathfrak{X}(\mathfrak{S}) = \sum\limits_{V' \in \mathcal{P}([6]\setminus R)} X(\mathfrak{S}\cup R)$ and similarly, $\mathfrak{Z}^R(\mathfrak{S}) = \sum\limits_{V' \in \mathcal{P}([6]\setminus R)} X(\mathfrak{S} \cup V')$ to obtain the constraints  
%
We break each $p^Q \in \mathcal{M}(Q)$ constraint into two cases. Since there are ${\binom{6}{3}}$ such constraints in the optimization problem%~\eqref{prob:opt}
, there are $2^{\binom{6}{3}}$ cases overall. We represent each case by a binary string of length 20 where a 0 or 1 at each position denotes whether the triplet $Q$ corresponding to that position is in \ref{Word:case1} or \ref{Word:case2}.

However, most of these $2^{\binom{6}{3}}$ cases are not unique up to the permutation of preferred budgets, i.e. when the preferred budgets of different voters are permuted, we may move from one case to another. Since these cases have the same objective value, we do not need to solve all the cases. 
Exploiting further symmetries, we have 2136 unique cases, each of which is formulated  as a linear program with precise details in Appendix \ref{pd_triadic_nash_relaxed_proof}.\remove{\S\ref{pd_triadic_nash_relaxed_proof}.} We obtain the optimal value for each case to be 0 hence, a contradiction. \qedhere %Therefore, $PD(\mathcal{M}) \leq 1.83$.
\end{hproof}

%However, when we solve the optimization problem under both cases, the result would be the same. Thus we remove all such cases which are not unique up to permutation amongst voters and we obtain 2136 such cases. %Also note that for the following optimization problem, $X(.)$ and $Z^R(.)$ are defined on the common budget space of budgets $P_1,P_2,\ldots,P_6$ i.e. $X(S)=g_P(S)$ and $Z^R(S) = g_{P\cup \{p^R\}}(S \cup \{p^R\})$
%Let us denote the set of all such cases by $\mathbb{K}$.
%
%
%Each case $\kappa \in \mathbb{K}$ reduces to a linear program described below:%, described in Appendix~\ref{LP}. %

\remove{
The linear program in each case is as follows with the optimization variables $\{X(S)\}_{S \in \mathcal{P}([6])},\{V(S)\}_{S \in \mathcal{P}([6])},\{Z^Q(S)\}_{S \in \mathcal{P}([6]); Q \in C([6],3)}$.\footnote{For simplicity of notation, we refer to budget $P_i$ by $i$ while constructing the sets $S$.}
%\sukolsak{TODO: Explain how to convert $d(\ldots, \ldots)$ to $u(\ldots, \ldots)$ to $X_{\ldots}$}
%
\begin{align*}{\label{opti_relaxed}}
&\text{maximize}~~~~ \frac{1}{60} \sum_{Q \in \combi{[6], 3}} \sum_{i \in [6] \setminus Q} d(p^Q, P_i) - 1.83 \cdot \frac{1}{6} \sum_{i \in [6]} d(v, P_i) \\
&\text{subject to} 
\end{align*}

\begin{align*}
& \sum_{S \in \power{[6]\backslash \{i\}}} X(S \cup i) =1 &\forall i\in [6],\\
& X(S) \geq 0  &\forall S \in \power{P}, \\
& \sum_{S \in \power{P}} Z^Q(S) = 1 &\forall Q \in \combi{[6], 3}, \\
& Z^Q(S) \geq 0 &\forall Q \in \combi{[6], 3}, \forall S \in \power{[6]}, \\
& X(S) \geq Z^Q(S) &\forall Q \in \combi{[6], 3}, \forall S \in \power{[6]},\\
& Z^Q(Q \cup V') = X(Q \cup V')& \forall Q \in \combi{[6], 3}, \forall V' \in \power{[6] \setminus Q}, \\  % p in Median
& Z^Q(V') = 0 &\forall Q \in \combi{[6], 3}, \forall V' \in \power{[6] \setminus Q}, \\ % p in Median
& \sum_{S \in \power{[6]}} V(S) = 1, &\\
& V(S) \geq 0 &\forall S \in \power{[6]}, \\
& X(S) \geq V(S)  &\forall S \in \power{[6]}.
\end{align*}
%Let $f$ map each such triplet in $\combi{[6], 3}$ to a distinct number in $\{0,1\ldots,19\}.$

%With the following additional constraints for each $Q \in \combi{[6], 3}:$
We also have following case-wise constraints for each $ Q \in \combi{[6], 3}$. If $Q$ is in \ref{Word:case1},
%\noindent If $\kappa_{f(Q)} == 0$ then,
\begin{gather*}
\sum_{i=2}^3 ~\sum_{Q' \in \combi{Q, i}} ~\sum_{V' \in \power{[6] \setminus Q}} X(Q' \cup V') \geq 1, \\
Z^Q(Q' \cup V') = 0 \ \ \ \ \ \ \ \ \ \ \ \ \ \ \ \ \ \ \ \ \ \ \ \ \forall Q' \in \combi{Q, 1}, \forall V' \in \power{[6] \setminus Q}.
\end{gather*}
Whereas if $Q$ is in \ref{Word:case2},
\begin{gather*}
\sum_{i=2}^3 ~\sum_{Q' \in \combi{Q, i}} ~\sum_{V' \in \power{[6] \setminus Q}} X(Q' \cup V') \leq 1, \\
Z^Q(Q' \cup V') = X(Q' \cup V') \ \ \ \ \ \ \ \ \forall Q' \in \combi{Q, 2}, \forall V' \in \power{[6] \setminus Q}.
\end{gather*}
%\}
%$\forall {Q \in \mathcal{C}([6],3)}$ 

%Note: We can show that optimization problem \ref{opti_relaxed} given by cases $\kappa$ and $\bar{\kappa}$ are identical for every case $\kappa \in \{0,1\}^{20}$. Note that $\bar{\kappa}$ toggles every bit in $\kappa$.

We solve all the linear programs for every case%in $\mathbb{K}$ ({\color{red} 2136} such cases)
and find that the objective value is always equal to $0$. Hence, a contradiction. %Therefore, $PD(\Triadic) < 1.93$.
} 

%\vspace{-2 em}
Using Lemmas~\ref{distortion_pd} and \ref{pd_triadic_nash_relaxed}, we get the following key result.

%We now give the main result of this section.
%\vspace{-0.5 em}
\begin{theorem}\label{thm:nash_rel}
$\Distortion_once(\mathcal{M}) \leq 1.80$.
\end{theorem}

\section{Distortion of  $\mathfrak{n}_{\text{rand}}$}{\label{sec:triadic_deliberation}}
%\input{project_substitution.tex}
%\subsection{Total Utility of Three Voters}
\remove{
\subsection{Bargaining under a triplet of budgets}

Recall the Nash bargaining scheme as discussed in Defn. \ref{def:nb}. We now characterize the outcomes of this bargaining \edit{in the common budget space}. Given budgets $a$, $b$, and $c$, there may be multiple outcomes of $\mathcal{A}(a, b, c)$ and we project each of these solutions on the common budget space of $\{a,b,c\}$. However, as we will see, all these outcomes correspond to a unique point in the common budget space. Recall that $X(S)$ is defined as $g_{a,b,c}(S)$ for every subset $S$ of three budgets, thus projecting on the common budget space of these three voters. Also, we define $Z(S)= g_{a,b,c,z}(\{z\}\cup S)$, where $z$ is the budget obtained after the bargaining scheme $\mathcal{A}$. Also, any budget $z$ must satisfy $\sum\limits_{S \subseteq \{a,b,c\}} Z(S)=1$ [from Lemma \ref{utility_common} and $u(z,z) = 1$] and $Z(S) \leq X(S)$ [from Corollary \ref{new_budget_z_ineq}].
%
% TODO: Explain more about Z_S = g(a, b, c, z)_{\{z\} \cup S} here.
%
%Recall that $X=g(a, b, c)$. Define $Z_S = g(a, b, c, z)_{\{z\} \cup S}$ for any $S \in \power{\{a, b, c\}}$. Then we have the following result for $Z.$
%
%
%\begin{theorem}
%For any budgets $a$, $b$, and $c$, and any outcome $z$ of $\Nash(a, b, c)$, $\langle u(a, z), u(b, z), u(c, z) \rangle$ is unique.
%\end{theorem}
%
%\begin{proof}
%This follows from Lemma~\ref{utility_common} and Lemma~\ref{nash_outcome}.
%\end{proof}
%
%
%
%\subsection{Total Utility Maximization}
%
%comment
% mention the knapsack voting paper ... idea of proof from that
%
%\edit{Further,} we show that outcomes of the Nash bargaining maximize the total utility among the three voters involved, \edit{two who are deliberating and one whose preferred budget is the disagreement point}. %The idea of the proof is that if we can show that the Nash bargaining allocates the budget to the portions that voters agree with most, then the total utility is maximized.

%\subsection{Technical Lemma~\ref{nash_outcome}}
%\label{Z_proof}
%Denote $\{a,b,c\}$ by $Q$. Recall that $X(S)=g_{Q}(S)$. Define the set function: $Z(S) = g_{Q \cup z}(S \cup z)$ for any subset $S$ of $Q$. Then we have the following result for $Z.$

%Recall the lemma corresponding to Nash Bargaining from Lemma~\ref{Nash_barg_outcome} which may be stated as follows
%The proof of this lemma is presented in Appendix~\ref{Z_nash_proof}. Note that this lemma has the following implication informally.

{\color{red} Recall that this lemma~\ref{Nash_barg_outcome} characterizes the bargaining solution in the common budget space $\{a,b,c\}$. Projects (Full or fractional) common to both budgets $a$ and $b$ definitely go to the budget of Bargaining outcome $z$ too. 

Suppose $\text{excess}$ as defined in Lemma~\ref{Nash_barg_outcome} is greater than 0 i.e. the sum of fractions of projects common to at least two budgets is less than 1. In this case, we definitely choose all projects common to at least two budgets among the three. However, we also choose some projects common to only $a$ (not $b$ nor $c$) of  quantity $\text{excess}$ and put into the bargaining solution. We follow a similar approach for budget $b$ as well.

However, when $\text{excess}$ as defined in Lemma~\ref{Nash_barg_outcome} is less than 0, i.e. the sum of fractions of projects common to at least two budgets is greater than 1. In this case, we first choose all projects common to at least two budgets out of three. However, we now choose projects of quantity $-\text{excess}$ from projects in both budgets $a$ and $b$ (but not in $c$) but do not put it into the Nash bargaining solution. A very similar approach can be followed for budget $c$ as well.

}

We use a randomized strategy to choose projects worth $|\text{excess}|$  from the desired set of projects. We present two randomizations (one over the project space) and one over the common budget space $X(S)$.

\subsection{Randomised Nash bargaining solution $\mathcal{N}_{\text{rand}}$}{\label{exact_soln_constr}}

We now propose a randomized Nash bargaining strategy denoted by $\mathcal{N}_{\text{rand}}$ to construct the bargaining budget $z$. We first look at projects (full or fractional) common to at least two budgets of the three budgets and all these projects are initially selected for the bargaining solution $z$ Recall that $X(S)$ and $Z(S)$ is defined over the common budget space $\{a,b,c\}$ [defined just above Lemma \ref{Nash_barg_outcome}]. We now consider the two following cases.

\begin{itemize}
    \item $X(ab) + X(bc) + X(ac) + X(abc)\leq 1$ - In this case the total quantity of projects in $z$ is still less than 1. Now we look at list of all projects (maybe fractional) which are present in only budget $a$ (neither in $b$ nor $c$). Amongst these projects, we choose projects worth $\text{excess}$ [defined in Lemma \ref{Nash_barg_outcome}] (without replacement) \footnote{to write about partition later so as to make integral} uniformly at random and put it into the bargaining solution $z$. A similar approach is followed for projects only in budget $b$ as well.

    \item $X(ab) + X(bc) + X(ac) + X(abc)\geq 1$ In this case the total quantity of projects in $z$ is greater than 1. Now we look at list of all projects (maybe fractional) which are present only in budgets $a$ and $c$ (not in $b$). Amongst these projects, we choose projects worth $-\text{excess}$ [defined in Lemma \ref{Nash_barg_outcome}] (without replacement) \footnote{to write about partition later so as to make integral} uniformly at random and remove it into the bargaining solution $z$. A similar approach is followed for projects only in budget $b$ as well.
\end{itemize}

Using this construction of the bargaining solution $z$, we now try to compute the Nash bargaining solution by computing $Z(S) = g_{(a,b,c,z)}(\{z\}\cup S)$.  
}

%In this section we define everything on the common budget space of $n$ budgets $P=(v_1,v_2,\ldots,v_n)$. 
%Recall $X_{j,P}(S)$ and $Z_{j,P}(S)$ from Definition ~\ref{proj_budget_defn} and ~\ref{new_budget_proj_budget_defn}. Define $X_P(S)$ and $Z_P(S)$ by summing $X_{j,P}(S)$ and $Z_{j,P}(S)$ respectively over all projects $j \in [m]$. 
%Note that $Z_P(S)$ must satisfy the Corollaries  \ref{new_budget_z_ineq},\ref{sum_budget_z} and  Lemma \ref{monotonicity_proj_budget_Z_space}.

Recall the randomized Nash bargaining scheme $\mathfrak{n}_{\text{rand}}$ explained in \S~\ref{exact_soln_constr}. In this section, we %analyse the distortion of $\mathfrak{n}_{\text{rand}}$ and 
derive an upper bound for it. Towards this, we first define a \emph{hypothetical} bargaining scheme $\tilde{\mathfrak{n}}_{\text{rand}}.$ This scheme is hypothetical because it assumes that the bargaining agents use some knowledge about the preferred budgets of the non-bargaining agents to break ties among potential outcomes. We then show in Lemma~\ref{distortion_comp_nash_alt_nash_lemma} that the Distortion of $\mathfrak{n}_{\text{rand}}$ is at most as much as that of $\tilde{\mathfrak{n}}_{\text{rand}}.$ We then bound the Distortion of $\tilde{\mathfrak{n}}_{\text{rand}}$ by its \emph{expected pessimistic distortion} (EPD), a quantity similar in essence to the PD. We define the EPD in Definition~\ref{exp_pess_dist_defn}. Our main technical contribution in this section is the analysis of the EPD of $\tilde{\mathfrak{n}}_{\text{rand}}$, which we do by expressing it as the solution of a bilinear optimization problem.

\subsection{Construction of % $\tilde{Z}_{j,P}(S)$
bargaining solution in $\tilde{\mathfrak{n}}_{\text{rand}}$}
{\label{sec_alt_barg_soln} }
%Note that we now define everything in the common budget space of $n$ voters with voting preference $(v_1,v_2,\ldots,v_n)$. Note that we now define $X_P(S) = g_P(S)$ as defined in Definition \ref{common_budget_defn} and we now directly give a construction of $\{\tilde{Z}_P(S)\}_{S \in \mathcal{P}(P)}$ which still satisfies the following conditions. 
%
Recall Definition~\ref{new_budget_proj_budget_defn} of $Z_{j,P}(\cdot)$ for an outcome budget $z$. Also recall that $Z_{j,P}(\cdot)$ satisfies Corollaries~\ref{new_budget_z_ineq} and~\ref{sum_proj_budget_z}.
% the following conditions [Corollaries~\ref{new_budget_z_ineq} and~\ref{sum_proj_budget_z}]:
\iffalse
\[
    0 \leq {Z}_{j,P}(S) \leq X_{j,P}(S) \text{~for all~} S \in \mathcal{P}(P) \text{~and all~} j \in [m].  \]
    %
\[  \sum_{j=1}^m ~~Z_{j,P}(S) =  Z_{P}(S) \ \ \text{~~and,~~}  \ \ \sum_{S \in \mathcal{P}(P)} ~~Z_{P}(S)  = 1. \]
\fi
%
 For $\tilde{\mathfrak{n}}_{\text{rand}},$ we characterize the outcome in the incremental allocation space; denoted by $\tilde{Z}_{j,P}(\cdot).$ 
Same as $Z_{j,P}(\cdot), \tilde{Z}_{j,P}(\cdot)$ also satisfies Corollaries~\ref{new_budget_z_ineq} and~\ref{sum_proj_budget_z}, i.e.,
%
%$\tilde{{Z}}_P(S)=\sum\limits_{j=1}^{m} \tilde{{Z}}_{j,P}(S)$
%
\begin{align}
    & 0 \leq \tilde{Z}_{j,P}(S) \leq X_{j,P}(S) \forall S \in \mathcal{P}(P) \text{~and all~} j \in [m].  \label{eq1_tilde_Z} \\
    &\sum_{j=1}^m ~~ \tilde{Z}_{j,P}(S) =   \tilde{Z}_{P}(S) \ \ \text{~~and,~~}  \ \ \sum_{S \in \mathcal{P}(P)} ~~ \tilde{Z}_{P}(S)  = 1. \label{eq2_tilde_Z}
\end{align}
%
%Also for any smaller subset of voters $Q\subseteq P$, we may define $\tilde{Z}_Q(\hat{S}) = \sum\limits_{\substack{\mathcal{S} \in \mathcal{P}(P\setminus Q); }} \tilde{Z}_P (\mathcal{S}\cup \hat{S})$ [similar to Lemma~\ref{proj_budget_new_space_lemma}] 
Before describing the construction of $\tilde{\mathfrak{n}}_{\text{rand}}$, we now give the following result on the overlap utility $u(a,z)$ of outcome budget $z$ and any budget $a \in P$ in terms of $Z_P(S)$.

\begin{lemma}{\label{utility_one_voter_proj_budget_Z_space}}
For a vote profile $P,$ a budget $a \in P,$ and any budget $z$, the overlap utility is $ u(a,z) = \sum_{S \in \mathcal{P}(P)|S \ni a} Z_P(S).$
\end{lemma}

\begin{proof}
In  Lemma~\ref{utility_common}, use $z$ for $b$, $a$ for $a,$ and $P\cup\{z\}$ for $P$.
% we have $u(a,z) = \sum_{S \in \mathcal{P}(P)\vert S \ni a}Z_P(S)$ %which equals $\sum\limits_{S \in \mathcal{P}(P)} |S| \cdot Z_P(S)$ 
\end{proof}

%Recall the representation of the overlap utility $u(v,z)$ in terms of $[Z_P(S)]_{S\in \mathcal{P}(P)}$ in Lemma~\ref{utility_one_voter_proj_budget_Z_space}.
%We use that representation to denote the overlap utility of any budget $v \in P$ with the outcome of $\tilde{\mathfrak{n}}_{\text{rand}}$. 
\vspace{-0.8 em}
By Lemma~\ref{utility_one_voter_proj_budget_Z_space},  %The overlap utility of $\{\tilde{Z}_{j,P}(.)\}_{j \in [m]}$ and a budget $a$ in vote profile $P$ is defined as
${u}(v, \tilde{Z}_P) = \sum_{S \in \mathcal{P}(P)| S \ni a} \tilde{Z}_P(S).$\footnote{Note the overload in the notation of the overlap utility; it was initially defined for a pair of budgets $v$ and $z$, here we define it for $v$ and $\tilde{Z}$ where $\tilde{Z}$ captures $z$.} %[similar to Lemma~\ref{utility_proj_budget_Z_space}] 
Similarly, the cost can be given by ${d}(v,\tilde{Z}_P) = 2 - 2u (v,\tilde{Z}_P).$

%and our construction of $\tilde{Z}_P(S)$ defined still satisfies  and , hence we can follow similar definitions and utility measures. Similarly, for any subset of voters say $Q \subseteq P$, we may define $\tilde{Z}_Q(S) = \sum\limits_{} $ 

%However, note that in this case the budget $\tilde{Z}$ may not be defined on the project space at all - only a representation in the common budget space. Also for this proof, we define $X_{j,P}(S) = g_{j,P}(S)$ [Definition \ref{proj_budget_defn}] informally capturing what fraction of project $j$ is allotted only to budgets in $S$ but no budget in $P\backslash S$. We now define its construction as follows. Our construction of $\tilde{Z}_P(S)$ \remove{selection mechanism }is very similar to that described in two cases in the previous section \ref{exact_soln_constr} just with a slight difference i.e. we directly propose a construction of $\{Z_{P}(S)\}_{S \in \mathcal{P}(P)}$

%We first construct $\tilde{Z}_{j,P}(S)$ as follows under the case where budgets $a,b,c$ are the bargaining budgets with $c$ being the disagreement point. As done previously in Section ~\ref{exact_soln_constr} we would first initialise $\tilde{Z}_{j,P}(S)$ to be $X_{j,P}(S)$ for every project $j$ and every subset $S \in \mathcal{P}(P)$ s.t $S$ contains at least two elements from the subset $Q=\{a,b,c\}$ and zero for all other sets. Note that $\tilde{Z}_{j,P}(S) = Z_{j,P}(S)$ as {\color{red} initialised} before {\color{red} Case \ref{less_1} and \ref{more_1}}. 

Let $c$ be the disagreement point, and $\{a,b\}$ be the preferred budgets of the agents chosen to bargain. Denote $Q = \{a,b,c\}.$ For the construction of $\tilde{Z}_P(\cdot),$ we first do \ref{Word:step1} and \ref{Word:step2} from \S~\ref{Sec:nash_outcomes}. We then have for all $j \in [m],$ $\tilde{Z}_{j,P}(S)= X_{j,P}(S)$ for all $ S \in \mathcal{P}(P) $ such that $S$ contains at least 2 elements of $Q$  and $\tilde{Z}_{j,P}(S)=0$ for all other $S \in \mathcal{P}(P)$. We then encounter either \ref{Word:case1} or \ref{Word:case2}, as in \S~\ref{Sec:nash_outcomes}. %$\tilde{Z}_P$ differs from $Z_P$ in the way these cases are handled.
%Now we construct two cases as done previously with a very similar allocation scheme but construction of $\tilde{Z}_{j,P}(S) $ is different.

%\textbf{Case 1:} $X_Q(ab) + X_Q(bc) + X_Q(ca) + X_Q(abc) \leq 1${\color{red} Case \ref{less_1}}
\ref{Word:case1}:  Here we need to allocate more funds to projects.
Recall the construction of $z$ for $\mathfrak{n}_{\text{rand}}$ in \S~\ref{exact_soln_constr}. Recall the random incremental allocations $r^a_j$ and $r^b_j$ used in $\mathfrak{n}_{\text{rand}}.$ For the incremental allocations in $\tilde{\mathfrak{n}}_{\text{rand}}$
%
%
    %In this case the total value of projects in $Z$ i.e. $\sum\limits_{j=1}^{m} \sum\limits_{S \in \mathcal{P}(P)} \tilde{Z}_{j,P}(S)$ is smaller than 1. We now select additional projects and assign them to $\tilde{Z}$ as follows.
    %Consider all projects $j$ which have incremental allocation in budget $a$ i.e., $X_{j,Q}(a) > 0.$ Now consider all projects $j$ with  $X_{j,Q}(a) > 0.$  [$\{X_{j,Q}(a)\}_{j \in [m]}$]. Amongst these projects, we choose projects worth $\frac{\textsc{excess}}{2}$ [defined in Lemma \ref{Nash_barg_outcome}] (without replacement) \footnote{to write about partition later so as to make integral} uniformly at random and put it into the bargaining solution $z$. Suppose we denote the quantity of such fractional projects selected by $[s^1_j]_{j \in [m]}$ (ensuring $\sum\limits_{j=1}s^{(1)}_j=\frac{\textsc{excess}}{2}$). 
 %   
 we construct $\alpha_{j,P}(S) = r^a_j \cdot (X_{j,P}(S)/X_{j,Q}(a))$\footnote{\textcolor{red}{\remove{Why is the fraction always $\leq 1$ and why does alpha always sum to excess/2?}} Note that $\alpha_{j,P}(S) \leq r^a_j$ since $X_{j,P}(S) \leq X_{j,Q}(a)$ [follows from Lemma ~\ref{proj_budget_new_space_lemma}] and $\sum_{j=1}^{m} \sum_{\substack{S \in \mathcal{P}(P) \\ S \ni a, S \not \ni b,c}}\alpha_{j,P}(S) = \frac{\textsc{excess}}{2}$ since $\sum_{\substack{S \in \mathcal{P}(P) \\ S \ni a, S \not \ni b,c}} X_{j,P}(S)= X_{j,Q}(S)$ [follows from Lemma ~\ref{proj_budget_new_space_lemma}] and the fact that $\sum_{j=1}^{m}r^a_j = \textsc{excess}/2$ [as defined in \ref{Word:case1} in \S \ref{exact_soln_constr}]. } for all $\{S~\vert~a \in S; b,c\not\in S\}$ for all projects $ j \in [m]$.  Intuitively, this may be thought of as a proportional selection of projects from every subset of budgets $S$.
 Similarly we construct $\beta_{j,P}(S) = r^b_j \cdot (X_{j,P}(S)/X_{j,Q}(b))$ for all $\{S ~\vert~b \in S; a,c \not\in S\}$ and all projects $j \in [m].$
 
 Now, set $\tilde{Z}_{j,P}(S)=\tilde{Z}_{j,P}(S)+\alpha_{j,P}(S)$  $ ~\forall~\{S~\vert~a\in S; b,c\not \in S\}$ and $\tilde{Z}_{j,P}(S)=\tilde{Z}_{j,P}(S)+\beta_{j,P}(S)$  $~\forall~\{S ~\vert~b\in S; a,c\not\in S\}~$and$ ~\forall~j \in [m].$

    %A similar approach is followed for projects only in budget $b$ as well.

    %Note that all these projects may be partitioned into $[[X_{j,P}(S)]_{j=1}^{m}]_{S \ni a, S \not\ni b,c}$ such that the sum of these quantities denotes the desired quantity. From these projects, we choose projects of quantity $|excess|$ (without replacement) \footnote{to write about partition to make it integral}, and the selected projects may also again be partitioned into $[[\alpha_{j,P}(S)]_{j=1}^{m}]_{S \ni a, S \not\ni b,c}$ \footnote{This can be thought of as sampling from a hyper-geometric distribution} with each $\alpha_{j,P}(S)$ denoting the precise fraction of project $j$ that was chosen from budgets in $S$. We can follow a very similar process for budget $b$ and obtain the sampled list to be $[[\beta_{j,P}(S)]_{j=1}^{m}]_{S \ni a; \not\ni b,c}$. Now, for every set $S \in \mathcal{P}(P)$ s.t $S \ni a$ but $\not \ni b,c$ and $j \in [1,m]$, we update $\tilde{Z}_{j,P}(S) = \tilde{Z}_{j,P}(S) + \alpha_{j,P}(S)$ and similarly for every set $S \in \mathcal{P}(P)$ s.t $S \ni b$ but $\not \ni a,c$ and $j \in [1,k]$, we update $\tilde{Z}_{j,P}(S) = \tilde{Z}_{j,P}(S) + \beta_{j,P}(S)$

    \ref{Word:case2}: 
    %$X_Q(ab) + X_Q(bc) + X_Q(ca) + X_Q(abc) \geq 1$ {\color{red} Case \ref{more_1}}
     In this case we need to remove allocations from projects.  Recall the construction of $z$ for $\mathfrak{n}_{\text{rand}}$ in \S~\ref{exact_soln_constr}. Recall the removals of allocations $r^a_j$ and $r^b_j$ used in $\mathfrak{n}_{\text{rand}}.$ For the removals of allocations in $\tilde{\mathfrak{n}}_{\text{rand}},$
    we construct $\alpha_{j,P}(S) = r^a_j \cdot (X_{j,P}(S)/X_{j,Q}(bc))$ for all $\{S~\vert~ b,c \in S , a \not\in S\}$ for all $ j \in [m]$.
 Similarly we construct $\beta_{j,P}(S) = r^b_j \cdot (X_{j,P}(S)/X_{j,Q}(ac)) $ for all $\{S~\vert~ a,c \in S; b \not\in S\}.$
    
 Now, set $\tilde{Z}_{j,P}(S)=\tilde{Z}_{j,P}(S) - \alpha_{j,P}(S)~\forall~\{S~\vert~ b,c \in S; a\not\in S\},$ and $\tilde{Z}_{j,P}(S)=\tilde{Z}_{j,P}(S) - \beta_{j,P}(S)~\forall~\{S~\vert~ a,c \in S; b \not\in S\}~\forall~j \in [m].$

We can now construct $\tilde{Z}_P(S)$ via $\tilde{Z}_P(S) = \sum_{j=1}^{m} \tilde{Z}_{j,P}(S).$ With this, we now construct $\tilde{Z}_Q$ as the outcome of the hypothetical bargaining process, via the projection from $P$ to $Q$ That is, $\tilde{Z}_Q(S) = \sum_{\hat{S} \in \mathcal{P}(P\setminus Q)} \tilde{Z}_P(S \cup \hat{S})$ [recall projection in Lemma~\ref{proj_budget_new_space_lemma}].

See that $\{\tilde{Z}_{j,P}(.)\}_{j \in [m]}$ satisfies Corollaries~\ref{new_budget_z_ineq} and~\ref{sum_proj_budget_z}. Further, $\tilde{Z}_Q(.)$ satisfies
%all the rules discussed in Equations \eqref{eq1_tilde_Z} and \eqref{eq2_tilde_Z} 
all equations of Lemma ~\ref{Nash_barg_outcome} [proof in Appendix \ref{results_tilde_Z}\remove{\S\ref{results_tilde_Z}}].
%when projected on the incremental common budget space of 3 voters $Q= \{a,b,c\}$. In other words $\tilde{Z}_Q(a) = \sum\limits_{\substack{S \ni a; S \not\ni b,c\\ S \in \mathcal{P}(P)}} \tilde{Z}_P(S)$ would satisfy all propositions on $\mathcal{Z}(a)$ and similarly for other sets $b,c,ab,\newline \ldots, abc$ can be defined and call this allocation into common budget space as $\tilde{\mathfrak{n}}_{\text{rand}}$. 

\remove{
We now state the following lemma using the constructions described above. This lemma essentially says that the utility of the allocation scheme described using $\tilde{Z}$ has a lower utility than that of our original construction. This bound is helpful as we would upper bound the distortion ratio of our construction $\tilde{\mathcal{N}}_{rand}$ and the same bound would hold for the original scheme $\tilde{N}_{rand}$ too.

\begin{lemma}{\label{alt_Nash_utility_bound}}
    $\sum\limits_{i=1}^{k} {E}[\tilde{u}(\tilde{\mathcal{N}}(a,b,c),v_j)] \leq \sum\limits_{i=1}^{k} {E}[{u}({\mathcal{N}}(a,b,c),v_j)]$  for all budgets $ a,b,c \in P$ and vote profiles given by $P=(v_1,v_2,\ldots,v_k)$.
\end{lemma}

}

%In the set of projects selected of quantity $|\text{excess}|$, we directly add or remove it from $\tilde{Z}_{j,P}(S)$ is the fraction selected is of project $j$ and it is precisely contained by voters in $P$. We do it similarly for both the cases described. 

%Note that using the construction mentioned above, we now define $\tilde{Z}_P(S)$, which satisfies the previous constraints discussed above Lemma ~\ref{Nash_barg_outcome} namely $\tilde{Z}(S) \leq X_P(S)$ and $\sum\limits_{S \in \mathcal{P}(P)} \tilde{Z}(S)=1$

%Projects common to all three budgets definitely belong to the bargaining solution as well and projects present in none don't go to the bargaining solution. However, when the total fraction of projects common to at least two budgets are larger than 1, we don't have any project (complete or fraction) present in exactly one budget going to the bargaining solution. Otherwise, every fractional project that is common to at least two budgets goes to the bargaining solution.}

%Note we now

\remove{
\begin{proof}
%Let $X=g(a, b, c)$ and $Z_S = g(a, b, c, z)_{\{z\} \cup S}$. %BLAH
From Lemma~\ref{nash_outcome} in Appendix~\ref{Z_proof}, we have: $$k_a - X(ac) = k_b - X(bc),$$ 
where $k_a = (1 - X(abc) - X(ab) + X(ac) - X(bc)) / 2$, \\and $k_b = (1 - X(abc) - X(ab) + X(bc) - X(ac)) / 2$.

\textbf{Case 1: }If $Z(a) = 0$, we have
$Z(a) = \max(0, k_a - X(ac)) = 0$. This implies 
$k_a - X(ac) \leq 0$ and 
$k_b - X(bc) \leq 0$.
Thus, $Z(b) = \max(0, k_b - X(bc)) = 0$.
%Therefore, by Lemma~\ref{knapsack}, the total utility is maximized.

\textbf{Case 2: }If $Z(a) > 0$, we have
$Z(a) = \max(0, k_a - X(ac)) > 0$. This implies $k_a - X(ac) > 0$ and
$k_b - X(bc) > 0$.
Thus, $Z(ac) = \min(k_a, X(ac)) = X(ac)$ and $Z(bc) = \min(k_b, X(bc)) = X(bc)$.

By Lemma~\ref{knapsack} in Appendix~\ref{proof_knapsack}, the sum of utilities $u(a, z) + u(b, z) + u(c, z)$ is maximized in both cases.% Case 1 and Case 2. 
\end{proof}
}
%The proof of Theorem~\ref{max_utility} is technical and is in Appendix~\ref{proof_max_utility}, along with supporting lemmas, Lemma~\ref{nash_outcome} in Appendix~\ref{Z_proof} and Lemma~\ref{knapsack} in Appendix~\ref{proof_knapsack}. 
%An important implication of this result is that the outcome of Nash bargaining is a \emph{median} of the three budgets $a,b,$ and $c.$ Therefore we can use tools applicable for median voting rules~\cite{comanor1976median} to our framework.
%Note that now we work in the common budget space of the set of preferred budgets over all voters $P$ and map every every subset of voters (given by $S$) to a number in $[0,1]$. This quantity informally captures the fraction of overlapping projects in $S$.   

\remove{
\subsection{Outcomes of Nash Bargaining}

%In this section, we look at bargaining processes and total utility of their outcomes among the three voters involved.
%We will now look at how bargaining processes can affect the utility. We define the social welfare of a budget to be the sum of the overlap utilities of voters and the budget. We show that, for three voters, outcomes from the Nash bargaining maximize the social welfare among the three voters.

%\subsection{Common Budget Space}

%With the result of Theorem~\ref{cs_main}, it is clear that Nash bargaining takes the project interactions into consideration and does not output budgets with $ \sum_i f(b)_i <1$. Going ahead we do not worry about project interactions and give results which are valid with and without them. To make it easy to prove some properties of the Nash bargaining, we primarily work in the \emph{common budget space}. 
\remove{

\edit{Recall that we refer to the set of preferred budgets of all voters as the vote profile $P$. In the common budget space, we map each subset of the vote profile to a number in the interval $[0,1]$. This quantity for a subset $S$ captures, the sum over all projects $j$, of how much do budgets in $S$ agree on increasing the allocation to $j$ above the maximum allocation to $j$ by any budget in $P \backslash S$. Informally, this quantity captures a notion of similarity of the budgets in a subset $S$ and their dissimilarity with budgets in $P \backslash S$.} Formally,

\begin{definition}
Given a vote profile $P=(v_1, v_2, \ldots, v_k),$ the \textbf{common budget function} $g_P: 2^{[k]} \rightarrow \mathbb{R}$ maps a subset of voters to a point in the common budget space
%whose coordinates are subsets of $\{v_1, v_2, \ldots, v_k\}$
where, for each subset $S$ of voters,
 $g_P(S) = \sum\limits_{j=1}^m \max \bigg(\Big(\min\limits_{s \in S} v_{s,j} \Big) - \Big(\max\limits_{\tilde{s} \in P \setminus S} v_{\tilde{s},j} \Big), 0 \bigg).$

%(Define $g(P)_\emptyset = 0$.)  % should probably be \infinite ? Doesn't matter as we never use it.
\end{definition}

%See that $g_P(S)$ tells us how much allocation budgets in $S$ have in common, over and above the allocation in $P\backslash S$. 
\edit{\textbf{Example 3.2:}} Consider \edit{an instance of PB with three projects and a vote profile $P$ with} three budgets $a = \langle 1, 0, 0 \rangle$, $b = \langle 0, 1, 0 \rangle$, and $c = \langle 0.25, 0.25, 0.5 \rangle$. Let $X(S) = g_{P}(S)$ \edit{for each subset $S$ of $P$}.\footnote{For brevity, we omit braces and commas in the argument of $X$.} Then, $X(a) = 0.75.$ This is because the budget $a$ has allocation $1$ to project $1,$ out of which $0.25$ is shared with budget $c$ and $0.75$ is \emph{exclusive}. Similarly $X(b) = 0.75,$ and $X(c) = 0.5$. Further, $X(ac) = 0.25$. This is because the subset $\{a,c\}$ has an allocation of $0.25$ to project $1$ common to both $a$ and $c$. It is also exclusive since $b$ has no allocation to project $1$. Further, we have $X(abc) = 0.$ This is because the group of all three budgets has no allocation that is common to all.
 
 \edit{Note that the number of points in the \emph{common budget space} is independent of the number of projects on the ballot. It is only a function of the number of voters. This helps in making the analysis of distortion in the next section computationally feasible.}
 
 \edit{We now give a formulation of the overlap utility of two budgets $a$ and $b$ in the common budget space. Recall that $\mathcal{P}(P)$ denotes the power set of $P$, i.e., the set of all subsets of the vote profile $P$. }
\begin{lemma}
\label{utility_common}
For any pair of budgets $a$ and $b$ and any vote profile $P$ that includes $a$ and $b$, we have $u(a, b) = \sum\limits_{ S \in \mathcal{P}(P \backslash \{a,b\}) }  g_P(  S \cup \{a,b\})$.
\end{lemma}
}
% FIXME: Type error? The subscripts refer to the variables, not the indices to the variables.

%\subsection{Outcomes of Nash Bargaining}

Recall Nash bargaining given in Definition~\ref{def:nb}. We now characterize the outcomes of the Nash bargaining \edit{in the common budget space}. Given budgets $a$, $b$, and $c$, there may be multiple outcomes of $\Nash(a, b, c)$. However, as we will see, all these outcomes correspond to a unique point in the common budget space.
%
% TODO: Explain more about Z_S = g(a, b, c, z)_{\{z\} \cup S} here.
%
%Recall that $X=g(a, b, c)$. Define $Z_S = g(a, b, c, z)_{\{z\} \cup S}$ for any $S \in \power{\{a, b, c\}}$. Then we have the following result for $Z.$
%
%
%\begin{theorem}
%For any budgets $a$, $b$, and $c$, and any outcome $z$ of $\Nash(a, b, c)$, $\langle u(a, z), u(b, z), u(c, z) \rangle$ is unique.
%\end{theorem}
%
%\begin{proof}
%This follows from Lemma~\ref{utility_common} and Lemma~\ref{nash_outcome}.
%\end{proof}
%
%
%
%\subsection{Total Utility Maximization}
%
%comment
% mention the knapsack voting paper ... idea of proof from that
%
\edit{Further,} we show that outcomes of the Nash bargaining maximize the total utility among the three voters involved, \edit{two who are deliberating and one whose preferred budget is the disagreement point}. %The idea of the proof is that if we can show that the Nash bargaining allocates the budget to the portions that voters agree with most, then the total utility is maximized.

\begin{theorem}
\label{max_utility}
For any budgets $a$, $b$, and $c$, any outcome $z$ of\\ $\mathcal{A}(a, b, c)$ maximizes the sum of utilities $u(a, z) + u(b, z) + u(c, z)$. 
\end{theorem}
\begin{comment}
\begin{proof}
%Let $X=g(a, b, c)$ and $Z_S = g(a, b, c, z)_{\{z\} \cup S}$. %BLAH
From Lemma~\ref{nash_outcome} in Appendix~\ref{Z_proof}, we have: $$k_a - X(ac) = k_b - X(bc),$$ 
where $k_a = (1 - X(abc) - X(ab) + X(ac) - X(bc)) / 2$, \\and $k_b = (1 - X(abc) - X(ab) + X(bc) - X(ac)) / 2$.

\textbf{Case 1: }If $Z(a) = 0$, we have
$Z(a) = \max(0, k_a - X(ac)) = 0$. This implies 
$k_a - X(ac) \leq 0$ and 
$k_b - X(bc) \leq 0$.
Thus, $Z(b) = \max(0, k_b - X(bc)) = 0$.
%Therefore, by Lemma~\ref{knapsack}, the total utility is maximized.

\textbf{Case 2: }If $Z(a) > 0$, we have
$Z(a) = \max(0, k_a - X(ac)) > 0$. This implies $k_a - X(ac) > 0$ and
$k_b - X(bc) > 0$.
Thus, $Z(ac) = \min(k_a, X(ac)) = X(ac)$ and $Z(bc) = \min(k_b, X(bc)) = X(bc)$.

By Lemma~\ref{knapsack} in Appendix~\ref{proof_knapsack}, the sum of utilities $u(a, z) + u(b, z) + u(c, z)$ is maximized in both cases.% Case 1 and Case 2. 
\end{proof}
\end{comment}
The proof of Theorem~\ref{max_utility} is technical and is in Appendix~\ref{proof_max_utility}, along with supporting lemmas, Lemma~\ref{nash_outcome} in Appendix~\ref{Z_proof} and Lemma~\ref{knapsack} in Appendix~\ref{proof_knapsack}. 
An important implication of this result is that the outcome of Nash bargaining is a \emph{median} of the three budgets $a,b,$ and $c.$ Therefore, we can use tools applicable for median voting rules~\cite{comanor1976median} to our framework.

}

\subsection{Distortion under $\mathfrak{n}_{\text{rand}}$}
\label{sec:distortion_Nash_Randomised}
%Note that Lemma~\ref{alt_Nash_utility_bound} would imply that the following lemma
%\input{fig}
We now bound the distortion of the triadic scheme with bargaining scheme ${\mathfrak{n}}_{\text{rand}}$ by that of the hypothetical scheme $\tilde{\mathfrak{n}}_{\text{rand}}$. A proof is in the Appendix \ref{Sec:alt_Nash_distortion_bound_proof}. \remove{~\ref{Sec:alt_Nash_distortion_bound_proof}}

\begin{lemma}{\label{distortion_comp_nash_alt_nash_lemma}}
$\Distortion_once({\mathfrak{n}}_{\text{rand}}) \leq \Distortion_once({\tilde{\mathfrak{n}}}_{\text{rand}})$.
\end{lemma}

We now follow a similar approach as in \S\ref{sec:distort_Nash_relaxed} and define expected pessimistic distortion under bargaining scheme $\tilde{\mathfrak{n}}_{\text{rand}}$ as follows. %

\begin{definition}{\label{exp_pess_dist_defn}}
The \textbf{expected pessimistic distortion} of  $\tilde{\mathfrak{n}}_{\text{rand}}$ with triadic scheme with 6 voters, $\text{EPD}(\tilde{\mathfrak{n}}_{\text{rand}}) $ is
%\edit{with 6 voters} (under any set of schemes $\mathcal{A}$) is:
% \[ \max_{P \in \mathbb{B}^5} \frac{ \frac{1}{20} \sum_{R \in \combi{P, 3}} \sum_{v \in P \setminus R}  d(\Nash(R), v) }
%{ \min_{p \in \mathbb{B}} \frac{1}{5} \sum_{v \in P} d(p, v) }. \]

$$\sup\limits_{P \in \mathbb{B}^6 } \frac{\displaystyle \frac{1}{60} \sum_{c \in [6]}\sum_{\scriptsize \substack{\{a,b\} \in \\ \combi{[6]\backslash \{c\}, 2}}} \frac{1}{3} \sum\limits_{i \in [6] \setminus \{a,b,c\}}  \mathbb{E}[d(\tilde{\mathfrak{n}}_{\text{rand}}(a,b,c), P_i)] }
{\displaystyle \min_{p \in \mathbb{B}} \frac{1}{6} \sum\limits_{i \in [6]} d(p, P_i) }.$$
\end{definition}

%by $\tilde{PD}_{\mathcal{N}_{\text{rand}}}(\Triadic)$ where we use the distance measure $\tilde{d}$ [defined in Sec \ref{sec_alt_barg_soln}] (also use expectation over $d$ as it is randomised) instead of $d$ and $\mathcal{N}$ instead of $\mathcal{A}$ Using a very similar approach as in Lemma ~\ref{distortion_pd} we can show

\begin{lemma}{\label{distortion_alt_Nash_bargaining}}
$\Distortion_once(\tilde{{\mathfrak{n}}}_{\text{rand}}) \leq \text{EPD}(\tilde{\mathfrak{n}}_{\text{rand}}).$
\end{lemma}

The proof is similar to Lemma ~\ref{distortion_pd} and is in Appendix \ref{distortion_alt_Nash_bargaining_proof}.
%\ref{distortion_alt_Nash_bargaining_proof}.
%We now use prove the following lemma 
\begin{lemma}{\label{distortion_Nash_bargaining_randomised}}
$\text{EPD}(\tilde{{\mathfrak{n}}}_{\text{rand}}) \leq 1.66.$
\end{lemma}

The proof is similar to that of Lemma \ref{pd_triadic_nash_relaxed} and is presented in Appendix \ref{sec_proof:distortion_Nash_bargaining_randomised}. \remove{\S ~\ref{sec_proof:distortion_Nash_bargaining_randomised}. }We present the key ideas of the proof here.
\begin{hproof}
    Recall the construction of $\tilde{Z}_{j,P}(.)\sim \tilde{\mathfrak{n}}_{\text{rand}}(a,b,c)$ and consider \ref{Word:case1}. A similar analysis holds for \ref{Word:case2} as well.
    
    We show in Appendix A.20 \remove{\S ~\ref{sec_proof:distortion_Nash_bargaining_randomised}} that $\mathbb{E}[\tilde{Z}_P(S)] = \gamma^1_a X_P(S) \ \text{ for all } \\ \{S: S \ni a; S \not\ni b,c\}$ and $\mathbb{E}[\tilde{Z}_P(S)] = \gamma^1_b X_P(S) \ \text{ for all }\{S: S \ni b; S \not\ni a,c\}$ for some variables $0 \leq \gamma^1_a, \gamma^1_b \leq 1$. Here, $\gamma^1_a$ and $\gamma^1_b$ denote what fraction of allocation from the incremental allocation $X_P(S)$ is ``accepted'' into $\tilde{Z}_P(S)$. % for each $S$ in $\{S: S \ni a; S \not\ni b,c\}$ and $\{S: S \ni b; S \not\ni a,c\}$ respectively. 
    In our optimization problem formulation equation \eqref{opt_bilinear} in Appendix \ref{sec_proof:distortion_Nash_bargaining_randomised} \remove{\S ~\ref{sec_proof:distortion_Nash_bargaining_randomised}, },we use  $\gamma^1_b,\gamma^1_a$ as variables of our optimization formulation, together with $X_P(S)$ and therefore we obtain a bilinear program. We solve it with the Gurobi solver \cite{gurobi}. Similar to the proof of Lemma~\ref{distortion_pd}, we remove the cases that are not unique to permutations of voters and use further symmetries of the problem %also use Lemma~\ref{toggle_redn} in Appendix \S\ref{sec_proof:distortion_Nash_bargaining_randomised} 
    to reduce number of bilinear programs from $2^{{{6\choose 3}}}$ to 1244.
\end{hproof}

    %where $\gamma^1_a$ and $\gamma^1_b$ are the proportionality 

\remove{
The proof is similar to that of Lemma \ref{pd_triadic_nash_relaxed} and uses a construction of a bilinear optimization problem. We give a proof sketch here.

 Consider a disagreement point $c$ and preferred budgets of bargaining agents, $a$ and $b$. Let $Q$ denote $\{a,b,c\}.$ Suppose $\tilde{Z}_{j,P}$ is sampled from $\tilde{\mathfrak{n}}_{\text{rand}}(Q).$  Consider the following two cases.
 
 %\textbf{Case 1:} $X_{j,Q}(ab)+X_{j,Q}(bc)+X_{j,Q}(ac)+X_{j,Q}(abc)\leq 1$. 
\ref{Word:case1}:
Recall from \S\ref{exact_soln_constr} that %since we choose projects uniformly at random without replacement, we have 
$\mathbb{E}[r^{a}_j]$ is proportional to  $s^a_j$ which is the same as $ X_{j,Q}(a)$. Using the fact that $\sum_{j=1}^{m} s^a_j = X_Q(a)$ and $\sum_{j=1}^{m} r^a_j = \textsc{Excess}/2$, we have  $\mathbb{E}[r^{a}_j] =\frac{X_{j,Q}(a)
\cdot\textsc{excess}/2}{X_{Q}(a)}.$ %\footnote{$r^a_j$ is proportional to $s^a_j = X_{j,Q}(a)$ [\ref{Word:case1} of \S \ref{exact_soln_constr}] and $\sum_{j=1}^{m} s^a_j = X_Q(a)$}
Now, from our construction of  $\tilde{Z}(S)$ using $\alpha_{j,P}(S)$ as described in Case 1 in  \S\ref{sec_alt_barg_soln}, we have the following for any set $\{S ~\vert~ S \subseteq P; a\in S; b,c\not \in S\}$.
\begin{align}
  \mathbb{E}[\tilde{Z}_{j,P}(S)]=\frac{X_{j,Q}(a)  \textsc{excess}/2}{X_{Q}(a)} \frac{X_{j,P}(S)}{X_{j,Q}(a)} = \frac{\textsc{excess} \cdot X_{j,P}(S)}{2 X_Q(a)}   \label{exp-1}
\end{align}
Taking the sum over all projects $j \in [m]$, we get  $\mathbb{E}[\tilde{Z}_{P}(S)] = \gamma\cdot X_P(S)$ for all  $\{S ~\vert~ S \subseteq P; a\in S, b,c\notin S\}$, for some $0 \leq \gamma \leq 1$. A similar result can be proven on the sets $\{S ~\vert~ S \subseteq P, b\in S, a,c\notin S\}$.

\ref{Word:case2}: %$X_{j,Q}(ab)+X_{j,Q}(bc)+X_{j,Q}(ac)+X_{j,Q}(abc) \geq 1$
Recall from \S\ref{exact_soln_constr} that %since we choose projects uniformly at random without replacement, we have 
$\mathbb{E}[r^{a}_j]$ is proportional to  $t^a_j$ which is the same as $ X_{j,Q}(bc)$. Using the fact that $\sum_{j=1}^{m} t^a_j = X_Q(bc)$ and $\sum_{j=1}^{m} r^a_j = \textsc{Excess}/2$, we have  $\mathbb{E}[r^{a}_j] =\frac{X_{j,Q}(bc)
\cdot\textsc{excess}/2}{X_{Q}(bc)}.$

%$\mathbb{E}[r^{a}_j]=\frac{X_{j,Q}(bc) \cdot \textsc{excess}/2}{X_{Q}(bc)}.$\footnote{$r^a_j$ is proportional to $t^a_j=X_{j,Q}(bc)$ [\ref{Word:case2} of \S\ref{exact_soln_constr}] and $\sum_{j=1}^{m} t^a_j = X_Q(bc).$}
%
From our construction of $\tilde{Z}(S)$ using $\alpha_{j,P}(S)$ as described in Case 2 in \S\ref{sec_alt_barg_soln}, we have that for any sets $\{S ~\vert~ S \subseteq P, S \not\ni a, S\ni b,c\}$:
\begin{align}
 \mathbb{E}[\tilde{Z}_{j,P}(S)] 
 %= X_{j,P}(S)-\frac{X_{j,Q}(bc) \cdot \textsc{excess}/2}{X_{Q}(bc)} \cdot \frac{X_{j,P}(S)}{X_{j,Q}(bc)} 
   =  X_{j,P}(S) \left(1-\frac{\textsc{excess}}{2 \cdot X_Q(bc)}\right).    \label{exp-2}
\end{align}
   Taking the sum over all projects $j \in [m]$, we get $\mathbb{E}[\tilde{Z}_{P}(S)] = \gamma\cdot X_P(S)$ for all  $\{S ~\vert~ S \subseteq P; a\notin S; b,c\in S\}$ for some $0 < \gamma < 1$. A similar result can be proven on sets $\{S ~\vert~ S \subseteq P; b \notin S; a,c \in S\}$.
   \iffalse
\begin{align*}
   \mathbb{E}[\tilde{Z}_{j,P}(S)] = & X_{j,P}(S)-\frac{X_{j,Q}() \cdot \textsc{excess}/2}{X_{Q}(a)} \cdot \frac{X_{j,P}(S)}{X_{j,Q}(a)}\\
   = & X_{j,P}(S) \left(1-\frac{\textsc{excess}}{2 \cdot X_Q(a)}\right).\\
\end{align*}
\fi

Using~\eqref{exp-1} and~\eqref{exp-2}, we formulate our objective function as the numerator of EPD minus 1.66 times its denominator. This optimization problem is a bilinear program with complete details in Appendix ~\ref{sec_proof:distortion_Nash_bargaining_randomised}. We solve it using the Gurobi package \cite{gurobi} in Python. 
 %
%Now using these equations, we can formulate it as a bilinear problem which we solved using the Gurobi optimization package in Python. The complete proof is given in Appendix ~\ref{sec_proof:distortion_Nash_bargaining_randomised}.\qedhere
%
}
Using Lemmas~\ref{distortion_comp_nash_alt_nash_lemma},~\ref{distortion_alt_Nash_bargaining}, and~\ref{distortion_Nash_bargaining_randomised}, we get the following result.

\begin{theorem}\label{distortion_Nash_bargaining_randomised_final_result}
$\Distortion_once({\mathfrak{n}}_{\text{rand}}) \leq 1.66.$
\end{theorem}

%Theorem~\ref{distortion_Nash_bargaining_randomised} bounds the ratio of total expected disutility of the bargaining after one round of bargaining under the scheme $\mathfrak{n}_{\text{red}}$ with all voters in a voting profile $P$ and the disutility of the optimum budget with all voters in the voting profile $P$ and bounds it by 1.66.

%However, note that whenever any set of budgets $a,b$ and $c$ are chosen for bargaining we get ${E}[\tilde{Z}_P(S)] = \alpha \times X_P(S)$ \footnote{Case - I as described in Sec~\ref{sec_alt_barg_soln}} when $S \ni a \text { but } \not \ni b,c$ for some $\alpha$ which does not depend on set $S$, which gives rise to a bilinear optimization problem solved using gurobi \footnote{ Describe } optimization package.  U
\section{Empirical Results}
\label{sec:exp}

Recall triadic scheme as described in \S \ref{sec:model_SD}. We now define a sequential deliberation mechanism that could run bargaining over multiple rounds by setting the disagreement point for each round as the outcome of the previous round as proposed in \cite{fain2017}.

\begin{enumerate}[leftmargin=0.5cm]
    \item Pick a voter $i$ uniformly at random. Set the disagreement point for the deliberation $c$ to their preferred budget $v_i.$
    \item Repeat the following process $T$ times, %\edit{where $T$ is the number of rounds of deliberations, set by the PB organizer} :
    \begin{enumerate}[leftmargin = 0.1 cm]
        \item Pick two voters \edit{independently and} uniformly at random with replacement. They bargain with $c$ as the disagreement point.
        \item Set the disagreement point $c$ to the outcome of the bargaining.
    \end{enumerate}
    \item The outcome of the process is $c$.
\end{enumerate}

\begin{figure}[h]
	\centering
	\begin{subfigure}{.32\textwidth}
		\centering
		\includegraphics[scale = 0.18]{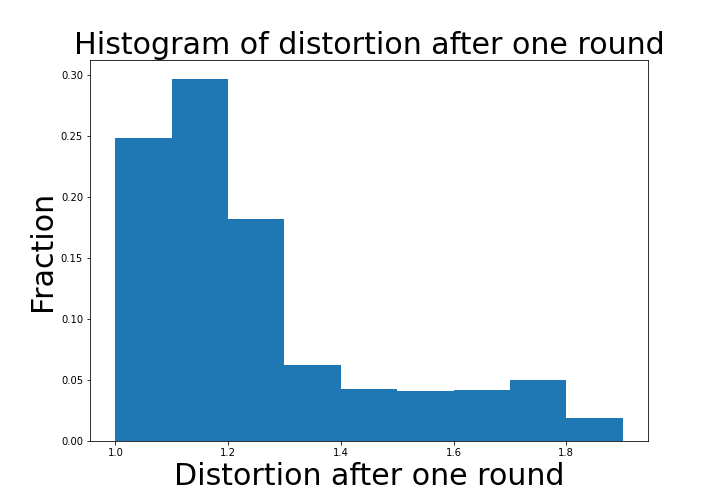}
		\caption{Histogram of triadic scheme}
		\label{histogram}
	\end{subfigure}%
        \centering
	\begin{subfigure}{.32\textwidth}
		\centering
		\includegraphics[scale = 0.33]{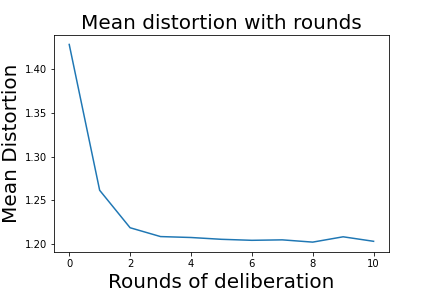}
		\caption{\hspace{ 1 em} Average distortion }
		\label{mean}
	\end{subfigure}%
 \centering
        \begin{subfigure}{.32\textwidth}
		\centering
		\includegraphics[scale = 0.33]{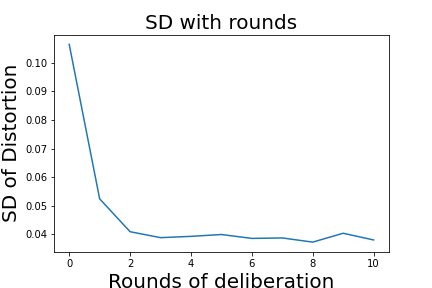}
		\caption{SD of distortion}
		\label{sd}
	\end{subfigure}%
 \caption{Distortion results on PB platform in Boston under the fractional allocation setup}
\end{figure}

Observe that on setting $T=1$, we exactly get triadic scheme as \S\ref{sec:model_SD}. To evaluate the distortion of sequential deliberation in PB empirically, we ran a simulation from the online participatory budgeting elections in Boston in $2016$ ($n=4,482$), Cambridge in $2015$ ($n=3,273$), Greensboro in $2019$ ($n=512$), and Rochester in $2019$ ($n=1,563$) where the data were obtained from \url{https://budget.pbstanford.org/}. In these elections, projects had a fixed cost, and voters participated in knapsack voting \cite{goel2019knapsack}, in which they could choose any number of projects as long as they fitted within the fund limits. {Note that in this simulation setup partial project funding is not allowed, unlike the setup in the theoretical model.}\remove{though this simulation setup is different from the setup in which we present our theoretical results, it supplements and generalises the same,} \remove{\color{red} with the key difference being that in this setup partial funding of projects is permitted unlike our theoretical setup.}We further present simulation results in Figures ~\ref{histogram}, \ref{mean},~\ref{sd} on real dataset from a PB (participatory budgeting) process run by a non-profit organisation in Boston in 2016 where they used a fractional allocation setting, more aligned with our theoretical work. 
%\vspace{-1 em}
\begin{figure*}[h!]
\centering
\begin{subfigure}{.5\textwidth}%%.5\textwidth
  \centering
  \includegraphics[scale = 0.71]{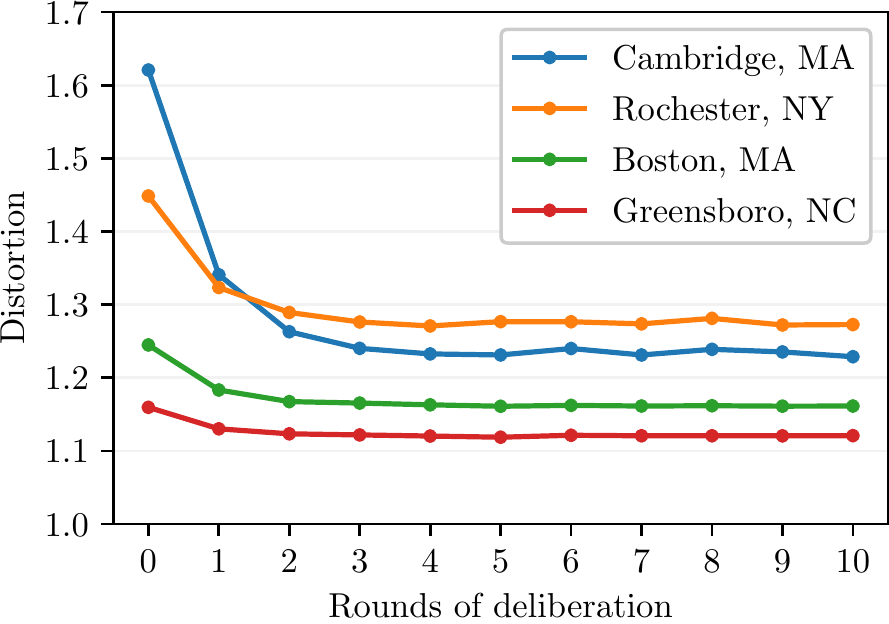} %[width=.95\linewidth]
  \caption{Average distortion}
 \label{experiment_distortion}
\end{subfigure}%
\begin{subfigure}{.5\textwidth}
  \centering
  \includegraphics[scale = 0.71]{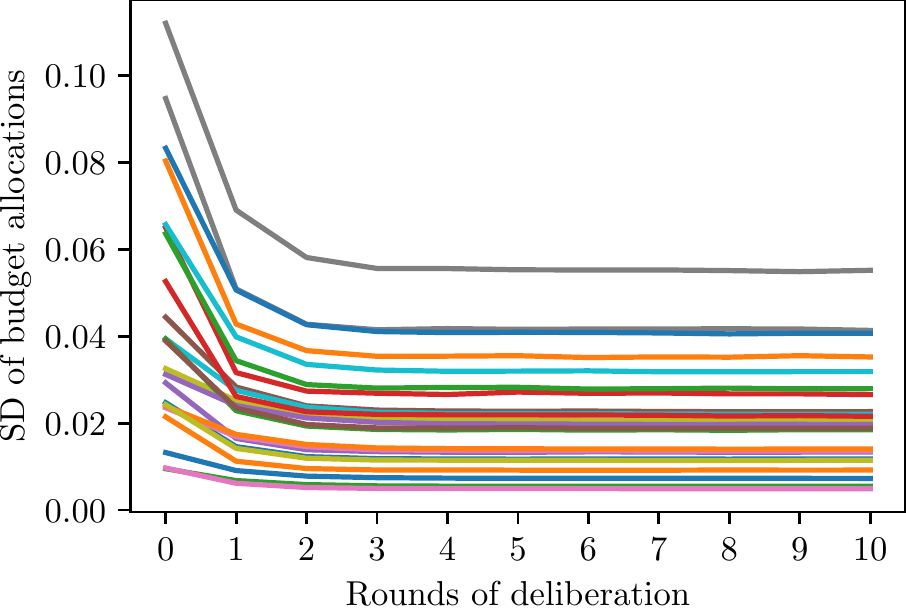}%[width=.95\linewidth]
  \caption{SD of distortion}
 \label{experiment_sd}
 \end{subfigure}
 %\vspace{-0.8cm}
\caption{(a) The average distortion after each round of sequential deliberation in a simulation using the data from PB elections in four cities. The simulation was run 10,000 times for each city.
(b) The standard deviation (SD) of the fund allocation to each project in the simulation of sequential deliberation in the PB election in Cambridge. Each line represents a project.}
\vspace{-0.25cm}
\label{fig:test}
\end{figure*}

%Stanford Participatory Budgeting Platform\footnote{\url{https://pbstanford.org/}}.  
\iffalse
\begin{figure*}[h!]
\centering
\begin{subfigure}{.5\textwidth}
  \centering
  \includegraphics[width=.95\linewidth]{plots/distortion}
  \caption{}
 \label{experiment_distortion}
\end{subfigure}%
\begin{subfigure}{.5\textwidth}
  \centering
  \includegraphics[width=.95\linewidth]{plots/sd}
  \caption{}
 \label{experiment_sd}
 \end{subfigure}
 \vspace{-0.8cm}
\caption{(a) The average distortion after each round of sequential deliberation in a simulation using the data from PB elections in four cities. The simulation was run 10,000 times for each city.
(b) The standard deviation (SD) of the fund allocation to each project in the simulation of sequential deliberation in the PB election in Cambridge. Each line represents a project.}
\label{fig:test}
\end{figure*}
\fi
To simulate sequential deliberation, we picked a voter \edit{uniformly at} random to set their preferred budget as the disagreement point. We then picked another two voters \edit{independently and} uniformly at random and calculated a Nash bargaining solution between them. We assumed that everyone voted truthfully. We then made the bargaining outcome the new disagreement point and repeated the deliberation process for $T = 10$ rounds. We repeated this entire simulation $10,000$ times for each PB election. The average distortion after each round of deliberation is shown in Figure~\ref{experiment_distortion}. The point corresponding to $0$ rounds of deliberation is the first disagreement point and is selected uniformly at random. \remove{In these elections, projects had a fixed cost, and voters participated in knapsack voting, in which they could choose any number of projects as long as they fitted within the fund limits.} %{\color{red} Note that though this simulation setup is different from the setup in which we present our theoretical results, it supplements and generalises the same. We further prese  }%Boston and Greensboro used knapsack voting as the main voting method, while Cambridge and Rochester used it as an experimental voting method with a disclaimer that stated to the voters that the results would not affect their actual vote.
Since voters did not have to use all the budget available, we added an `unspent' project and allocated the unspent budget of each voter to this project. We normalized the budget to sum to $1$ in each election. 

%Since the ground-truth project interactions are not known for these elections, we show the distortion results without considering any project interactions. Recall that the distortion of sequential deliberation is no worse in the presence of project interactions than without them. This follows from Theorem~\ref{cs_main}. 

{The mean and standard deviation of the distortion after each round of sequential deliberation for the fractional allocation setting as in the PB process in Boston is shown in Figures~\ref{mean} and~\ref{sd}, respectively. A histogram plot of the distortion after one round of deliberation is in Figure~\ref{histogram}. As before, we observe a quick convergence within three rounds of sequential deliberation with the point corresponding to zero rounds of deliberation being the first disagreement point.}

The results from all the PB elections show that the average distortion is quite low, even after only two rounds of deliberation. It also shows that the distortion converges quickly within three rounds. %Therefore, running it for more rounds would not improve the results much.
%\begin{figure}[t]
%  \centering
%  \includegraphics[width=0.5\textwidth]{plots/distortion}
%  \caption{The average distortion after each round of sequential deliberation in a simulation using the data from PB elections in various cities. The simulation was run $10,000$ times for each city. }
%  \label{experiment_distortion}
%\end{figure}
Further, we measured the stability of the fund allocation to the projects after each round of deliberation. We simulated sequential deliberation on the data from the PB in Cambridge $1,000,000$ times, each time with $10$ rounds of deliberation. The fund allocation to each project after each round was recorded. The fund allocation's standard deviation (SD) is shown in Figure~\ref{experiment_sd}. We can see that the SD stabilizes after only three rounds of deliberation.

%\begin{figure}[t]
%  \centering
%  \includegraphics[width=0.5\textwidth]{plots/sd}
%  \caption{The standard deviation of the budget allocation to each project in the PB election in Cambridge, MA. Each line represents a project.}
%  \label{experiment_sd}
%\end{figure}

%%%%%%%%%%%%%%%%%%%%%%%%%%%%%%%%%%%%%%

%%%%%%%%%%%
\newpage
\section{Triadic Scheme With Project Interactions}
\label{sec:TD_project_interaction}
%In this section, we analyze the outcome of triadic deliberation.
%We start with a motivating example.

%We model complementarity and substitutions as follows. 
\edit{Mathematically, we model project interactions as follows:} if projects in group $q$ are \textbf{perfect complements} of each other, then the overlap utility that voters can derive from \emph{each} project in $q$ is the minimum funding of any project in $q$. For example, consider a proposal of buying some computers for the community. Within this, one project is for buying hardware and another one is for buying software. If the software and hardware projects are funded $0.2$ and $0.5,$ then the community members can only use $0.2$ each, and the extra funding of $0.3$ for the hardware project is wasted \footnote{This is a stylized model and in general, the scale of the funds required for each project can be very different.}.

On the other hand, if the projects in group $r$ are \textbf{perfect substitutes}, then the utility that voters can derive from group $r$ is the maximum funding of a project in $r$. Thus, if two companies are paid $0.2$ and $0.5$ to do the same work, only $0.5$ will be used, and $0.2$ is wasted. 

We now give a formal model of the set of projects.
%\edit{\textbf{Example 3.1.}} Suppose that there are two voters $i_1$ and $i_2$ and two projects that are \edit{perfect} substitutes for each other, and that $i_1$ has a preferred budget $\langle 1, 0 \rangle$ and $i_2$ has a preferred budget $\langle 0, 1 \rangle$. If the outcome of the PB is $\langle 0.5, 0.5 \rangle$ (which can be a result of social welfare-maximizing aggregation with proportional tie-breaking), then the \edit{overlap} utility of each voter is $0.5$. However, if the outcome is $\langle 1, 0 \rangle$, then the \edit{overlap} utility of each voter is $1$ since the projects are \edit{perfect} substitutes. Note that even though the former outcome appears fairer to someone who does not know the project interactions, it is not optimal. This example illustrates the need to take project interaction into account while designing a voting scheme for PB.
%We now provide a formal model. 
Let $m_c$ denote the number of groups of \emph{perfect complementary} projects, $m_s$ denote the number of groups of \emph{\edit{perfect} substitute} projects, and $m_r$ denote the number of \emph{regular} projects. Let $s(q)$ denote the number of projects in group $q$. For groups of perfect complementary and \edit{perfect} substitute projects, $s(q) \ge 2$ and for regular projects $s(q) = 1.$ The total number of projects is $m = \left( \sum_{q=1}^{m_c + m_s} s(q) \right)+ m_r$.
For simplicity, %assume that project groups are arranged in the order such that complementary projects come first, followed by \edit{perfect} substitute projects, and then regular projects. 
project groups are arranged such that groups $1, \ldots, m_c$ are perfect complementary, groups $m_c+1, \ldots, m_c+m_s$ are \edit{perfect} substitutes, and $m_c+m_s+1, \ldots, m_c+m_s+m_r$ are regular projects.
%Let $u(x, y)$ denote the \emph{utility function} of budgets $x$ and $y$. Define $u(x, y) = u'(f(x), f(y))$ where $u'(p, q) = \sum_i \min(p_i, q_i)$ is the overlap utility between $p$ and $q$.

Let $f(b)$ be the \emph{efficiency function} which quantifies how much budget $b$ respects the project interactions. Specifically, $f(b)$ takes a budget $b \in \mathbb{R}^m$ and outputs a vector in $\mathbb{R}^{m_c+m_s+m_r}$, where
\begin{align*}
\hspace{-1 em} f(b)_q = \begin{cases}
s(q) \cdot \min(\{b_j ~|~ j \in \mbox{group }q\}) &\text{if } q \in [1, m_c]  \ \ \ \text{(perfect complementary \remove{project} groups)}, \\ 
\max(\{b_j ~|~ j \in \mbox{group }q\}) & \hspace{-2.5 em}\text{if } q \in [m_c+1, m_c+m_s], \ \ \text{(perfect substitute \remove{project} groups)} \\
\{b_j ~|~ j \in \mbox{group~}q\} & \hspace{-2 em}\mbox{otherwise.} \ \ \ \ \ \ \ \ \ \ \text{(regular projects).} %\mbox{if } i \in [m_c+m_s+1, m_c+m_s+m_n]
\end{cases}
\end{align*}
For a group of perfect complementary projects, the corresponding output element is the bottle-neck allocation in the group, multiplied by the number of projects in the group. For a group of \edit{perfect} substitute projects, the corresponding output element is the largest allocation in that group. For regular projects, the corresponding output elements are the same as the allocation to the project. We now give a modified definition of the overlap utility, accounting for project interactions.
%For example, suppose that $m_c = 1$, $m_s = 1$, and $m_n = 1$. If $b = \langle .4, .3, .2, .0, .1 \rangle$, then $f(b) = \langle 2\min(.4, .3), \max(.2, .0), .1 \rangle = \langle .6, .2, .1 \rangle$.

\begin{definition} \label{def:ou_new}
The \textbf{overlap utility} of budgets $a$ and $b$, accounting for project interactions is $u(a, b) = \sum\limits_{q=1}^{m_c+m_s+m_r} \min(f(a)_q, f(b)_q)$.
\end{definition}

\edit{In the following definition we formally state the requirements for a budget to be consistent with the project interactions.}
\begin{definition}
A budget $b$ \textbf{respects the project interactions} if and only if projects in each perfect complementary group are all funded equally, and at most one project in each perfect substitute group is funded at all.
\end{definition}

The following lemma states that the efficiency function $f(b)$ sums to $1$ if and only if the budget $b$ \emph{respects the project interactions}.

\begin{lemma}
\label{cs_sum}
Budget $b$ respects the project interactions iff the efficiency function $f(b)$ satisfies $\sum_q f(b)_q = 1$. Otherwise, $\sum_q f(b)_q < 1.$
\end{lemma}
%The proof of this lemma is described in Appendix A.5 in the extended version \cite{goyal2023low}. \remove{\S\ref{cs_sum_proof}.}

We now give a result that a Pareto improvement exists over a budget that does not respect the project interactions.
\begin{lemma}
\label{cs_new_b}
If $\sum_q f(b)_q < 1$, then for some $k \in [m_c+m_s+m_r]$, there exists a budget $b'$ for which $f(b')_k > f(b)_k$ and $f(b')_q \geq f(b)_q$  for all project groups $q$.
\end{lemma}

The proofs of lemmas \ref{cs_sum} and \ref{cs_new_b} are presented in Appendix \ref{cs_sum_proof} and \ref{cs_new_b_proof}. \remove{\S\ref{cs_new_b_proof}.}

%The proof of Lemma~\ref{cs_sum} is simple and is in Appendix~\S\ref{cs_sum_proof}.

We now give the main result of this section. We show that if either of the budgets of the bargaining agents respect the project interactions (which will be true for rational agents), then the outcome of any median scheme respects project interactions. Since the class of median schemes contains the class of Nash bargaining schemes (Theorem~\ref{max_utility_Nash}), this result also applied to $\mathcal{N}$ and therefore also to our randomized bargaining scheme $\mathfrak{n}_{\text{rand}}.$

%The following lemma implies that the outcome of triadic deliberation respects the project interactions.
\begin{theorem}
\label{cs_main}
%Let $x$, $y$, and $a$ be budgets. 
If budget $a$ or $b$ respects the project interactions, then for any budget $c \in \mathbb{B},$ $\mathcal{M}(a,b,c)$ respects the project interactions.
\end{theorem}
\begin{proof}
Let $z$ be an outcome from $\mathcal{M}(a,b,c)$.
Assume without loss of generality that budget $a$ respects the project interactions, and suppose that outcome $z$ does not. By Lemma~\ref{cs_sum},
$\sum_q f(a)_q = 1$ and $\sum_q f(z)_q < 1$.
Thus, there exists some $k$ where $f(a)_k > f(z)_k$.
By Lemma~\ref{cs_new_b},
there exists a budget $z'$ \edit{which respects the project interactions and} $f(z')_q \geq f(z)_q$ for all project groups $q$ and $f(z')_k > f(z)_k$. The overlap utility functions satisfy:
\begin{align*}
&u(a, z') = \sum_q \min( f(a)_q, f(z')_q ) > \sum_q \min(f(a)_q, f(z)_q) = u(a, z), \\
&u(b, z') = \sum_q \min( f(b)_q, f(z')_q ) \geq \sum_q \min(f(b)_q, f(z)_q) = u(b, z). \\
&u(c, z') = \sum_q \min( f(c)_q, f(z')_q ) \geq \sum_q \min(f(c)_q, f(z)_q) = u(c, z).
\end{align*}
%Since u'(x, z') > u'(x,z) and u'(y,z') \geq u'(y,z), z is not an outcome from a Pareto-efficient bargaining, a contradiction.
%Since $u(f(x), f(z')) > u(f(x), f(z)),$ $u(f(y), f(z')) \geq u(f(y), f(z))$, 
This implies that the sum of overlap utilities of $a,b,$ and $C$ with $z'$ is higher than that with $z$, a contradiction for an outcome of $\mathcal{M}$.
\end{proof}
%The proof is in Appendix~\ref{proof_cs_main}.

Theorem~\ref{cs_main} implies that if every voter has a preferred budget that respects the project interactions, then the outcome of the sequential deliberation mechanism will also respect the project interactions, no matter how many rounds it runs. 
\vspace{-0.05cm}
\section{Conclusion}{\label{sec:conclusion}}
%\label{sec:conclusion}
%We propose two methods for handling project interactions in PB. First, we show that sequential deliberation can handle two types of project interactions: perfect complementarity and \edit{perfect} substitution, without explicit input  from the PB organizer \edit{regarding the project interactions}. The single-round restriction of sequential deliberation maximizes the total utility of the voters involved in the bargaining, and it has a distortion bounded below $1.93$. We empirically evaluate the performance of sequential deliberation on data from real participatory budgeting elections and show that the average distortion is small and the budget converges within $3$ rounds of deliberation. Additionally, we propose a menu-based ballot design where the PB organizer provides the partition of project into groups and analyze the strategic behavior of voters in our model. %We argue that the proposed ballot makes it cognitively easier for voters to indicate their preferences on interacting projects. 
%We also show that when at most one project from a group can be funded, we get strategy-proof preference elicitation and aggregation methods.

 We study low sample-complexity mechanisms for PB, which are particularly attractive when the policymakers are interested in obtaining a quick estimate of the voter's preferences or when a full-fledged PB election is difficult or costly to conduct. In our PB setup, the distortion of mechanisms that obtain and use the votes of only one  uniformly randomly sampled voter is $2$. Extending this result, we show that when two voters are sampled, and a convex combination of their votes is used by the mechanism, the distortion cannot be made smaller than $2$. We then show that with $3$ samples, there is a significant improvement in the distortion -- we give a PB mechanism that obtains a distortion of $1.66$. Our mechanism builds on the existing works on Nash bargaining between two voters with a third voter's preferred outcome as the disagreement point and give a lower bound of $1.38$ for our mechanism.    %{\color{red} An interesting future direction of work could be to bound the distortion ratio after multiple rounds of bargaining where the disagreement point of every round could be the outcome of the previous round of bargaining. For future work, it will be interesting to study models for deliberation in bigger groups. Closing the gap of distortion of Nash bargaining also remains an interesting open problem.}

%We also show that there exist PB instances where a randomised dictator strategy i.e. one that selects a budget uniformly at random might attain a dis-utility ratio of 2. 

\remove{
We also do simulations over four real-world PB instances and show that distortion converges after 4-5 rounds of deliberation to a value less than 1.30.

For future work, it will be interesting to study models for deliberation in bigger groups. %Deliberation, in the real world, often leads to change of preferences of voters and also may produce innovative solutions to collective problems \cite{fishkin1991democracy}. Developing a model that incorporates these observations is an interesting avenue for future work.
Closing the gap of the distortion of Nash bargaining and analysing the distortion for more rounds of deliberation are also interesting open problems.

}

%We characterise and define the distortion ratio of a outcome budget in sequential deliberation as the ratio of the expected cumulative dis-utility of the outcome budget with the budgets in vote profile and the cumulative dis-utility of the optimal budget with the budgets with the budgets in the vote profile. We propose a randomised Nash bargaining strategy which achieves a distortion ratio of 1.66 and a class of relaxed Nash Bargaining schemes each of which attain a distortion of at most 1.83 after one round of bargaining. We also show that a randomised dictator strategy one that selects a budget at random attains a dis-utility ratio of at least 2 and a strategy that is constrained to choose a budget on a line segment joining any two samples budgets attains 

\bibliography{references}

\begin{thebibliography}{10}

\bibitem{anshelevich2017randomized}
Elliot Anshelevich and John Postl.
\newblock Randomized social choice functions under metric preferences.
\newblock {\em Journal of Artificial Intelligence Research}, 58:797--827, 2017.

\bibitem{arrow2010handbook}
Kenneth~J Arrow, Amartya Sen, and Kotaro Suzumura.
\newblock {\em Handbook of social choice and welfare}, volume~2.
\newblock Elsevier, 2010.

\bibitem{aziz2022approximate}
Haris Aziz, Bo~Li, and Xiaowei Wu.
\newblock Approximate and strategyproof maximin share allocation of chores with
  ordinal preferences.
\newblock {\em Mathematical Programming}, pages 1--27, 2022.

\bibitem{aziz2021participatory}
Haris Aziz and Nisarg Shah.
\newblock Participatory budgeting: Models and approaches.
\newblock In {\em Pathways Between Social Science and Computational Social
  Science}, pages 215--236. Springer, 2021.

\bibitem{benade2021preference}
Gerdus Benade, Swaprava Nath, Ariel~D Procaccia, and Nisarg Shah.
\newblock Preference elicitation for participatory budgeting.
\newblock {\em Management Science}, 67(5):2813--2827, 2021.

\bibitem{binmore1986nash}
Ken Binmore, Ariel Rubinstein, and Asher Wolinsky.
\newblock The nash bargaining solution in economic modelling.
\newblock {\em The RAND Journal of Economics}, pages 176--188, 1986.

\bibitem{brams2002voting}
Steven~J Brams and Peter~C Fishburn.
\newblock Voting procedures.
\newblock {\em Handbook of social choice and welfare}, 1:173--236, 2002.

\bibitem{ewens2019organizational}
Hendrik Ewens and Joris van~der Voet.
\newblock Organizational complexity and participatory innovation: participatory
  budgeting in local government.
\newblock {\em Public Management Review}, 21(12):1848--1866, 2019.

\bibitem{fain2020concentration}
Brandon Fain, William Fan, and Kamesh Munagala.
\newblock Concentration of distortion: The value of extra voters in randomized
  social choice.
\newblock {\em IJCAI}, 2020.

\bibitem{fain2019}
Brandon Fain, Ashish Goel, Kamesh Munagala, and Nina Prabhu.
\newblock Random dictators with a random referee: Constant sample complexity
  mechanisms for social choice.
\newblock {\em AAAI}, 2019.

\bibitem{fain2017}
Brandon Fain, Ashish Goel, Kamesh Munagala, and Sukolsak Sakshuwong.
\newblock Sequential deliberation for social choice.
\newblock In {\em Web and Internet Economics}, pages 177--190. Springer, 2017.

\bibitem{freeman2021truthful}
Rupert Freeman, David~M Pennock, Dominik Peters, and Jennifer~Wortman Vaughan.
\newblock Truthful aggregation of budget proposals.
\newblock {\em Journal of Economic Theory}, 193:105234, 2021.

\bibitem{ganuza2012}
Ernesto Ganuza and Gianpaolo Baiocchi.
\newblock The power of ambiguity: How participatory budgeting travels the
  globe.
\newblock {\em Journal of Public Deliberation}, 8, 2012.

\bibitem{garg2019iterative}
Nikhil Garg, Vijay Kamble, Ashish Goel, David Marn, and Kamesh Munagala.
\newblock Iterative local voting for collective decision-making in continuous
  spaces.
\newblock {\em Journal of Artificial Intelligence Research}, 64:315--355, 2019.

\bibitem{goel2019knapsack}
Ashish Goel, Anilesh~K. Krishnaswamy, Sukolsak Sakshuwong, and Tanja Aitamurto.
\newblock Knapsack voting for participatory budgeting.
\newblock {\em ACM Transactions on Economics and Computation}, 7(2), 2019.

\bibitem{goel2016towards}
Ashish Goel and David~T Lee.
\newblock Towards large-scale deliberative decision-making: Small groups and
  the importance of triads.
\newblock In {\em Proceedings of the 2016 ACM Conference on Economics and
  Computation}, pages 287--303, 2016.

\bibitem{gurobi}
{Gurobi Optimization, LLC}.
\newblock {Gurobi Optimizer Reference Manual}, 2022.
\newblock URL: \url{https://www.gurobi.com}.

\bibitem{jain2020}
Pallavi Jain, Krzysztof Sornat, and Nimrod Talmon.
\newblock Participatory budgeting with project interactions.
\newblock In {\em Proceedings of the Twenty-Ninth International Joint
  Conference on Artificial Intelligence, {IJCAI-20}}, pages 386--392, 2020.

\bibitem{meir2021representative}
Reshef Meir, Fedor Sandomirskiy, and Moshe Tennenholtz.
\newblock Representative committees of peers.
\newblock {\em Journal of Artificial Intelligence Research}, 71:401--429, 2021.

\bibitem{peters2020proportional}
Dominik Peters, Grzegorz Pierczy{\'n}ski, and Piotr Skowron.
\newblock Proportional participatory budgeting with cardinal utilities.
\newblock {\em arXiv preprint arXiv:2008.13276}, 2020.

\bibitem{Pythonscripts}
Sahasrajit Sarmasarkar, Mohak Goyal, Sukolsak Sakshuwong, and Ashish Goel.
\newblock Python scripts for the optimization problems, 2023.
\newblock URL: \url{https://github.com/Sahasrajit123/Low-sample-complexity-PB}.

\bibitem{wainwright2003}
Hilary Wainwright.
\newblock Making a people's budget in porto alegre.
\newblock {\em NACLA Report on the Americas}, 36:37--42, 03 2003.

\bibitem{wampler2007}
Brian Wampler.
\newblock A guide to participatory budgeting.
\newblock In {\em Participatory Budgeting}. World Bank, 2007.

\end{thebibliography}

\appendix

\newpage
\onecolumn
%\usepackage{geometry}
%\newgeometry{left=3cm,bottom=0.1cm}
\begin{changemargin}{-1.5cm}{-1.5cm}
\deffootnote[-1.5 cm]{-1.5 cm}{-1.5cm}{\textsuperscript{\thefootnotemark}\,}

   \renewcommand\footnoterule{%
   \parindent=-1.5cm\hrulefill
   \vspace{1 em}
  } 
\nolinenumbers
\appendix
\section{Appendix}

\subsection{Proof of Lemma~\ref{dist2}}
\label{proof_dist2}
[Restatement of  Lemma~\ref{dist2}]
{\it Any aggregation method constrained to choose outcomes from a uniformly at random sampled voter's preferred budgets has distortion $2$.}

\begin{proof}
Let $n$ be the number of voters, and $2n$ be the number of projects with no project interactions.
Let the preferred budget of each voter $i$ be $v_i$
where $v_{i,j} = 1/n$ if $i=j$ or 0 otherwise for all $j \in [n]$. Further, $v_{i,j} = 0$ if $i=j-n$ or $1/n$ otherwise for all $j \in [2n] \backslash [n]$.
Therefore, the preferred budgets are:
\begin{gather*}
\left\langle \overbrace{\frac{1}{n}, 0, 0, \ldots, 0}^\text{$n$ elements}, \ \ \ \   \overbrace{0, \frac{1}{n}, \frac{1}{n}, \ldots, \frac{1}{n}}^\text{$n$ elements} \right\rangle \\
\left\langle 0, \frac{1}{n}, 0, \ldots, 0, \ \ \ \   \frac{1}{n}, 0, \frac{1}{n}, \ldots, \frac{1}{n} \right\rangle \\
\ldots \\
\left\langle 0, 0, \ldots,0, \frac{1}{n}, \ \ \ \   \frac{1}{n}, \frac{1}{n},  \ldots,\frac{1}{n}, 0 \right\rangle
\end{gather*}
The welfare-maximizing outcome, which can be found using greedy aggregation, is 
$ \langle \overbrace{0, 0, 0, \ldots, 0}^\text{$n$ elements}, \overbrace{\frac{1}{n}, \frac{1}{n}, \frac{1}{n}, \ldots, \frac{1}{n}}^\text{$n$ elements} \rangle $.
Therefore, the social cost between each voter and the optimal outcome is $\frac{2}{n}$.
Suppose that the mechanism chooses the preferred budget of voter $i$.
Then any voter other than $i$ has social cost $\frac{4}{n}$.
Therefore, the distortion is $\frac{(n-1) \cdot (4/n)}{n \cdot (2/n)} = 2 - \frac{2}{n},$ which supremum over $n$ is 2. 

We thus showed that {\it random dictator} has a lower bound on distortion to be 2.

However, we know from \cite{anshelevich2017randomized} that under any metric space where distances denote the costs incurred by a voter, a {\it random dictator} has distortion upper bounded by 2 especially because our metric space is 0-decisive as the preferred budget has zero social cost for a voter. Combining the two results prove our theorem.

\end{proof}

\subsection{Proof of Lemma~\ref{distortion_two_bargaining}}{\label{distortion_two_bargaining_proof_section}}
[Restatement of Lemma~\ref{distortion_two_bargaining}]
{\it For Random Diarchy $\mathcal{Q}$,
$\inf_{\mathfrak{q} \in \mathcal{Q}} \Distortion_once(\mathfrak{q}) = 2.$}
\begin{proof}
Let $n$ be the number of voters, and $2n$ be the number of projects with no project interactions.
Let the preferred budget of each voter $i$ be $v_i$
where $v_{i,j} = 1/n$ if $i=j$ or 0 otherwise for all $j \in [n]$. Further, $v_{i,j} = 0$ if $i=j-n$ or $1/n$ otherwise for all $j \in [2n] \backslash [n]$.
Therefore, the preferred budgets are:
\begin{gather*}
\left\langle \overbrace{\frac{1}{n}, 0, 0, \ldots, 0}^\text{$n$ elements}, \ \ \ \   \overbrace{0, \frac{1}{n}, \frac{1}{n}, \ldots, \frac{1}{n}}^\text{$n$ elements} \right\rangle \\
\left\langle 0, \frac{1}{n}, 0, \ldots, 0, \ \ \ \   \frac{1}{n}, 0, \frac{1}{n}, \ldots, \frac{1}{n} \right\rangle \\
\ldots \\
\left\langle 0, 0, \ldots,0, \frac{1}{n}, \ \ \ \   \frac{1}{n}, \frac{1}{n},  \ldots,\frac{1}{n}, 0 \right\rangle
\end{gather*}

The welfare-maximizing outcome, which can be found using greedy aggregation, is
$ \langle \overbrace{0, 0, 0, \ldots, 0}^\text{$n$ elements}, \frac{1}{n}, \frac{1}{n}, \frac{1}{n}, \ldots, \frac{1}{n} \rangle $.
Therefore, the cost for each voter from the optimal outcome is $\frac{2}{n}$ and hence the social cost for the optimal outcome is given by $2$.
%
%Let us break this problem into two cases. 
%
%\begin{itemize}
 %   \item \textbf{Case 1}: 
    % We  consider the case when the two budgets chosen at random are distinct. 
    Without loss of generality, we can assume the voters chosen to deliberate are $v_1$ and $v_2$. Consider any mechanism $\mathfrak{q} \in \mathcal{Q}$, the outcome of $\mathfrak{q}(v_1,v_2)$ will be $$z = \left\langle \overbrace{\frac{1}{n}.\alpha(P), \frac{1}{n}.(1-\alpha(P)), 0, \ldots, 0}^\text{$n$ elements}, \ \ \ \   \overbrace{\frac{1}{n}(1-\alpha(P)), \frac{1}{n}\alpha(P), \frac{1}{n}, \ldots, \frac{1}{n}}^\text{$n$ elements} \right\rangle$$ for some $0<\alpha(P)<1$. The cost of the outcome $z$ for any budget $\{v_i\}_{i>2}$  is $\frac{1}{n}+\frac{1}{n}\alpha(P)+\frac{1}{n}(1-\alpha(P))+ \frac{1}{n}(1-\alpha(P))+\frac{1}{n}\alpha(P) + \frac{1}{n}=\frac{4}{n}$. However, the cost of outcome $z$ for budget $v_1$ is  $\frac{2}{n}(1-\alpha(P))+\frac{2}{n}(1-\alpha(P))=\frac{4}{n}(1-\alpha(P)).$  Similarly the cost of outcome $z$ for budget $v_2$ is $\frac{4}{n}\alpha(P)$. Thus, the social cost of the outcome $z$ is $\frac{4}{n}(n-2)+\frac{4\alpha(P)}{n}+\frac{4(1-\alpha(P))}{n} = \frac{4(n-2)+4}{n}$.

  %  \item \textbf{Case 2}: Suppose both the budgets chosen are $v_i$. Well in this case, the outcome of any deliberation process $\mathfrak{q} \in \mathcal{Q}$ would be budget $v_i$. The total social cost in this case is given by $\frac{4}{n}$.

%\end{itemize}

Thus, the expected social cost is given by $\left(\frac{4(n-2)+2}{n}\right)$ and we observe the supremum of the distortion is 2.

However, we may observe that {\it randomized dictator} mechanism which samples a preferred budget uniformly at random would also belong to the class of {\it random diarchy} ($\mathcal{Q}$) mechanisms as $\alpha(P)$ may be optimized over the entire space of preferred budgets. Thus, we can further upper bound $\inf_{\mathfrak{q} \in \mathcal{Q}} \Distortion_once(\mathfrak{q})$ by 2 following the result in Lemma \ref{dist2}. This proves our result

\end{proof}

\subsection{Proof of Lemma~\ref{distortion_random_referee}}{\label{distortion_random_referee_proof_section}}
[Restatement of Lemma~\ref{distortion_random_referee}]
{\it For Random Referee $\mathcal{R}$,
$\inf_{\mathfrak{q} \in \mathcal{R}} \Distortion_once(\mathfrak{q}) \geq 2.$}
\begin{proof}
Let $n$ be the number of voters, and $2n$ be the number of projects with no project interactions.
Let the preferred budget of each voter $i$ be $v_i$
where $v_{i,j} = 1/n$ if $i=j$ or 0 otherwise for all $j \in [n]$. Further, $v_{i,j} = 0$ if $i=j-n$ or $1/n$ otherwise for all $j \in [2n] \backslash [n]$.
Therefore, the preferred budgets are:
\begin{gather*}
\left\langle \overbrace{\frac{1}{n}, 0, 0, \ldots, 0}^\text{$n$ elements}, \ \ \ \   \overbrace{0, \frac{1}{n}, \frac{1}{n}, \ldots, \frac{1}{n}}^\text{$n$ elements} \right\rangle \\
\left\langle 0, \frac{1}{n}, 0, \ldots, 0, \ \ \ \   \frac{1}{n}, 0, \frac{1}{n}, \ldots, \frac{1}{n} \right\rangle \\
\ldots \\
\left\langle 0, 0, \ldots,0, \frac{1}{n}, \ \ \ \   \frac{1}{n}, \frac{1}{n},  \ldots,\frac{1}{n}, 0 \right\rangle
\end{gather*}

The welfare-maximizing outcome, which can be found using greedy aggregation, is
$ \langle \overbrace{0, 0, 0, \ldots, 0}^\text{$n$ elements}, \frac{1}{n}, \frac{1}{n}, \frac{1}{n}, \ldots, \frac{1}{n} \rangle $.
Therefore, the cost for each voter from the optimal outcome is $\frac{2}{n}$ and hence the social cost for the optimal outcome is given by $2$.
%
%Let us break this problem into two cases. 
%
%\begin{itemize}
 %   \item \textbf{Case 1}: 
    % We  consider the case when the two budgets chosen at random are distinct. 
    Without loss of generality, we can assume the voters chosen to deliberate are $1$ and $2$ with the random referee being the preferred budget $v_3$. Consider any mechanism $\mathfrak{q} \in \mathcal{R}$, the outcome of $z=\mathfrak{q}(v_1,v_2,v_3)$ will be either $v_1$ or $v_2$. The cost of the outcome $z$ for any budget $\{v_i\}_{i>2}$  is $\frac{4}{n}$. However, the cost of outcome $z$ for budget $v_1$ is  $0$ if $z=v_1$ or $\frac{4}{n}$ otherwise.  Similarly the cost of outcome $z$ for budget $v_2$ is $0$ if $z=v_2$ or $\frac{4}{n}$ otherwise. Thus, the social cost of the outcome $z$ is $\frac{4}{n}(n-2)+\frac{4}{n} = \frac{4(n-2)+4}{n}$.

  %  \item \textbf{Case 2}: Suppose both the budgets chosen are $v_i$. Well in this case, the outcome of any deliberation process $\mathfrak{q} \in \mathcal{Q}$ would be budget $v_i$. The total social cost in this case is given by $\frac{4}{n}$.

%\end{itemize}

Thus, the expected social cost is given by $\left(\frac{4(n-2)+4}{n}\right)$ and we observe the supremum of the distortion is 2.

\end{proof}

\subsection{Proof of Theorem ~\ref{thm:lower_bound}}
\label{app:lower_bnd}
[Restatement of Theorem~\ref{thm:lower_bound}]
{\it  $\Distortion_once(\mathcal{M}) \geq \Distortion_once(\mathcal{N})> 1.38.$ Also, $\Distortion_once(\mathfrak{n}_{\text{rand}}) > 1.38.$}
\begin{proof}
The proof is by the following example. Suppose there are $n_A + n_B$ voters
and $n_A + 1$ projects for some $n_A, n_B \geq 1$. Let $o_i$ denote the budget where the $i$-th project receives allocation $1$, and all the other projects receive allocation $0$. 
Each voter $i$ in group A ($1 \leq i \leq n_A$) prefers budget $o_i$.
Each voter $i$ in group B ($n_A+1 \leq i \leq n_A+n_B$) prefers budget $o_{n_A+1}$.
Let $c$ be the preferred budget randomly chosen as the disagreement point and $a$ and $b$ be the preferred budgets of the voters randomly chosen to bargain.

Since every voter prefers a single project, the outcome budget of the bargaining, under a scheme in $\mathcal{N}$ can be written as $(o_i + o_j) / 2$ for some $i, j \in [n_A + 1]$ ($i$ can equal $j$). We analyze the cases of the outcome budget as follows:

\textbf{Event 1:} the outcome is $o_{n_A+1}$. \\
The social cost, in this case, is $2n_A$. This event happens if at least two of $a$, $b$, and $c$ are in group $B;$ its probability is $p_1=(n_B^3 + 3 n_A n_B^2) / (n_A+n_B)^3.$

\textbf{Event 2:} the outcome is $(o_i + o_{n_A+1}) / 2$ where $i \neq n_A + 1$. \\
In this case, the total social cost is $2n_A + n_B -1$.
This happens if either $a$ or $b$ is in group $B$ and $a, b, c$ all have different preferred budgets.
The probability of this event is $p_2= 2 n_A (n_A - 1) n_B / (n_A+n_B)^3.$

\textbf{Event 3:} the outcome is $(o_i + o_j) / 2$ where $i,j \neq n_A + 1$. \\
The total social cost, in this case, is $2n_A + 2n_B -1$, and the probability of this event is $1 -p_1 - p_2 = p_3$.

Since the $(n_A+1)$-th project receives the most votes,
$o_{n_A+1}$ is an optimal outcome for maximizing social welfare. The optimal social cost is $2n_A$.
Therefore, the distortion is:
\begin{gather*}
\frac{2n_A p_1 + (2n_A + n_B -1) p_2 + (2n_A + 2n_B -2)p_3}{ 2n_A}.
\end{gather*}
For $n_A = 2200$ and $n_B = 3000$,
we obtain that the distortion is at least $1.38$.
\end{proof}

[Restatement of  Lemma~\ref{dist2}]
{\it Any aggregation method constrained to choose outcomes from a uniformly at random sampled voter's preferred budgets has distortion $2$.}

\subsection{Proof of Lemma \ref{cs_sum}}{\label{cs_sum_proof}}

[Restatement of Lemma \ref{cs_sum}]
{\it Budget $b$ respects the project interactions iff the efficiency function $f(b)$ satisfies $\sum_q f(b)_q = 1$. Otherwise, $\sum_q f(b)_q < 1.$}

\begin{proof}
We use the fact that for any set of non-negative numbers $S$, $|S|\cdot \min(S) = \sum_{s\in S} s$ if and only if all the numbers in $S$ are equal. Otherwise, $|S| \cdot \min(S) < \sum_{s\in S} s$. We also use the fact that for any set of non-negative numbers $S$, $ \max(S) = \sum_{s\in S} s$ if and only if at most one number in $S$ is greater than $0$. Otherwise, $ \max(S) < \sum_{s\in S} s$.

From the definition of $f(b)$ and using the facts above, it is easy to check that if $b$ respects the project interactions, we get $\sum_q f(b)_q = b_1 + b_2 + \ldots + b_m = 1$. If $b$ does not respect the project interactions, then, from the above facts, we have $\sum_q f(b)_q < 1$. 
\end{proof}

\subsection{Proof of Lemma \ref{cs_new_b}}{\label{cs_new_b_proof}}

[Restatement of Lemma \ref{cs_new_b}]
{\it If $\sum_q f(b)_q < 1$, then for some $k \in [m_c+m_s+m_r]$, there exists a budget $b'$ for which $f(b')_k > f(b)_k$ and $f(b')_q \geq f(b)_q$  for all project groups $q$.}

\begin{proof}

Since $\sum_q f(b)_q \neq 1$, from Lemma~\ref{cs_sum}, either the perfect complementary condition is violated (a project gets funded more than the others in a perfect complementary group) or the \edit{perfect} substitute condition is violated (more than one project in a \edit{perfect} substitute group gets funded). In either case, there is funding, which we will call $F$, that can be reallocated to other projects without decreasing any $f(b)_q$. In the case of violation of the perfect complementary condition, $F$ is the excess funding to the more funded project in a group \edit{of perfect complementary projects}. In the case of violation of the \edit{perfect} substitute condition, $F$ is the funds allocated to the projects which are not the highest funded project in a group  \edit{of perfect substitute projects}.

If $k$ represents a group \edit{of perfect complementary projects}, we construct $b'$ from $b$ by reallocating $F$ equally to all projects in group $k$. If $k$ represents a group \edit{of perfect substitute projects}, we construct $b'$ from $b$ by reallocating $F$ to the most funded project in group $k$. %If $k$ represents a regular project, we construct $b'$ from $b$ by reallocating $F$ to the project $k$.
This increases $f(b)_k$ and does not decrease $f(b)_q$ for any other project group $q$. 
\end{proof}

%Another important implication of this result is that if sequential deliberation works \edit{well} for PB without project interactions, then it will also work \edit{well} for PB with project interactions. %\sukolsak{I'm not sure anymore if this theorem is that useful.}
 
%Note that Theorem~\ref{cs_main} holds not only with the Nash bargaining, but also with any bargaining solution that is Pareto-efficient, such as the Kalai-Smorodinsky solution \cite{kalai1975} or the egalitarian solution \cite{kalai1977}. Although we use the Nash product in the proof, the last paragraph in the proof is simply the definition of Pareto efficiency.

%Now we characterize the total utility of the three voters, two involved in the bargaining and one whose preferred budget is chosen as the disagreement point, under the Nash bargaining solution. 

%\subsection{Proof of Lemma~\ref{}}
\subsection{Results on incremental allocation space $X_{j,P}(S)$ and $Z_{j,P}(S)$}

Here we present Lemma~\ref{total_allocation_budget_space} which captures, in the incremental allocation space, the fact that the total allocation by budget $v_i$ is 1. 

\begin{lemma}{\label{total_allocation_budget_space}}
For any vote profile $P= (v_1,v_2,\ldots,v_n)$, $\sum_{S \in \mathcal{P}(P)| S \ni v_i}  X_{j,P}(S) = v_{i,j}$ for all $ i \in [n]$ and all $j \in [m]$. Summing over $j \in [m]$, we get  $\sum_{S \in \mathcal{P}(P)| S \ni v_i}  X_{P}(S) = 1$ for all $i \in [n]$.% since the total allocation in any budget is $1$.
\end{lemma}

\begin{proof}
    This follows from $v_{i,j} = X_{j,\{v_i\}}(\{v_i\})$ [Definition \ref{proj_budget_defn}] and then applying Lemma~\ref{proj_budget_new_space_lemma} with $Q=\{v_i\}.$
\end{proof}

%Lemma~\ref{total_allocation_budget_space} captures, in the incremental allocation space, the fact that the total allocation by budget $v_i$ is 1.

%\subsection{Proof of Lemma \ref{utility_common}}

\subsection{Proof of Corollary \ref{sum_each_proj}}
\label{app_sum_each_proj}
[Restatement of Corollary \ref{sum_each_proj}]
{\it $\sum_{S \in \mathcal{P}(P)} X_{j,P} (S) = 1,$ for all projects $j \in [m].$}
\begin{proof}
    Recall that $v_{i,j}$ denotes the allocation of budget $v_i \in \mathbb{B}$ to project $j$  for all $i \in [n]$. Let us sort these elements in an increasing order such that $v_{l_1,j} \leq v_{l_2,j} \ldots \leq v_{l_n,j}$. Note, by definition $X_{j,P}(S) = v_{l_p,j} - v_{l_{p-1},j}$ for $S = \{v_{l_n},v_{l_{n-1}}, \ldots, v_{l_p}\}$ for
    all $1<p\leq n$. $X_{j,P}(P) = v_{l_1,j},$ and $X_{j,P}(\emptyset) = 1- v_{l_n,j}$. For every other set $S$, we have $X_{j,P}(S)=0$ which follows since $\min\limits_{i \in S} v_{i,j} \leq \max\limits_{i \in P\setminus S} v_{i,j}$ for every such $S$. Summing over all  sets $S \in \mathcal{P}(P)$, we get the desired result.
\end{proof}

\remove{
\begin{corollary*}%%{\label{monotonicity_proj_budget_space}}
    For every project $j \in [m],$ and two subsets of $P$ satisfying $ S \subset \hat{S},$ and $X_{j,P}(S)>0$
    %s.t. $\hat{S},S \in \mathcal{P}(P)$, 
    we have $X_{j,P}(\hat{S}) \leq X_{j,P}(S)$.
\end{corollary*}

\begin{proof}
Recall that $v_{i,j}$ denotes the allocation of project $j$ to budget $v_i \in \mathbb{B}$ for every $i \in \{1,2,\ldots,n\}$. Let us sort these elements in an increasing order s.t. $v_{l_1,j} \leq v_{l_2,j} \ldots v_{l_n,j}$. Note, by definition $X_{j,P}(S) = v_{l_p,j} - v_{l_{p-1},j}$ when $S = \{v_{l_n},v_{l_{n-1}}, \ldots, v_{l_p}\}$ for every $p>1$ and $X_{j,P}(S) = v_{l_1,j}$ whenever $S = \{v_{l_n},v_{l_{n-1}}, \ldots, v_{l_1}\}$ and $X_{j,P}(\emptyset) = 1- v_{l_n,j}$. $S$ must be one of these sets as $X_{j,P}(S)>0$, however any set $\hat{S}\supset S$ must either be zero or have a smaller h
\end{proof}
}

\subsection{Proof of Lemma~\ref{proj_budget_new_space_lemma}}
\label{proj_budget_new_space_proof}

%The \textbf{common budget function} $g: \mathbb{B}^k \rightarrow \mathbb{R}^{(2^k)}$ is a function that maps a vote profile $P=(v_1, v_2, \ldots, v_k)$ to a point in the common budget space
%where $g(P)_S = \sum_{i=1}^m \max((\min_{s \in S} v_{s,i}) - (\max_{s \in P \setminus S} v_{s,i}), 0)$
%for each $S \in \mathcal{P}(P)$.

%For any budgets $a$ and $b$ and any vote profile $P$ that includes $a$ and $b$, $u(a, b) = \sum_{ S \in \mathcal{P}(P \setminus \{a,b\} ) }  g(P)_{ \{a,b\} \cup S }$.

[Restatement of Lemma~\ref{proj_budget_new_space_lemma}]
{\it    For any vote profile $P$ and $Q \subseteq P$, we have 
    $X_{j,Q}({S}) = \sum_{\hat{S} \in \mathcal{P}(P \setminus Q)}X_{j,P}(\hat{S} \cup {S}) $ for all ${S} \in \mathcal{P}(Q)$, $j \in [m]$. Summing over projects $j \in [m]$, we get $X_{Q}({S}) = \sum_{\hat{S} \in \mathcal{P}(P \setminus Q)}X_{P}(S \cup \hat{S}).$
}

\begin{proof}
We use the identity
$\max(\min(A) - \max(B), 0) = \max(\min(A \cup \{x\}) - \max(B), 0) + \max(\min(A) - \max(B \cup \{x\}), 0),$ where $A$ and $B$ are sets of real numbers and $x$ is a real number. This identity can be proved by using case analysis on the order of $\min(A)$, $\max(B)$, and $x$.
From the definition of $X_{j,Q}(S)$, for any vote profile $P$ and any subset $S \subseteq P$, we have:
\begin{align*}
& X_{j,Q}(S) \\
&= \max \left(\left(\min_{i \in S} v_{i,j}\right) - \left(\max_{i \in P \setminus S} v_{i,j}\right), 0 \right), \\
&= \max \left(\left(\min_{i \in S \cup \{x\}} v_{i,j}\right) - \left(\max_{i \in P \setminus S} v_{i,j}\right), 0 \right)  + \max \left( \left(\min_{i \in S} v_{i,j}\right)-\left(\max_{i \in (P \setminus S)\cup \{x\}} v_{i,j}\right), 0 \right), \\
&= X_{(P \cup \{x\})}(S \cup \{x\}) ~+~ X_{(P \cup \{x\})}(S).
\end{align*}

On applying this step inductively over every element in $P\setminus Q$, we get $X_{j,Q}(S)= \sum\limits_{\hat{S} \in \mathcal{P}(P\setminus Q)}X_{j,P}(\hat{S}\cup S)$.
\remove{
Let $P = (a, b, x_1, x_2, \ldots, x_n)$. We have:
\begin{align*}
u(a, b)
&= g_{(a, b)}(a, b), \\
%
% &= g(a, b, x_1)_{ \{a, b\} \cup \{x_1\} } + g(a, b, x_1)_{ \{a, b\} }, \\
%
&= g_{(a, b, x_1)}(\{a, b, x_1\} ) + g_{(a, b, x_1)}( \{a, b\} ), \\
&= \sum_{S \in \power{x_1}} g_{(a, b, x_1)}(   S \cup \{a, b\} ), \\
&= \sum_{S \in \power{ \{x_1, x_2\} }} g_{(a, b, x_1, x_2)}(   S \cup \{a, b\} ), \\
%
%&\ \ \vdots \\
&= \sum_{S \in \power{ \{x_1, x_2, \ldots, x_n\} }} g_{(a, b, x_1, x_2, \ldots, x_n)}(   S \cup \{a, b\} ), \\
&= \sum_{ S \in \power{P \backslash \{a,b\}}}  g_{P}(   S \cup \{a, b\} ). \qedhere
\end{align*}
}
\end{proof}

\subsection{Proof of Theorem ~\ref{median_outcome}}{\label{median_outcome_proof_section}}

[Restatement of Theorem ~\ref{median_outcome}]
  {\it  Any budget $z \in \mathcal{M}(a,b,c)$ if and only if it satisfies the following conditions.
    \begin{enumerate}[leftmargin=0.2in, start = 0]
    \item $\mathcal{Z}(abc)= \mathfrak{X}(abc)$ and $ \mathcal{Z}(\emptyset) = 0$ are always satisfied.
    \item 
    If in \ref{Word:case1}~(eq.~\ref{case1}) then,
     $\mathcal{Z}(ab)=\mathfrak{X}(ab),$ $~~\mathcal{Z}(bc)=\mathfrak{X}(bc),$ and $~~\mathcal{Z}(ca)=\mathfrak{X}(ca).$
     \item 
     If in \ref{Word:case2}~(eq.~\ref{case2}) then, $\mathcal{Z}(a)=\mathcal{Z}(b)=\mathcal{Z}(c)=0.$
    
    %\item $\mathcal{Z}(ab)=\mathfrak{X}(ab)$; $\mathcal{Z}(bc)=\mathfrak{X}(bc)$; $\mathcal{Z}(ca)=\mathfrak{X}(ca)$ \text{ otherwise }
\end{enumerate}
}
\begin{proof}

Consider any set of 3 voters $Q=\{a,b,c\}$.
By Lemma~\ref{utility_one_voter_proj_budget_Z_space}, we have
\begin{align}
&u(a,z) + u(b,z) + u(c,z)  \nonumber\\
&= (\mathcal{Z}(a) + \mathcal{Z}(ac) + \mathcal{Z}(ab) + \mathcal{Z}(abc)) + (\mathcal{Z}(b) + \mathcal{Z}(bc) + \mathcal{Z}(ab) + \mathcal{Z}(abc)) \nonumber\\
&+ (\mathcal{Z}(c) + 
\mathcal{Z}(ac) + \mathcal{Z}(bc) + \mathcal{Z}(abc)), \nonumber \\
&= 3\mathcal{Z}(abc) + 2(\mathcal{Z}(ac) + \mathcal{Z}(bc) + \mathcal{Z}(ab))+ (\mathcal{Z}(a) + \mathcal{Z}(b) + \mathcal{Z}(c)). \label{eq:coff}
\end{align}

Recall from Corollary~\ref{sum_proj_budget_z} that
$\sum_{S \in \power{Q}} \mathcal{Z}(S) = 1$. We use this constraint in the arguments below.

%$u(a,z) + u(b,z) + u(c,z)
%= (Z_a + Z(ac) + Z(ab) + Z(abc)) +
%(Z_b + Z(bc) + Z(ab) + Z(abc)) +
%%(Z_c + Z(ac) + Z(bc) + Z(abc))
%= 3Z(abc) + 2(Z(ac) + Z(bc) + Z(ab)) + 1(Z(a) + Z(b) + Z(c)) + 0Z_\emptyset$,
%and $u(z,z) = \sum_{S \in \power{\{a, b, c\}}} Z(S) = 1$.
Since $\mathfrak{X}(S) \geq \mathcal{Z}(S) \geq 0$ for any subset $S$, the only way to maximize $u(a,z) + u(b,z) + u(c,z)$ is to \edit{select $\mathcal{Z}$ which maximizes the elements from the highest coefficient. That is, we first maximize $\mathcal{Z}(abc)$ which has the highest coefficient in Equation~\eqref{eq:coff}, i.e., $3.$ We then maximize $\mathcal{Z}(ac) + \mathcal{Z}(bc) + \mathcal{Z}(ab)$ which has a coefficient $2$ in Equation~\eqref{eq:coff} followed by  $\mathcal{Z}(a) + \mathcal{Z}(b) + \mathcal{Z}(c)$. The result of the lemma follows. }
%The last expression contains all the coordinates of $Z$, grouped into strata according to the number of voters who benefit from each coordinate. Since $Z_S \leq X_S$ for any $S \in \power{\{a, b, c\}}$, and the conditions in the lemma fill the strata in order from the highest coefficient to the lowest coefficient, the total utility is maximized.
\end{proof}

\subsection{Proof of Theorem ~\ref{max_utility_Nash}}{\label{max_utility_Nash_proof_section}}
[Restatement of Theorem~\ref{max_utility_Nash}]
{\it Every scheme in $\mathcal{N}$ is also a median scheme i.e. $\mathcal{N} \subseteq \mathcal{M}$. }
\begin{proof}
Now consider the outcomes in Lemma~\ref{Nash_barg_outcome}. Thus, we have $\mathcal{Z}(abc)=\mathcal{X}(abc)$ and $\mathcal{Z}(\emptyset)=0$.

Now consider two cases. 

\begin{itemize}
    \item in \ref{Word:case1}~(eq.~\ref{case1}) i.e. $\textsc{excess} \geq 0$ in Lemma \ref{Nash_barg_outcome}\\

    In this case, we have $\mathcal{Z}(ab)=\mathfrak{X}(ab)$, $\mathcal{Z}(ac)=\mathfrak{X}(ac)$ and $\mathcal{Z}(bc)=\mathfrak{X}(bc)$ from Lemma ~\ref{Nash_barg_outcome}.

    \item in \ref{Word:case2}~(eq.~\ref{case2}) i.e. $\textsc{excess} \leq 0$ in Lemma \ref{Nash_barg_outcome}\\

    In this case, we have $\mathcal{Z}(a) = 0$, $\mathcal{Z}(b)=0$ and $\mathcal{Z}(c)=0$ from Lemma ~\ref{Nash_barg_outcome}.
    
\end{itemize}

Now, note that using Lemma~\ref{median_outcome}, we can show that $z \in \mathcal{M}(a,b,c)$.
\end{proof}

\remove{
\begin{corollary*}%{\label{new_budget_z_ineq}}
    $g_P(S) \geq g_{P\cup z} (S \cup z)$ $\forall$ vote profiles $P$ and a budget $z \in \mathbb{B}$.
\end{corollary*}

\begin{proof}
Recall that $g_{j,P}(S) = \max\Bigl(\min\limits_{s \in S} v_{s,j} - \max\limits_{\tilde{s} \in P\setminus S} v_{\tilde{s},j},0\Bigr)$ and $g_{j,P \cup z} (S\cup z) = \max\Bigl(\min\limits_{s \in S \cup z} v_{s,j} - \max\limits_{\tilde{s} \in P\setminus S} v_{\tilde{s},j},0\Bigr)$. Now, since $\min\limits_{s \in S} v_{s,j}\geq \min\limits_{s \in S \cup z} v_{s,j}$, we can say that $g_{j,P}(S) \geq g_{j,P \cup z} (S\cup z)$
and thus, $g_P(S) \geq g_{P \cup z} (S \cup z)$ as $g_P(S) = \sum\limits_{j=1}^{m}g_{j,P}(S)$
\end{proof}

}

\begin{comment}
\subsection{Proof of Theorem \ref{cs_main}}
\label{proof_cs_main}
\begin{proof}
Let $z$ be an outcome from $\Nash(x, y, a)$.
Assume w.l.o.g. that $x$ respects the project interactions, and suppose that $z$ does not. By Lemma~\ref{cs_sum},
$\sum_i f(x)_i = 1$ and $\sum_i f(z)_i < 1$.
Thus, there exists some $k$ where $f(x)_k > f(z)_k$.
By Lemma~\ref{cs_new_b} in Appendix~\ref{cs_new_b},
there exists a budget $z'$  where $f(z')_i \geq f(z)_i$ for all $i$ and $f(z')_k > f(z)_k$. The overlap utility functions satisfy
\begin{align*}
&u(f(x), f(z'))
%&= \min( f(x)_k, f(z')_k ) + \sum_{i; i \neq k} \min( f(x)_i, f(z')_i ) \\  % Not enough space
%&> \min( f(x)_k, f(z)_k ) + \sum_{i; i \neq k} \min( f(x)_i, f(z)_i ) \\  % Not enough space
= \sum_i \min( f(x)_i, f(z')_i ) \\
&> \sum_i \min( f(x)_i, f(z)_i ) = u(f(x), f(z)), \text{~and}\\
&u(f(y), f(z')) = \sum_i \min( f(y)_i, f(z')_i ) \\
&\geq \sum_i \min( f(y)_i, f(z)_i ) = u(f(y), f(z)).
\end{align*}
%Since u'(x, z') > u'(x,z) and u'(y,z') \geq u'(y,z), z is not an outcome from a Pareto-efficient bargaining, a contradiction.
Since $u(f(x), f(z')) > u(f(x), f(z))$ and $u(f(y), f(z')) \geq u(f(y), f(z))$, the Nash product of $z'$ is higher than that of $z$, a contradiction. 
\end{proof}
\end{comment}

\subsection{Technical Lemma~\ref{Nash_barg_outcome}}
\label{Z_nash_proof}
[Restatement of Lemma~\ref{Nash_barg_outcome}]
{\it 
%\label{nash_outcome}
For any preferred budgets of bargaining agents $a$ and $b$, disagreement point $c$, and outcome $z$ of $\mathcal{N}(a, b, c)$, we have,
% if $\mathfrak{X}=g(a, b, c)$ and $\mathcal{Z}_S = g(a, b, c, z)_{\{z\} \cup S}$, then %BLAH
\begin{align*}
\mathcal{Z}(abc) &= \mathfrak{X}(abc), \ \ \ \mathcal{Z}(ab) = \mathfrak{X}(ab), \\
\mathcal{Z}(ac) &= \mathfrak{X}(ac) + \min(\textsc{Excess}/2, 0), \\
\mathcal{Z}(bc) &= \mathfrak{X}(bc) + \min(\textsc{Excess}/2, 0), \\
\mathcal{Z}(a) &= \mathcal{Z}(b) = \max(0, \textsc{Excess}/2), \\
\mathcal{Z}(c) &= \mathcal{Z}(\emptyset) = 0.
\end{align*}
Where, $\textsc{Excess} = (1 - \mathfrak{X}(abc) - \mathfrak{X}(ab) - \mathfrak{X}(ac) - \mathfrak{X}(bc)).$
}
\begin{proof}
We note that the portions of $\mathcal{Z}$ that benefit neither $a$ nor $b$ must be 0. That is, $\mathcal{Z}(c) = \mathcal{Z}(\emptyset) = 0$. Otherwise,
%$u(a, z)$ and $u(b, z)$ must be less than 1, and
we could construct a new outcome $z'$ that reallocated the budget from $\mathcal{Z}(c)$ or $\mathcal{Z}(\emptyset)$ to $a$ and $b$. This would increase $u(a, z)$ and $u(b, z)$ and thus increase the Nash product, a contradiction.

Next, we note the portions of $\mathcal{Z}$ that benefit both $a$ and $b$ must be maximized. That is, $\mathcal{Z}(abc) = \mathfrak{X}(abc)$ and $\mathcal{Z}(ab) = \mathfrak{X}(ab)$. Otherwise, we could construct a new outcome $z'$ that reallocated the budget from any other portion to the portion that benefited both $a$ and $b$.
%, which could not be greater than 1.
This would increase the Nash product, a contradiction.

We now find the maximum of the Nash product.
By Lemma~\ref{utility_common}, %we have
\begin{align*}
%u(a,z) &= \mathcal{Z}_a + \mathcal{Z}(ac) + \mathcal{Z}(ab) + \mathcal{Z}(abc) = \mathcal{Z}(a) + \mathcal{Z}(ac) + \mathfrak{X}(ab) + \mathfrak{X}(abc), \\
%u(b,z) &= \mathcal{Z}(b) + \mathcal{Z}(bc) + \mathcal{Z}(ab) + \mathcal{Z}(abc) = \mathcal{Z}(b) + \mathcal{Z}(bc) + \mathfrak{X}(ab) + \mathfrak{X}(abc), \\
u(a,z) &= \mathcal{Z}(a) + \mathcal{Z}(ac) + \mathfrak{X}(ab) + \mathfrak{X}(abc), \\
u(b,z) &= \mathcal{Z}(b) + \mathcal{Z}(bc) + \mathfrak{X}(ab) + \mathfrak{X}(abc), \\
u(a,c) &= \mathfrak{X}(ac) + \mathfrak{X}(abc), \\
u(b,c) &= \mathfrak{X}(bc) + \mathfrak{X}(abc).
\end{align*}
Let $p_a = \mathcal{Z}(a) + \mathcal{Z}(ac)$ and $p_b = \mathcal{Z}(b) + \mathcal{Z}(bc)$.
%\begin{align*}
%p_a &= \mathcal{Z}(a) + \mathcal{Z}(ac) \\
%p_b &= \mathcal{Z}(b) + \mathcal{Z}(bc)
%\end{align*}
Since $\sum_i z_i = 1$,
%we have
\begin{align*}
\mathcal{Z}(a) + \mathcal{Z}(b) + \mathcal{Z}(ab) + \mathcal{Z}(ac) + \mathcal{Z}(bc) + \mathcal{Z}(abc) &= 1, \\
p_a + p_b + \mathfrak{X}(ab) + \mathfrak{X}(abc) &= 1.
%p_b &= 1 - p_a - \mathfrak{X}(ab) - \mathfrak{X}(abc)
\end{align*}
We rewrite the Nash product as:
\begin{align*}
&(u(a,z) - u(a,c))\cdot (u(b,z) - u(b,c)) \\
&= (p_a + \mathfrak{X}(ab) - \mathfrak{X}(ac))\cdot (p_b + \mathfrak{X}(ab) - \mathfrak{X}(bc)), \\
&= (p_a + \mathfrak{X}(ab) - \mathfrak{X}(ac))\cdot (1 - p_a - \mathfrak{X}(abc) - \mathfrak{X}(bc)), \\
&= -p_a^2 + p_a(1 - \mathfrak{X}(abc) - \mathfrak{X}(bc) - \mathfrak{X}(ab) + \mathfrak{X}(ac)) \\
&+ (\mathfrak{X}(ab) - \mathfrak{X}(ac))\cdot (1 - \mathfrak{X}(abc) - \mathfrak{X}(bc)).
\end{align*}

Let us define $k_a = (1 - \mathfrak{X}(abc) - \mathfrak{X}(ab) + \mathfrak{X}(ac) - \mathfrak{X}(bc)) / 2$ and $k_b = (1 - \mathfrak{X}(abc) - \mathfrak{X}(ab) + \mathfrak{X}(bc) - \mathfrak{X}(ac)) / 2$ and thus, we have $k_a = \frac{\textsc{excess}}{2} + \mathfrak{X}(ac)$ and $k_b = \frac{\textsc{excess}}{2} + \mathfrak{X}(bc)$

We now find the value of $p_a$ which maximizes the Nash product:
\begin{align*}
%0 &= -2p_a + 1 - \mathfrak{X}(abc) - \mathfrak{X}(bc) - \mathfrak{X}(ab) + \mathfrak{X}(ac) \\
p_a &= (1 - \mathfrak{X}(abc) - \mathfrak{X}(ab) + \mathfrak{X}(ac) - \mathfrak{X}(bc)) / 2 = k_a, \\
p_b &= (1 - \mathfrak{X}(abc) - \mathfrak{X}(ab) + \mathfrak{X}(bc) - \mathfrak{X}(ac)) / 2 = k_b.
\end{align*}
We verify that the solution satisfies individual rationality:
\begin{align*}
%1 &= \mathfrak{X}(c) + \mathfrak{X}(ac) + \mathfrak{X}(bc) + \mathfrak{X}(abc) \\
%1 &\geq \mathfrak{X}(ac) + \mathfrak{X}(bc) + \mathfrak{X}(abc) \\
\mathfrak{X}(c) + \mathfrak{X}(ac) + \mathfrak{X}(bc) + \mathfrak{X}(abc) &= 1, \\
1 - \mathfrak{X}(abc) - \mathfrak{X}(bc) &\geq \mathfrak{X}(ac), \\
1 - \mathfrak{X}(abc) + \mathfrak{X}(ab) - \mathfrak{X}(bc) &\geq \mathfrak{X}(ac), \\
1 - \mathfrak{X}(abc) - \mathfrak{X}(ab) + \mathfrak{X}(ac) - \mathfrak{X}(bc) + 2\mathfrak{X}(ab) &\geq 2\mathfrak{X}(ac), \\
%(1 - \mathfrak{X}(abc) - \mathfrak{X}(ab) + \mathfrak{X}(ac) - \mathfrak{X}(bc)) / 2 + \mathfrak{X}(ab) &\geq \mathfrak{X}(ac), \\
p_a + \mathfrak{X}(ab) &\geq \mathfrak{X}(ac), \\
% \mathcal{Z}_a + \mathcal{Z}(ac) + \mathcal{Z}(ab) &\geq \mathfrak{X}(ac), \\
\mathcal{Z}(a) + \mathcal{Z}(ac) + \mathcal{Z}(ab) &\geq \mathcal{Z}(ac), \\
% \mathcal{Z}_a + \mathcal{Z}(ac) + \mathcal{Z}(ab) + \mathcal{Z}(abc) &\geq \mathcal{Z}(ac) + \mathcal{Z}(abc), \\
u(a,z) &\geq u(a,c).
\intertext{Using the symmetry of $a$ and $b$, we also find that:}
u(b,z) &\geq u(b,c).
\end{align*}
% Thus
% \begin{align*}
% \mathcal{Z}_a + \mathcal{Z}(ac) &= k_a \\
% \mathcal{Z}_b + \mathcal{Z}(bc) &= k_b
% \end{align*}
Thus, $\mathcal{Z}(a) + \mathcal{Z}(ac) = p_q = k_a$ and $\mathcal{Z}(b) + \mathcal{Z}(bc) = p_b = k_b$.
Since $\mathcal{Z}(ac) \leq \mathfrak{X}(ac)$, $\mathcal{Z}(bc) \leq \mathfrak{X}(bc)$, and we favor $c$ in case of a tie, we have
$\mathcal{Z}(ac)$, $\mathcal{Z}(bc)$, $\mathcal{Z}(a)$, and $\mathcal{Z}(b)$ as stated in the lemma.
%\begin{align*}
%\mathcal{Z}(ac) &= \min(k_a, \mathfrak{X}(ac)) \\
%\mathcal{Z}(bc) &= \min(k_b, \mathfrak{X}(bc)) \\
%\mathcal{Z}_a &= \max(0, k_a - \mathfrak{X}(ac)) \\
%\mathcal{Z}_b &= \max(0, k_b - \mathfrak{X}(bc))
%\end{align*}
\end{proof}

\remove{
\subsection{Technical Lemma \ref{knapsack}} \label{proof_knapsack}
\begin{lemma}
\label{knapsack}
For budgets $a$, $b$, and $c$, denote $Q = \{a,b,c\}.$
A budget $z$ maximizes
$u(a, z) + u(b, z) + u(c, z)$ iff 
 $$Z(abc) = X(abc)~,~ Z(\emptyset) = 0,$$ and,
\begin{itemize}
\item $Z(a) = Z(b) = Z(c) = 0, \text{~~or,}$
\item $Z(ac) = X(ac)~, ~Z(bc) = X(bc)~, ~Z(ab) = X(ab).$
\end{itemize}
where $X(S)=g_{Q}(S)$ and $Z(S) = g_{Q \cup z}(S \cup z)$ for any subset $S$ of $Q$. % BLAH
\end{lemma}

\begin{proof}
By Lemma~\ref{utility_common},
\begin{align}
&u(a,z) + u(b,z) + u(c,z)  \nonumber\\
&= (Z_a + Z(ac) + Z(ab) + Z(abc)) + (Z_b + Z(bc) + Z(ab) + Z(abc))  \nonumber\\
&+ (Z_c + Z(ac) + Z(bc) + Z(abc)), \nonumber \\
&= 3Z(abc) + 2(Z(ac) + Z(bc) + Z(ab))+ (Z(a) + Z(b) + Z(c)), \label{eq:coff} \\
&\text{and, } u(z,z) = \sum_{S \in \power{Q}} Z(S) = 1. \nonumber
\end{align}
%$u(a,z) + u(b,z) + u(c,z)
%= (Z_a + Z(ac) + Z(ab) + Z(abc)) +
%(Z_b + Z(bc) + Z(ab) + Z(abc)) +
%%(Z_c + Z(ac) + Z(bc) + Z(abc))
%= 3Z(abc) + 2(Z(ac) + Z(bc) + Z(ab)) + 1(Z(a) + Z(b) + Z(c)) + 0Z_\emptyset$,
%and $u(z,z) = \sum_{S \in \power{\{a, b, c\}}} Z(S) = 1$.
Since $X(S) \geq Z(S) \geq 0$ for any subset $S$, the only way to maximize $u(a,z) + u(b,z) + u(c,z)$ is to \edit{select $Z$ which maximizes the elements from the highest coefficient. That is, we first maximize $Z(abc)$ which has the highest coefficient in Equation~\eqref{eq:coff}, i.e., $3.$ We then maximize $Z(ac) + Z(bc) + Z(ab)$ which has a coefficient $2$ in Equation~\eqref{eq:coff} followed by  $Z(a) + Z(b) + Z(c)$. The result of the lemma follows. }
%The last expression contains all the coordinates of $Z$, grouped into strata according to the number of voters who benefit from each coordinate. Since $Z_S \leq X_S$ for any $S \in \power{\{a, b, c\}}$, and the conditions in the lemma fill the strata in order from the highest coefficient to the lowest coefficient, the total utility is maximized.
\end{proof}
}
%\begin{comment}

\remove{
\subsection{Proof of Theorem~\ref{max_utility} }
\label{proof_max_utility}
\begin{proof}
%Let $X=g(a, b, c)$ and $Z_S = g(a, b, c, z)_{\{z\} \cup S}$. %BLAH
From Lemma~\ref{nash_outcome} in Appendix~\ref{Z_proof}, we have: $$k_a - X(ac) = k_b - X(bc).$$ 

\textbf{Case 1: }If $Z(a) = 0$, we have
$Z(a) = \max(0, k_a - X(ac)) = 0$. This implies 
$k_a - X(ac) \leq 0$ and 
$k_b - X(bc) \leq 0$.
Thus, $Z(b) = \max(0, k_b - X(bc)) = 0$.
%Therefore, by Lemma~\ref{knapsack}, the total utility is maximized.

\textbf{Case 2: }If $Z(a) > 0$, we have
$Z(a) = \max(0, k_a - X(ac)) > 0$. This implies $k_a - X(ac) > 0$ and
$k_b - X(bc) > 0$.
Thus, $Z(ac) = \min(k_a, X(ac)) = X(ac)$ and $Z(bc) = \min(k_b, X(bc)) = X(bc)$.

By Lemma~\ref{knapsack} in Appendix~\ref{proof_knapsack}, the total overlap utility of $a,b,$ and $c$ is maximized in both Case 1 and Case 2. 
\end{proof}

}
%\end{comment}
\subsection{Proof of Lemma~\ref{disutility}}
\label{disutility_proof}
[Restatement of Lemma~\ref{disutility}]
{\it For budgets $a,b \in \mathbb{B}$, $d(a,b) = 2 - 2u(a,b)$.
}
\begin{proof}
Using the facts that $\sum_i a_i = 1$, $\sum_i b_i = 1$ and, $x + y - |x - y|= 2\min(x, y)$, we have:
\begin{align*}
&d(a, b) = \sum_i |a_i - b_i| = (2 - \sum_i a_i - \sum_i b_i) + \sum_i |a_i - b_i|, \\
&= 2 - \sum_i (a_i + b_i - |a_i - b_i|) = 2 - 2\sum_i \min(a_i, b_i) = 2 - 2u(a, b). \qedhere
\end{align*}
%This completes the proof. 
\end{proof}

%Note that this lemma says that for any deliberation scheme $\mathfrak{q} \in \mathcal{Q}$, there exists a vote-profile $P$ s.t. the distortion ratio goes all the way to 2. We prove this lemma in Appendix \ref{distortion_two_bargaining_proof_section}

\subsection{Proof of Lemma~\ref{distortion_pd}}
\label{proof_distortion}

[Restatement of Lemma~\ref{distortion_pd}]
{\it $\Distortion_once(\mathcal{M}) \leq PD(\mathcal{M})$.}
\begin{proof}
Denote the set of budgets $\{v_1, v_2, \ldots, v_n\},$ %each in $\mathbb{B},$ 
by $P$ such that $v_i \in \mathbb{B}$ for all $i \in [n].$ For class of bargaining schemes $\mathcal{M},$
\begin{align*}
\Distortion_once(\mathcal{M}) &= \sup_{n \in \mathbb{Z}^+,~P \in \mathbb{B}^n,~h \in \mathcal{H}} \frac{ \frac{1}{n^4} \sum\limits_{i,j,k,l \in [n]} d(h(v_i, v_j, v_k), v_l)}
{ \min\limits_{p \in \mathbb{B}} \frac{1}{n} \sum\limits_{i \in [n]} d(p, v_i) }, \\
&= \sup_{n \in \mathbb{Z}^+,~P \in \mathbb{B}^n,~h \in \mathcal{H}} \frac{ \sum\limits_{Q \in \combiwithr{[n], 6}}  \!\!\frac{\Prob(Q)}{60} \!\!\sum\limits_{R \in \combi{Q, 3}} \sum\limits_{i \in Q \setminus R}  d(h(R), v_i) }
{ \min\limits_{p \in \mathbb{B}} \frac{1}{n} \sum\limits_{i \in [n]} d(p, v_i) }, \\
% FIXME: Explain why
% FIXME: Explain why
% FIXME: Explain why
% FIXME: Explain why
&\leq \sup_{n \in \mathbb{Z}^+,~P \in \mathbb{B}^n,~h \in \mathcal{H}} \frac{ \sum\limits_{Q \in \combiwithr{[n], 6}} \!\!\frac{\Prob(Q)}{60} \!\!\sum\limits_{Q \in \combi{P, 3}} \sum\limits_{i \in Q \setminus R}  d((h(R), v_i) }
{ \sum\limits_{Q \in \combiwithr{[n], 6}} \Prob(Q) \min\limits_{p \in \mathbb{B}} \frac{1}{6} \sum\limits_{i \in Q} d(p, v_i) }, \\
&\leq \sup_{n \in \mathbb{Z}^+,~P \in \mathbb{B}^n,~h \in \mathcal{H}} \max_{Q \in \combiwithr{[n], 6}} \frac{ \frac{1}{60} \sum\limits_{R \in \combi{Q, 3}} \sum\limits_{i \in Q \setminus R}  d((h(R), v_i) }
{ \min\limits_{p \in \mathbb{B}} \frac{1}{6} \sum\limits_{i \in Q} d(p, v_i) }, \\
&= PD(\mathcal{M}). \qedhere
\end{align*}
%This completes the proof.
\end{proof}

\subsection{Proof of Lemma~\ref{pd_triadic_nash_relaxed}}{\label{pd_triadic_nash_relaxed_proof}}

[Restatement of Lemma~\ref{pd_triadic_nash_relaxed}]
 {\it   $PD(\mathcal{M}) \leq 1.80$.}
\begin{proof}
Let $p^Q$ denote a budget obtained on bargaining with budgets in set $Q$ using a bargaining scheme in $\mathcal{M}$. Note that outcomes of $\mathcal{M}$ are \ref{Word:symmetric} in  $Q$ therefore we do not need to designate a disagreement point in $Q$ for the purpose of analysis.

%maximize the total utility of three voters in set $R$.
%That is, $$\Median(R) = \operatorname*{argmax}_{p \in \mathbb{B}} \sum_{v \in R} u(p, v).$$

%By Theorem~\ref{max_utility}, $\Nash(R) \subseteq \Median(R)$. Therefore,
\begin{align*}
PD(\mathcal{M}) &= \sup_{P \in \mathbb{B}^6;~\mathfrak{n} \in \mathcal{M}} \frac{ \displaystyle \frac{1}{60} \sum\limits_{Q \in \combi{[6], 3}} \sum\limits_{i \in [6] \setminus Q}  d(\mathfrak{n}(Q), P_i) }
{   \frac{1}{6} \min\limits_{v \in \mathbb{B}} \sum\limits_{i \in [6]} d(v, P_i) }, \\
&\leq \sup_{P \in \mathbb{B}^6} \frac
{ \displaystyle \frac{1}{60} \sum\limits_{Q \in \combi{[6], 3}} \sup\limits_{p^Q \in \mathcal{M}(Q)} \sum\limits_{i \in [6] \setminus Q} d(p^Q, P_i) }
{ \frac{1}{6} \min\limits_{v \in \mathbb{B}}  \sum\limits_{i \in [6]} d(v, P_i) }.
\end{align*}

Suppose that $PD(\mathcal{M}) > 1.80$.
Then the following optimization problem has an optimal objective value strictly greater than $0$. 

%maximize $\frac{1}{20} \sum_{R \in \combi{P, 3}} \sum_{v \in P \setminus R} d(p^R, v)
%-
%1.93 \cdot \frac{1}{5} \sum_{v \in P} d(p_{\text{all}}, v)$ \\
%\-\ \ \ \ s.t.
%\begin{align*}
%P &\in \mathbb{B}^5 \\
%p^R &\in \Median(R) \ \ \ \ \forall R \in \combi{P, 3} \\
%p_{\text{all}} &\in \mathbb{B}
%\end{align*}
\begin{align}
\text{maximize } ~~~&\frac{1}{60} \sum_{Q \in \combi{[6], 3}} \sum_{i \in [6] \setminus Q} d(p^Q, P_i) - 1.80 \cdot \frac{1}{6} \sum_{i \in [6]} d(v, P_i), \nonumber\\
\text{subject to } ~~~&P \in \mathbb{B}^6, \nonumber\\
~~~&p^Q \in \mathcal{M}(Q) \ \ \ \ \ \ \ \ \ \ \ \ \ \ \ \ \ \ \ \ \ \ \ \ \ \ \ \ \ \ \ \ \ \ \ \ \ \ \ \ \ \ \ \forall Q \in \combi{[6], 3}, \nonumber \\
~~~&v \in \mathbb{B}.  \label{prob:opt}
\end{align}
To convert this problem into a linear program, we map it to the incremental allocation space of the set of 6 budgets $P=\{P_1,P_2,\ldots,P_6\}$.
% FIXME: Maybe define these later.
Denote $X(S) = X_{P}(S)$, the projection
$V(S) = X_{(P \cup \{v\})}( S \cup \{v\})$ and,
$Z^Q(S) = X_{(P \cup p^Q)}( S \cup p^Q)$ for each $Q \in \combi{[6], 3}$.
Using Lemma~\ref{disutility},
we rewrite $d(p^Q, P_i)$ as
$2 - 2u(p^Q, P_i),$ which, by Lemma~\ref{utility_common}, equals $2 - 2\sum_{ S \in \power{P\backslash P_i } }  Z^Q(S \cup P_i)$.
The same goes for $d(v, P_i)$.

% FIXME: Confusing
% FIXME: Confusing
To make the $p^Q \in \mathcal{M}(Q)$ constraints linear, we use case analysis.
% to make it manageable ?
Consider a given $Q=\{ q_1, q_2, q_3 \} \in  \combi{[6],3}$ and a budget $p^Q \in \mathbb{B}$.
Let $\mathfrak{X}(S) = X_{Q}(S)$ and $\mathcal{Z}(S) = X_{(Q \cup p^Q)}( S \cup p^Q)$. % BLAH
\S~\ref{relax_Nash_defn} implies that $p^Q \in \mathcal{M}(Q)$ if and only if the following holds: %$Z(123) = Y(123)$, $Z(\emptyset) = 0$, and
\begin{itemize}
%\item (Case 0) If $X_{123} \geq 1$, then $p \in \Median(R)$ if and only if $Z_{12} = Z_{13} = Z_{23} = Z_1 = Z_2 = Z_3 = Z_\emptyset = 0$. %$Z_{123}=1$ and other $Z$'s are 0.
% \item (Case 1) If $X_{123} \leq 1$ and $X_{123} + X_{12} + X_{13} + X_{23} \geq 1$, then $Z_1 = Z_2 = Z_3 = 0$.
\item \ref{Word:case1}: If $\mathfrak{X}(q_1q_2q_3) + \mathfrak{X}(q_1q_2) + \mathfrak{X}(q_1q_3) + \mathfrak{X}(q_2q_3) \geq 1$, then $\mathcal{Z}(q_1) = \mathcal{Z}(q_2) = \mathcal{Z}(q_3) = 0$.
\item \ref{Word:case2}: If $\mathfrak{X}(q_1q_2q_3) + \mathfrak{X}(q_1q_2) + \mathfrak{X}(q_1q_3) + \mathfrak{X}(q_2q_3) \leq 1$, then,  $\mathcal{Z}(q_1q_2)=\mathfrak{X}(q_1q_2), \ \ \mathcal{Z}(q_1q_3)=\mathfrak{X}(q_1q_3),  \text{and~} \ \   \mathcal{Z}(q_2q_3)=\mathfrak{X}(q_2q_3)$.
\end{itemize}
%
%Now we apply Lemma~\ref{proj_budget_new_space_lemma} to argue that $\mathfrak{X}(\mathfrak{S}) = \sum\limits_{V' \in \mathcal{P}([6]\setminus R)} X(\mathfrak{S}\cup R)$ and similarly, $\mathfrak{Z}^R(\mathfrak{S}) = \sum\limits_{V' \in \mathcal{P}([6]\setminus R)} X(\mathfrak{S} \cup V')$ to obtain the constraints  
%
For each $p^Q \in \mathcal{M}(Q)$ constraint, we break it into two cases. Since there are ${\binom{6}{3}}$ such sets in the optimization problem~\eqref{prob:opt}, there are $2^{\binom{6}{3}}$ instances overall. Let us represent each such instance by a binary string of length 20 where a 0 or 1 at each position would denote whether the triplet $Q$ corresponding to that position belongs in \ref{Word:case1} or \ref{Word:case2}.
However, most of these $2^{20}$ instances are not unique up to the permutation of voter preferences, i.e. when the voting preferences of different voters are permuted, we may move from one instance to another. We do not need to solve all these cases. In all, we have 2136 unique instances. \footnote{We can show a similar result as Lemma~\ref{toggle_redn} here for this optimization problem and can reduce number of instances further to 1244.}
%However, when we solve the optimization problem under both cases, the result would be the same. Thus we remove all such cases which are not unique up to permutation amongst voters and we obtain 2136 such cases. %Also note that for the following optimization problem, $X(.)$ and $Z^R(.)$ are defined on the common budget space of budgets $P_1,P_2,\ldots,P_6$ i.e. $X(S)=g_P(S)$ and $Z^R(S) = g_{P\cup \{p^R\}}(S \cup \{p^R\})$
%Let us denote the set of all such cases by $\mathbb{K}$.
%
%
%Each case $\kappa \in \mathbb{K}$ reduces to a linear program described below:%, described in Appendix~\ref{LP}. %
The linear program in each case is as follows with the optimization variables $\{X(S)\}_{S \in \mathcal{P}([6])},\{V(S)\}_{S \in \mathcal{P}([6])},\{Z^Q(S)\}_{S \in \mathcal{P}([6]); Q \in \combi{[6],3}}$.\footnote{For simplicity of notation, we refer to budget $P_i$ by $i$ while constructing the sets $S$.} Note that \eqref{allocation_V_postive},\eqref{allocation_Z_positive},\eqref{postive_allocation_i} follow from the fact that allocations in the incremental allocation space is non-negative. Note \eqref{total_allocation_i} follows from Lemma ~\ref{total_allocation_budget_space},\eqref{total_allocation_Z} and \eqref{total_allocation_V} follows from Corollary ~\ref{sum_proj_budget_z}. Also \eqref{allocation_Z_bound},\eqref{allocation_V_bound} follow from Corollary \ref{new_budget_z_ineq}. Also, \eqref{allocation_common_none} and \eqref{allocation_common_none} follow from Lemma ~\ref{median_outcome} and applying the projection operation [Lemma ~\ref{proj_budget_new_space_lemma}] from the space of 6 budgets $P$ to 3 budgets $Q=\{q_1,q_2,q_3\}$. For the case wise constraints in the optimization problem, we use the cases described above and again apply the projection operation from the space of 6 budgets $P$ to 3 budgets $Q=\{q_1,q_2,q_3\}$
%\sukolsak{TODO: Explain how to convert $d(\ldots, \ldots)$ to $u(\ldots, \ldots)$ to $X_{\ldots}$}
%

\begin{align}{\label{opti_relaxed}}
&\text{maximize}~~~~ \frac{1}{60} \sum_{Q \in \combi{[6], 3}} \sum_{i \in [6] \setminus Q} d(p^Q, P_i) - 1.80 \cdot \frac{1}{6} \sum_{i \in [6]} d(v, P_i) \\
\text{where} & ~~~~~~~~d(p^Q, P_i) = 2 - 2\sum_{ S \in \power{P\backslash P_i } } 
Z^Q(S \cup P_i)\\ & ~~~~~~~~ d(v, P_i) = 2 - 2\sum_{ S \in \power{P\backslash P_i } } 
V(S \cup P_i).\\
&\text{subject to} \nonumber
\end{align}

The variables in this optimization problem is $\{X(S)\}_{S \in \mathcal{P}([6])},\{V(S)\}_{S \in \mathcal{P}([6])},\{Z^Q(S)\}_{S \in \mathcal{P}([6]); Q \in \combi{[6],3}}$.

\begin{align}
& \sum_{S \in \power{[6]\backslash \{i\}}} X(S \cup i) =1 &\forall i\in [6], \label{total_allocation_i}\\
& X(S) \geq 0  &\forall S \in \power{P}, \label{postive_allocation_i}\\
& \sum_{S \in \power{P}} Z^Q(S) = 1 &\forall Q \in \combi{[6], 3},\label{total_allocation_Z} \\
& Z^Q(S) \geq 0 &\forall Q \in \combi{[6], 3}, \forall S \in \power{[6]}, \label{allocation_Z_positive}\\
& X(S) \geq Z^Q(S) &\forall Q \in \combi{[6], 3}, \forall S \in \power{[6]}, \label{allocation_Z_bound}\\
& Z^Q(Q \cup V') = X(Q \cup V')& \forall Q \in \combi{[6], 3}, \forall V' \in \power{[6] \setminus Q}, \label{allocation_common_three}\\  % p in Median
& Z^Q(V') = 0 &\forall Q \in \combi{[6], 3}, \forall V' \in \power{[6] \setminus Q}, \label{allocation_common_none}\\ % p in Median
& \sum_{S \in \power{[6]}} V(S) = 1, \label{total_allocation_V} &\\
& V(S) \geq 0 &\forall S \in \power{[6]} \label{allocation_V_postive}\\ ,
& X(S) \geq V(S)  &\forall S \in \power{[6]}\label{allocation_V_bound}.
\end{align}
%Let $f$ map each such triplet in $\combi{[6], 3}$ to a distinct number in $\{0,1\ldots,19\}.$

%With the following additional constraints for each $Q \in \combi{[6], 3}:$
We also have following case-wise constraints for each $ Q \in \combi{[6], 3}$. If $Q$ is in \ref{Word:case1},
%\noindent If $\kappa_{f(Q)} == 0$ then,
\begin{gather*}
\sum_{i=2}^3 ~\sum_{Q' \in \combi{Q, i}} ~\sum_{V' \in \power{[6] \setminus Q}} X(Q' \cup V') \geq 1, \\
Z^Q(Q' \cup V') = 0 \ \ \ \ \ \ \ \ \ \ \ \ \ \ \ \ \ \ \ \ \ \ \ \ \forall Q' \in \combi{Q, 1}, \forall V' \in \power{[6] \setminus Q}.
\end{gather*}
Whereas if $Q$ is in \ref{Word:case2},
\begin{gather*}
\sum_{i=2}^3 ~\sum_{Q' \in \combi{Q, i}} ~\sum_{V' \in \power{[6] \setminus Q}} X(Q' \cup V') \leq 1, \\
Z^Q(Q' \cup V') = X(Q' \cup V') \ \ \ \ \ \ \ \ \forall Q' \in \combi{Q, 2}, \forall V' \in \power{[6] \setminus Q}.
\end{gather*}
%\}
%$\forall {Q \in \mathcal{C}([6],3)}$ 

%Note: We can show that optimization problem \ref{opti_relaxed} given by cases $\kappa$ and $\bar{\kappa}$ are identical for every case $\kappa \in \{0,1\}^{20}$. Note that $\bar{\kappa}$ toggles every bit in $\kappa$.

We solve all the linear programs for each of 2136 instances%in $\mathbb{K}$ ({\color{red} 2136} such cases)
 and find that the objective value is always equal to $0$. Hence, a contradiction. %Therefore, $PD(\Triadic) < 1.93$. 

 The optimization problem has been solved in Gurobi\cite{gurobi} and the scripts for the same can be found in \cite{Pythonscripts}.
\end{proof}

\subsection{Some results on $\{\tilde{Z}_{j,P}(.)\}_{j \in [m]} \sim \tilde{\mathfrak{n}}_{\text{rand}}(a,b,c)$ and distinction from $Z_{j,P}(.)$ where $z \sim \mathfrak{n}(a,b,c)$} \label{results_tilde_Z}

Consider $\{\tilde{Z}_{j,P}(.)\}_{j \in [m]} \sim \tilde{\mathfrak{n}}_{\text{rand}}(a,b,c)$ and $z \sim \mathfrak{n}(a,b,c)$ as described in \S \ref{sec_alt_barg_soln} and \S \ref{exact_soln_constr} respectively. Recall Definition \ref{new_budget_proj_budget_defn}, of $Z_{j,P}(S)=X_{j,P\cup\{z\}}(S \cup \{z\})$. 
We now restate and prove the corollaries corresponding to Corollaries \ref{sum_proj_budget_z} and \ref{new_budget_z_ineq} on $\tilde{Z}_{j,P}(.)$ instead of $Z_{j,P}(.)$.

\begin{corollary}{\label{new_budget_z_tilde_ineq}}
   $\tilde{Z}_{j,P}(S) \leq X_{j,P} (S)$ for all $j \in [m].$ Summing over $j\in [m]$, $\tilde{Z}_{P}(S) \leq X_{P} (S)$.
\end{corollary}

\begin{proof}

Recall the construction of $\tilde{Z}_{j,P}(S)$ as described in \S \ref{sec_alt_barg_soln} as described before \ref{Word:case1} and \ref{Word:case2} in \S \ref{sec_alt_barg_soln}.

Now consider we arrive at \ref{Word:case1} i.e. $\mathfrak{X}(ab)+\mathfrak{X}(bc)+\mathfrak{X}(ca)+\mathfrak{X}(abc) \leq 1$. In this case, recall that the construction of $\alpha_{j,P}(S) = r_j^a \frac{X_{j,P}(S)}{X_{j,Q}(a)}$ where $r_j^a$ was the randomised excess allocation to project $j$ as defined in \S \ref{exact_soln_constr}. Since $r_j^a \leq s_j^a = X_{j,Q}(a)$ as defined in \S \ref{exact_soln_constr}, we must have $\alpha_{j,P}(S) \leq X_{j,P}(S)$ for $S \in \{S| a \in S; b,c \notin S\}$. Similarly, we must have $\beta_{j,P}(S) \leq X_{j,P}(S)$ for all $S \in \{S| b \in S; a,c \notin S\}$. However, after \ref{Word:step1} and \ref{Word:step2} in \S \ref{sec_alt_barg_soln}, we initially assigned $\tilde{Z}_{j,P}(S)= X_{j,P}(S)$ for all $\{S: S \text{ contains at least 2 elements of $\{a,b,c\}$}\}$ and 0 otherwise. Thus, final assigned values to $\tilde{Z}_{j,P}(S)$ after \ref{Word:case1} would satisfy $\tilde{Z}_{j,P}(S) \leq X_{j,P}(S) \ \forall S \in \mathcal{P}(P)$.

Now consider we arrive at \ref{Word:case2} i.e. $\mathfrak{X}(ab)+\mathfrak{X}(bc)+\mathfrak{X}(ca)+\mathfrak{X}(abc) \geq 1$. In this case, recall that the construction of $\alpha_{j,P}(S) = r_j^a \frac{X_{j,P}(S)}{X_{j,Q}(bc)}$ where $r_j^a$ was the randomised allocation removed from project $j$ as defined in \S \ref{exact_soln_constr}. Since $r_j^a \leq t_j^a = X_{j,Q}(bc)$ as defined in \S \ref{exact_soln_constr}, we must have $\alpha_{j,P}(S) \leq X_{j,P}(S) \text{ for all } \{S| a,c \in S; b \notin S\}$. Similarly, we must have $\beta_{j,P}(S) \leq X_{j,P}(S) \text{ for all } \{S| b,c \in S; a \notin S\}$. However, after \ref{Word:step1} and \ref{Word:step2} in \S \ref{sec_alt_barg_soln}, we initially assigned $\tilde{Z}_{j,P}(S)= X_{j,P}(S)$ for $\{S: S \text{ contains at least 2 elements of $\{a,b,c\}$}\}$ and 0 otherwise. Thus, final assigned values to $\tilde{Z}_{j,P}(S)$ after \ref{Word:case2} would satisfy $0 \leq \tilde{Z}_{j,P}(S) \leq X_{j,P}(S) \ \forall S \in \mathcal{P}(P)$ after decrementation by $\alpha_{j,P}(.)$ and $\beta_{j,P}(.)$
%Thus, after incrementing $\{S: a \in S; b,c \notin S\}$ by $\alpha_{j,P}(S)$ and  
\end{proof}
%Note that this results says the incremental project allocation common to a new budget $z$ is bounded by the incremental project allocation for every subset of voters $S$.

\begin{corollary}{\label{sum_proj_budget_z_tilde}}
$\sum_{S \in \mathcal{P}(P)} \tilde{Z}_{j,P} (S)=z_j$ for all vote profiles $P$ and $z \in \mathbb{B}.$ Summing over all projects $j \in [m]$, we get $\sum_{S \in \mathcal{P}(P)} \tilde{Z}_{P} (S)=1.$
\end{corollary}

\begin{proof}

    Recall the construction of $\tilde{Z}_{j,P}(S)$ as described in \S \ref{sec_alt_barg_soln} as described before \ref{Word:case1} and \ref{Word:case2} in \S \ref{sec_alt_barg_soln}. Let us denote the initial assignment of $\tilde{Z}_{j,P}(.)$ by $\tilde{Z}^{\text{init}}_{j,P}(.)$ after \ref{Word:step1} and \ref{Word:step2} in \S \ref{sec_alt_barg_soln}. Recall that we denote $Q=\{a,b,c\}$. Also let us denote the allocation to project $j$ by outcome $z$ after \ref{Word:step1} and \ref{Word:step2} in \S\ref{Sec:nash_outcomes} by $z^{\text{init}}_j$.
    Now, 

    \begin{align*}
        \sum\limits_{{S \in \mathcal{P}(P)}}\tilde{Z}^{\text{init}}_{j,P}(S) \overset{(a)}{=}  \sum\limits_{\substack{S \in \mathcal{P}(P)\\ S \ni a,b \text{ or } S \ni b,c \\ \text{ or } S \ni c,a}}X_{j,P}(S) & \overset{(b)}{=} X_{j,Q} (ab)+ X_{j,Q} (bc) + X_{j,Q} (ca) + X_{j,Q}(abc) \\ & \overset{(c)}{=} \max(\min(a_j,b_j),\min(b_j,c_j),\min(a_j,c_j)) \overset{(d)}{=} z^{\text{init}}_j
    \end{align*}
    %$$\sum\limits_{{S \in \mathcal{P}(P)}}\tilde{Z}^{\text{init}}_{j,P}(S) \overset{(a)}{=}  \sum\limits_{\substack{S \in \mathcal{P}(P)\\ S \ni a,b \text{ or } S \ni b,c \\ \text{ or } S \ni c,a}}X_{j,P}(S) \overset{(b)}{=} X_{j,Q} (ab)+ X_{j,Q} (bc) + X_{j,Q} (ca) + X_{j,Q}(abc) \overset{(c)}{=} \max(\min(a_j,b_j),\min(b_j,c_j),\min(a_j,c_j)) \overset{(d)}{=} z^{\text{init}}_j$$.

Note $(a)$ follows since $\tilde{Z}_{j,P}(S)= X_{j,P}(S)$ for all $S$ containing 2 elements of $Q$. $(b)$ follows from the Lemma ~\ref{proj_budget_new_space_lemma} and $(c)$ follows from Definition \ref{proj_budget_defn} and note that this expression is precisely equal to the median of $a_j,b_j$ and $c_j$ and $(d)$ follows from the \ref{Word:step1} and \ref{Word:step2} in \S \ref{Sec:nash_outcomes}.

    Now, let us first consider \ref{Word:case1} in \S \ref{sec_alt_barg_soln}. Now, we have 
    
    $$ \sum\limits_{\substack{S \in \mathcal{P}(P)\\ S \ni a; S \not\ni b,c}}\alpha_{j,P}(S) = r^a_j \text { as } \sum\limits_{\substack{S \in \mathcal{P}(P)\\ S \ni a; S \not \ni b,c}} X_{j,P}(S) = X_{j,Q}(a)\text{ [from Lemma \ref{proj_budget_new_space_lemma}]}$$

    $$ \sum\limits_{\substack{S \in \mathcal{P}(P)\\ S \ni b; S \not\ni a,c}}\alpha_{j,P}(S) = r^b_j \text { as } \sum\limits_{\substack{S \in \mathcal{P}(P)\\ S \ni b; S \not \ni a,c}} X_{j,P}(S) = X_{j,Q}(b)\text{ [from Lemma \ref{proj_budget_new_space_lemma}]}$$.

    \[
    \sum\limits_{{S \in \mathcal{P}(P)}}\tilde{Z}_{j,P}(S) = \left\{\begin{array}{lr}
        z^{\text{init}}_j + r^a_j, & z_j<a_j\\
        z^{\text{init}}_j + r^b_j, & z_j<b_j\\
        z^{\text{init}}_j, & \text{ otherwise}
        \end{array}\right\} \overset{(f)}{=} z_j
  \]

  Note $(f)$ follows from the construction described in \ref{Word:case1} described in \S \ref{Sec:nash_outcomes}

Now, let us first consider \ref{Word:case2} in \S \ref{sec_alt_barg_soln}. Now, we have 
    
    $$ \sum\limits_{\substack{S \in \mathcal{P}(P)\\ S \ni b,c; S \not\ni a}}\alpha_{j,P}(S) = r^a_j \text { as } \sum\limits_{\substack{S \in \mathcal{P}(P)\\ S \ni b,c; S \not \ni a}} X_{j,P}(S) = X_{j,Q}(bc)\text{ [from Lemma \ref{proj_budget_new_space_lemma}]}$$

    $$ \sum\limits_{\substack{S \in \mathcal{P}(P)\\ S \ni a,c; S \not\ni b}}\alpha_{j,P}(S) = r^b_j \text { as } \sum\limits_{\substack{S \in \mathcal{P}(P)\\ S \ni a,c; S \not \ni b}} X_{j,P}(S) = X_{j,Q}(ac)\text{ [from Lemma \ref{proj_budget_new_space_lemma}]}$$.

    \[
    \sum\limits_{{S \in \mathcal{P}(P)}}\tilde{Z}_{j,P}(S) = \left\{\begin{array}{lr}
        z^{\text{init}}_j - r^a_j, & z_j>a_j\\
        z^{\text{init}}_j - r^b_j, & z_j>b_j\\
        z^{\text{init}}_j, & \text{ otherwise}
        \end{array}\right\} \overset{(g)}{=} z_j
  \]

  Note $(g)$ follows from the construction described in \ref{Word:case1} described in \S \ref{Sec:nash_outcomes}
    %Now, $$\sum\limits_{{S \in \mathcal{P}(P)}}\tilde{Z}_{j,P}(S) \overset{a}{=} \max(\min(a_j,b_j),\min(b_j,c_j),\min(a_j,c_j))+ r^a_j \text{ if }$$ 
%Howwever, we have $\sum\limits_{S \}$
\end{proof}

Recall that we constructed $\tilde{Z}_Q(S) = \sum_{\hat{S} \in \mathcal{P}(P\setminus Q)} \tilde{Z}_P(S \cup \hat{S})$ for any subset $Q \subseteq P$.

We now show that our construction satisfies all lemmas of Corollary \ref{Nash_barg_outcome} for $\tilde{Z}_Q$.

\begin{lemma}
\label{Nash_barg_outcome_tilde_Z}
Any construction $\tilde{Z} \sim \tilde{\mathfrak{n}}_{\text{rand}}(a,b,c)$ satisfies the following conditions.
\begin{align*}
\tilde{Z}_Q(abc) &= \mathfrak{X}(abc), \ \ \ \tilde{Z}_Q(ab) = \mathfrak{X}(ab), \\
\tilde{Z}_Q(ac) &= \mathfrak{X}(ac) + \min(\textsc{Excess}/2, 0), \\
\tilde{Z}_Q(bc) &= \mathfrak{X}(bc) + \min(\textsc{Excess}/2, 0), \\
\tilde{Z}_Q(a) &= \tilde{Z}_Q(b) = \max(0, \textsc{Excess}/2), \\
\tilde{Z}_Q(c) &= \tilde{Z}_Q(\emptyset) = 0.
\end{align*}
Where, $\textsc{Excess} = (1 - \mathfrak{X}(abc) - \mathfrak{X}(ab) - \mathfrak{X}(ac) - \mathfrak{X}(bc)).$
\end{lemma}

\begin{proof}
    Note that $\tilde{Z}_{j,P}(S)$ is unchanged for $S$ s.t. $a,b \in S$ and $a,b \notin S$ both in \ref{Word:case1} and \ref{Word:case2} of \S \ref{sec_alt_barg_soln}. Thus, summing over all such sets and applying Lemma \ref{proj_budget_new_space_lemma}, we obtain $\tilde{Z}_Q(abc) = \mathfrak{X}(abc)$, $\tilde{Z}_Q(ab) = \mathfrak{X}(ab)$ , $\tilde{Z}_Q(c) = \mathfrak{X}(c)$ and $\tilde{Z}_Q(\emptyset) = \mathfrak{X}(\emptyset)$.

    Now consider \ref{Word:case1} in \S \ref{sec_alt_barg_soln} i.e. $\textsc{excess} \geq 0$.

    In this case, we have\\
    
    $$\sum\limits_{j=1}^{m} \sum\limits_{\substack{S \in \mathcal{P}(P)\\ S \ni a; S \not \ni b,c}}\alpha_{j,P}(S) \overset{(a)}{=} \sum_{j=1}^{m} r^a_j \overset{(b)}{=}\frac{\textsc{excess}}{2} \text{ and similarly } \sum\limits_{j=1}^{m} \sum\limits_{\substack{S \in \mathcal{P}(P)\\ S \ni b; S \not \ni a,c}}\beta_{j,P}(S) = \frac{\textsc{excess}}{2}$$

Note that $(a)$ follows from the fact that $\sum\limits_{\substack{S \in \mathcal{P}(P)\\ S \ni a; S \not \ni b,c}} X_{j,P}(S) = X_{j,Q}(a)$ [Lemma ~\ref{proj_budget_new_space_lemma}] and $(b)$ follows from \ref{Word:case1} in \S \ref{exact_soln_constr}. Thus, we obtain $\tilde{Z}_Q(a)=\mathfrak{X}(a) + \frac{\textsc{excess}}{2}$ ; $\tilde{Z}_Q(b)=\mathfrak{X}(b) + \frac{\textsc{excess}}{2}$ ;$\tilde{Z}_Q(ac) = \mathfrak{X}(ac)$ and $\tilde{Z}_Q(bc) = \mathfrak{X}(bc)$

    Now consider \ref{Word:case2} in \S \ref{sec_alt_barg_soln} i.e. $\textsc{excess} \leq 0$.

In this case, we have\\
    
    $$\sum\limits_{j=1}^{m} \sum\limits_{\substack{S \in \mathcal{P}(P)\\ S \ni b,c; S \not \ni a}}\alpha_{j,P}(S) \overset{(a)}{=} \sum_{j=1}^{m} r^a_j \overset{(b)}{=}\frac{-\textsc{excess}}{2} \text{ and similarly } \sum\limits_{j=1}^{m} \sum\limits_{\substack{S \in \mathcal{P}(P)\\ S \ni a,c; S \not \ni b}}\beta_{j,P}(S) = \frac{-\textsc{excess}}{2}$$

Note that $(a)$ follows from the fact that $\sum\limits_{\substack{S \in \mathcal{P}(P)\\ S \ni b,c; S \not \ni a}} X_{j,P}(S) = X_{j,Q}(bc)$ [Lemma ~\ref{proj_budget_new_space_lemma}] and $(b)$ follows from \ref{Word:case1} in \S \ref{exact_soln_constr}. Thus, we obtain $\tilde{Z}_Q(bc)=\mathfrak{X}(bc) + \frac{\textsc{excess}}{2}$ and similarly, $\tilde{Z}_Q(ac)=\mathfrak{X}(ac) +\frac{\textsc{excess}}{2}$ but $\tilde{Z}_Q(a) = \tilde{Z}_Q(b)=0$. 
%Hence proved.
\end{proof}

Note, that Corollaries \ref{sum_proj_budget_z}, \ref{new_budget_z_ineq} and Lemma ~\ref{Nash_barg_outcome} are satisfied by both $Z_{j,P}(.)$ and $\tilde{Z}_{j,P}(.)$ So there might be a question regarding what condition is satisfied by $Z_{j,P}(.)$ but not $\tilde{Z}_{j,P}$. We now state and prove a lemma regarding the same.

\begin{lemma}{\label{monotonicity_proj_budget_Z_space}}
    For all projects $j \in [m]$ and sets $S \subseteq P$, we have $0<Z_{j,P}(S) < X_{j,P}(S)$ implies $Z_{j,P}(\hat{S}) = X_{j,P}(\hat{S})$ for all $\{ \hat{S} \mid \hat{S} \supset S, \hat{S} \subseteq P.\}$
\end{lemma}

\begin{proof}
    Consider a voting profile $P=(v_1,v_2,\ldots,v_n)$. Let these budgets in non-decreasing order of fractional allocation of project $j$ be $v_{l_1},v_{l_2},\ldots,v_{l_n}$ with their respective allocations being $v_{l_1,j}\geq v_{l_2,j}\ldots v_{l_p,j}$. Observe that $X_{j,P}(S)$ is non-zero only when $S$ is precisely the top $q$ budgets from this list for some $0\leq q\leq n$, else the $\min_{s \in S} v_{s,j} \leq \max_{s \in P\setminus S} v_{s,j}$ implying $X_{j,P}(S)$ =0

    Now, $0<X_{j,P\cup \{z\}}(S\cup \{z\}) < X_{j,P}(S)$ for some $S\in \mathcal{P}(P)$ would imply that $S$ precisely consists of top $q$ elements of $v_{l_1},v_{l_2},\ldots,v_{l_p}$ as $X_{j,P}(S)$ is non-zero for only these values of $S$. Suppose the allocation of project $j$ in budget $z$ is given by $z_j$ in which case $v_{l_q,j}> z_j\geq v_{l_{q+1},j}$ from definition of $X_{j,P}(S)$. Now consider any other set $\hat{S}$ which consists of top $q'>q$ budgets of $v_{l_1},v_{l_2},\ldots,v_{l_n}$ in which case $X_{j,P\cup \{z\}}(\hat{S}\cup \{z\}) = X_{j,P}(\hat{S})$. Note these sets $\hat{S}$ is a superset of $S$. However, any other set which is a superset of $S$ but is not precisely top $q'$ budgets from the list $v_{l_1},v_{l_2},\ldots,v_{l_n}$ for any $q'>0$ would have $X_{j,P\cup \{z\}}(\hat{S}\cup \{z\}) = X_{j,P}(\hat{S})=0$
\end{proof}

Let us give an example to illustrate the same and describe the construction of outcome budget $z$ and $\tilde{Z}_{j,P}$ after the triadic mechanism  and show how they differ.

\begin{example}
    Let us consider four budgets $a = \langle 0.4, 0.1 ,0.4 ,0.1\rangle$, $b= \langle 0.1, 0.4, 0.1, 0.4 \rangle$, $c = \langle 0.2, 0.05, 0.2,0.55 \rangle$,  and $d = \langle 0.25, 0.25, 0.25,0.25 \rangle$. 

    Now let us construct $z \sim \tilde{\mathfrak{n}}_{\text{rand}}(a,b,c)$. As observed after \ref{Word:step1} and \ref{Word:step2}, we assign $z = \langle0.2, 0.1, 0.2, 0.4\rangle$ [by allocating to outcome $z$, the median element of $a_j,b_j$ and $c_j$ for project $j$]. Since we still have funds of of 0.1 left, we have case \ref{Word:case1} from \S \ref{exact_soln_constr}. %Note that $\textsc{excess}=0.1$ 
    
    We now consider projects 1 and 3 since $a_1>z_1$ and $a_2>z_2$. As described in \S\ref{exact_soln_constr}, we may now have a selection $r^a_1=0.02$ and $r^a_3=0.03$, satisfying $r^a_j \leq s^a_j \ \forall j \in [4]$. Recall that $r^a$ and $r^b$ are random variables. However, we have $b_j>z_j$ for project $j=2$ and thus choose $r^b_2=0.05$. Hence, the allocation to projects to outcome $z$ under this sampling is given by $z = \langle 0.22, 0.15, 0.23,0.4\rangle$.

    Now, let us consider $X_{1,P}(S)$ for $S \in \mathcal{P}(\{a,b,c,d\})$. Observe that $X_{1,P}(\emptyset) = 0.6$, $X_{1,P}(a)=0.15$, $X_{1,P}(ad)=0.05$, $X_{1,P}(adc)=0.1$ and $X_{1,P}(abcd)=0.1$ and zero for all other sets. 
    Also observe that $Z_{1,P}(\emptyset)=Z_{2,P}(a)=0$, $Z_{1,P}(ad)=0.02$;
    $Z_{1,P}(adc)=0.1$ and $Z_{2,P}(abcd)= 0.1$ and $Z_{2,P}(S)$ is $0$ for every other $S$. 
    However, $\tilde{Z}_{1,P}(a)= \frac{r^a_1. X_{j,P}(a)}{X_{j,P}(ad)+X_{j,P}(ad)} = 0.02*\frac{3}{4}=0.015$, $\tilde{Z}_{1,P}(ad)=0.005$ but, $\tilde{Z}_{1,P}(adc) = X_{1,P}(adc)=0.1$ and $\tilde{Z}_{1,P}(abcd)= X_{1,P}(abcd)=0.1$.
    Thus, we may observe that $\tilde{Z}_{1,P}(.)$ does not  satisfy Lemma ~\ref{monotonicity_proj_budget_Z_space} but $Z_{1,P}(.)$ does. Also observe that $\tilde{Z}_{1,P}(.)$ satisfies all statements of Corrollary~\ref{sum_proj_budget_z_tilde} and ~\ref{new_budget_z_tilde_ineq}.

\end{example}
%other se%us co_{j}nsid

%ying 

\iffalse

\subsection{Proof of Lemma~\ref{cs_sum}}
\label{cs_sum_proof}
\begin{lemma*}[Restatement]
%\label{cs_sum}
Budget $b$ respects the project interactions iff the efficiency function $f(b)$ satisfies $\sum_i f(b)_i = 1$. Otherwise, $\sum_i f(b)_i < 1.$
\end{lemma*}

\begin{proof}
We use the fact that for any set of non-negative numbers $S$, $|S| \min(S) = \sum_{i\in S} {\color{red} i}$ iff all the numbers in $S$ are equal. Otherwise, $|S| \min(S) < \sum_{i\in S} i$. We also use the fact that for any set of non-negative numbers $S$, $ \max(S) = \sum_{i\in S} i$ iff at most one number in $S$ is greater than $0$. Otherwise, $ \max(S) < \sum_{i\in S} i$.

From the definition of $f(b)$ and using the facts above, it is easy to check that if $b$ respects the project interactions, we get $\sum_i f(b)_i = b_1 + b_2 + \ldots + b_m = 1$. If $b$ doesn't respect the project interactions, then, from the above facts, we have $\sum_i f(b)_i < 1$. 
\end{proof}

\fi

\remove{
\subsection{Proof of Lemma ~\ref{monotonicity_proj_budget_space}}{\label{sec:monotonicity_proj_budget_space_proof}}

\begin{proof}
By definition $X_{j,P}(S) = \max(\min_{s \in S} v_{s,j}-\max_{s \in P\setminus S} v_{s,j},0)$ and $X_{j,P}(\hat{S}) = \max(\min_{s \in \hat{S}} v_{s,j}-\max_{s \in P\setminus \hat{S}} v_{s,j},0)$ Since $\hat{S} \supset S$, we have $\min_{s \in \hat{S}} v_{s,j} \leq \min_{s \in {S}} v_{s,j}$ and  $\min_{s \in P \setminus \hat{S}} v_{s,j} \geq \min_{s \in P \setminus {S}} v_{s,j}$, thus proving the lemma.

\end{proof}

}
\subsection{Technical Lemma ~\ref{utility_proj_budget_Z_space}}{\label{sec:monotonicity_proj_budget_space_Z_proof}} 
We now state and prove Lemma~\ref{utility_proj_budget_Z_space}. 

\remove{
\begin{lemma*}
    For every project $j \in [m]$, we have $0<g_{j,P\cup \{z\}}(S\cup \{z\}) < g_{j,P}(S)$ implies $g_{j,P\cup \{z\}}(\hat{S}\cup \{z\}) = g_{j,P}(\hat{S}) \forall \hat{S} \supset S$  
\end{lemma*}
}

\begin{lemma}{\label{utility_proj_budget_Z_space}}
The sum of overlap utilities of all budgets in vote profile $P$ with a budget $z$ is $\sum\limits_{v \in P} u(v,z) = \sum\limits_{S \in \mathcal{P}(P)} |S| \cdot Z_P(S)$ and with an outcome $\{\tilde{Z}_P(S)\}_{S \in \mathcal{P}(P)}$ in the incremental allocation space is given by $\sum\limits_{v \in P}u(\tilde{Z}_P,v)= \sum\limits_{S \in \mathcal{P}(P)}|S|\tilde{Z}_P(S)$.
\end{lemma}

\begin{proof}
By Lemma~\ref{utility_one_voter_proj_budget_Z_space}, we have $\sum\limits_{v \in P} u(z,v) = \sum\limits_{v \in P} \sum\limits_{S \in \mathcal{P}(P)\vert S \ni v}Z_P(S)$ which equals $\sum\limits_{S \in \mathcal{P}(P)} |S| \cdot Z_P(S)$. For the construction in the incremental allocation space, from the definition of utility in \S \ref{sec_alt_barg_soln},  $\sum\limits_{v \in P} u(\tilde{Z}_P,v) = \sum\limits_{v \in P} \sum\limits_{S \in \mathcal{P}(P)\vert S \ni v}\tilde{Z}_P(S)$ which equals $\sum\limits_{S \in \mathcal{P}(P)} |S| \cdot \tilde{Z}_P(S)$ 
\end{proof}

\subsection{Proof of Lemma ~\ref{distortion_comp_nash_alt_nash_lemma}}{\label{Sec:alt_Nash_distortion_bound_proof}}

[Restatement of Lemma~\ref{distortion_comp_nash_alt_nash_lemma}]
{\it $\Distortion_once({\mathfrak{n}}_{\text{rand}}) \leq \Distortion_once({\tilde{\mathfrak{n}}}_{\text{rand}})$.}

To prove Lemma~\ref{distortion_comp_nash_alt_nash_lemma}, we first give the following intermediate result where the schemes $\mathfrak{n}_{\text{rand}}$ and $\tilde{\mathfrak{n}}_{\text{rand}}$ defined in Sections \ref{exact_soln_constr} and \ref{sec_alt_barg_soln} respectively with the utility function $u(z,a)$ for budget $z\sim\mathfrak{n}_{\text{rand}}$ and $u(\tilde{Z}_P,a)$ for $\tilde{Z}_P \sim \tilde{\mathfrak{n}}_{\text{rand}}$ defined in \S\ref{utility_common} and \S\ref{sec_alt_barg_soln} respectively. %
\begin{lemma}{\label{alt_Nash_utility_bound}}
    $\sum\limits_{i=1}^{n} \mathbb{E}[{u}(\tilde{\mathfrak{n}}_{\text{rand}}(a,b,c),v_i)] \leq \sum\limits_{i=1}^{n} \mathbb{E}[{u}({\mathfrak{n}}_{\text{rand}}(a,b,c),v_i)]$  for all budgets $ a,b,c \in P$ and vote profiles given by $P=(v_1,v_2,\ldots,v_n)$.
\end{lemma}
%
%Recall that we denote the  bargaining solution constructed in budget space for $\tilde{N}$ by $\tilde{Z}_P(S)$ and Nash bargaining solution constructed in project space by $\mathcal{N}$.
%
\begin{proof}
Let us first define budget $z \in \mathbb{B}$ as $z \sim \mathfrak{n}_{\text{rand}}(a,b,c)$ and $\{\tilde{Z}_{j,P}(S)\}^{S \in \mathcal{P}(P)}_{j \in [m]} \sim \tilde{\mathfrak{n}}_{\text{rand}}(a,b,c)$.\footnote{Note that we use a coupling argument here in since the randomness in $z$ while sampling under $\mathfrak{n}_{\text{rand}}$ is used for the construction of $\tilde{Z}_P$ as well. }

Note that $\sum_{S \in \mathcal{P}(P)}\tilde{Z}_{j,P}= z_j$ since $\tilde{Z}_{j,P}(.)$ satisfies conditions of Corollary~\ref{sum_proj_budget_z}, as shown in the construction of $\tilde{Z}$ in \S \ref{sec_alt_barg_soln}. Also from Corollary~\ref{sum_proj_budget_z}, we have $\sum_{S \in \mathcal{P}(P)} {Z}_{j,P}= z_j$. Now we characterize the following difference: $\sum_{i=1}^{n} u(z, v_i) - \sum_{i=1}^{n} u(\tilde{Z}_P, v_i)$.

%The event that the allocation  of project $j$ is $\hat{u}_j$ in the bargaining solution $z$ (i.e. $z_j=\hat{u}_j$) is the same as the probability of the event that $\sum\limits_{S \in \mathcal{P}(P)}\tilde{Z}_{j,P}$ is $\hat{u}_j$. 

%This directly follows from the construction in Section \ref{sec_alt_barg_soln}.

Let us order the budgets in a decreasing order of the allocation to project $j$. Denote these budgets by $v_{l_1},v_{l_2},\ldots,v_{l_n}$ and let these budgets have allotment to projects $j$ given by $v_{l_1,j} \geq v_{l_{2},j} \ldots \geq v_{l_n,j}$ We additionally define $v_{l_{n+1},j}=1.$ Let $i*$ be the smallest element of the set  $\{i \mid v_{l_{{i}},j} \leq z_j\}$. Now note that $X_{j,P}(S)$ may be non-zero only for $S = \phi  \text{ or } \{v_{l_1}\} \text{ or } \{v_{l_1},v_{l_{2}}\} \text{ or } \{v_{l_1},\ldots,v_{l_{n-1}},v_{l_n}\}$. Also, the values would respectively be $1-v_{l_1,j}, v_{l_1,j}-v_{l_2,j}, v_{l_2,j}-v_{l_3,j} \ldots, v_{l_{n-1},j}-v_{l_n,j},v_{l_n,j}$ and similarly $Z_{j,P}(S)$ would be $X_{j,P}(S)$ for $S = \{v_{l_1},\ldots,v_{l_n}\},  \{v_{l_1},\ldots,v_{l_{n-1}}\}, \ldots , \{v_{l_{1}},\ldots,v_{l_{i*}}\}$ and $Z_{j,P}(S)= z_j- v_{l_{i*},j}$ for $S=\{v_{l_{1}},\ldots,v_{l_{i*-1}}\}$ and zero for every other set $S$ if $i* \leq n$. Else $Z_{j,P}(S) = z_j$ for $S=\{v_{l_{1}},\ldots,v_{l_n}\}$ and zero for every other set $S$.

Note let us denote the set $\phi,\{v_{l_1}\},\{v_{l_1},v_{l_2}\},\ldots,\{v_{l_1},\ldots,v_{l_n}\}$ by $S_0,S_1,\ldots,S_n$. Note that sets $\tilde{Z}_{j,P}(S)$ can be non-zero only for $S$ = $S_0,S_1,\ldots,S_n$ as $\tilde{Z}_{j,P}(S)\leq X_{j,P}(S)$[as $\tilde{Z}_{j,P}$ satisfies Corollary \ref{new_budget_z_ineq}]. The net overlap utility (summing over overlap utility w.r.t every budget in $P$) in both bargaining solutions $Z$ and $\tilde{Z}$ may be shown equivalently as $\sum\limits_{S \in \mathcal{P}(P)} |S|Z_{P}(S)$ and $\sum\limits_{S \in \mathcal{P}(P)} |S|\tilde{Z}_{P}(S)$ [follows from Lemma \ref{utility_proj_budget_Z_space} and definition of $u(\tilde{Z},a)$ in \S \ref{sec_alt_barg_soln}]. We now compare the two of them by taking their difference below.

Now, we have 

\begin{align}
    & \sum\limits_{S \in \mathcal{P}(P)} |S|(Z_{j,P}(S) - \tilde{Z}_{j,P}(S))\\
    & =  \sum\limits_{S = S_0}^{S_{i*}-2} |S|(Z_{j,P}(S) - \tilde{Z}_{j,P}(S)) + |S_{i*-1}|.(Z_{j,P}(S_{i*-1}) - \tilde{Z}_{j,P}(S_{i*-1}))+ \sum\limits_{S = S_{i*}}^{S_n} |S|(Z_{j,P}(S) - \tilde{Z}_{j,P}(S)) \\
    & \overset{(a)}{\geq} |S_{i*-1}| \sum\limits_{S = S_0}^{S_{i*-2}}(Z_{j,P}(S) - \tilde{Z}_{j,P}(S)) + |S_{i*-1}|.(Z_{j,P}(S) - \tilde{Z}_{j,P}(S))+ |S_{i*-1}| \sum\limits_{S = S_{i*}}^{S_n}(Z_{j,P}(S) - \tilde{Z}_{j,P}(S)) \\
    & \overset{(b)}{=} 0
\end{align}

Note that first term in $(a)$ follows from the fact that $|S| \leq |S_{i*-1}|  \ \forall S= S_0,S_2\ldots, S_{i*-2}$ and the fact that $(Z_{j,P}(S) - \tilde{Z}_{j,P}(S)) \leq 0 \ \forall S= S_0,S_1,S_2\ldots, S_{i*-1}$ since $Z_{j,P}(S) = 0$ and $\tilde{Z}_{j,P}(S) \leq X_{j,P}(S)$ for these values of $S$. Similarly the second term in $(a)$ follows from $|S| \geq |S_{i*-1}|  \ \forall S= S_{i*},\ldots, S_{n}$ but $(Z_{j,P}(S) - \tilde{Z}_{j,P}(S)) \geq 0 \ \forall S= S_{i*-1},\ldots, S_{n}$ since $Z_{j,P}(S) = X_{j,P}(S)$ for these values of $S$.

Equation $(b)$ follows from $\sum\limits_{S \in \mathcal{P}(P)} Z_{j,P}(S) = \sum\limits_{S \in \mathcal{P}(P)} \tilde{Z}_{j,P}(S)= z_j$ and that $Z_{j,P}(S)=\tilde{Z}_{j,P}(S)=X_j(S)= 0$ for any $S \not\in \{S_0,S_1,\ldots,S_n\}$. Summing over all projects $j \in [m]$, we get $\sum\limits_{S \in \mathcal{P}(P)} |S| Z_{P}(S) \geq \sum\limits_{S \in \mathcal{P}(P)} |S|\tilde{Z}_{P}(S)$ for every $z \in \mathbb{B}$. Thus, we get the desired result.
%
%Since the inequality is true for every $z \in \mathbb{B}$, we would get the desired ineq
%
%Thus conditioned on the event $A_{\hat{u}_j}$, we would have the total overlap utility with all voters in $P$ under the first bargaining scheme $\mathfrak{n}_{\text{rand}}$ at least as large as the second bargaining scheme $\tilde{\mathfrak{n}}_{\text{rand}}$ by summing over all projects $j \in [m]$. Now we remove the conditional expectation by summing over probabilities of events $A_{\hat{u}_j}$, to obtain the desired result. 
%
%However, since the the 
%
%
\end{proof}

Now, since the distance $d(z,a) = 2- 2u(z,a) \text{ }\forall a \in P$, the result on expected distortion in Lemma \ref{distortion_comp_nash_alt_nash_lemma} follows.

\subsection{Proof of Lemma \ref{distortion_alt_Nash_bargaining}} {\label{distortion_alt_Nash_bargaining_proof}}
[Restatement of Lemma~\ref{distortion_alt_Nash_bargaining}]
{\it $\Distortion_once(\tilde{{\mathfrak{n}}}_{\text{rand}}) \leq \text{EPD}(\tilde{\mathfrak{n}}_{\text{rand}}).$
}
\begin{proof} Denote the set of budgets $\{v_1, v_2, \ldots, v_n\},$ %each in $\mathbb{B},$ 
by $P$ such that $v_i \in \mathbb{B}$ for all $i \in [n].$ For the bargaining schemes $\mathfrak{n}_{\text{rand}}$
\begin{align*}
& \Distortion_once(\tilde{\mathfrak{n}}_{\text{rand}}) \\
&= \sup_{n \in \mathbb{Z}^+,~P \in \mathbb{B}^n} \frac{ \displaystyle \frac{1}{n^4} \sum\limits_{i,j,k,l \in [n]} \mathbb{E}[d(\tilde{\mathfrak{n}}_{\text{rand}}(v_i, v_j, v_k), v_l)]}
{ \min\limits_{p \in \mathbb{B}} \frac{1}{n} \sum\limits_{i \in [n]} d(p, v_i) }, \\
&= \sup_{n \in \mathbb{Z}^+,~P \in \mathbb{B}^n} \frac{ \displaystyle \sum\limits_{Q \in \combiwithr{[n], 6}}  \!\!\frac{\Prob(Q)}{180} \!\!\sum\limits_{c \in Q} \sum\limits_{\substack{\{x,y\} \\\in \combi{Q\setminus\{c\}, 2}}} \sum\limits_{i \in Q \setminus \{x,y,c\}}  \mathbb{E}[d(\tilde{\mathfrak{n}}_{\text{rand}}(\{x,y\},c), v_i)] }
{ \min\limits_{p \in \mathbb{B}} \frac{1}{n} \sum\limits_{i \in [n]} d(p, v_i) }, \\
% FIXME: Explain why
% FIXME: Explain why
% FIXME: Explain why
% FIXME: Explain why
&\leq \sup_{n \in \mathbb{Z}^+,~P \in \mathbb{B}^n} \frac{ \displaystyle \sum\limits_{Q \in \combiwithr{[n], 6}}  \!\!\frac{\Prob(Q)}{180} \!\!\sum\limits_{c \in Q} \sum\limits_{\substack{\{x,y\} \\\in \combi{Q\setminus\{c\}, 2}}} \sum\limits_{i \in Q \setminus \{x,y,c\}}  \mathbb{E}[d(\tilde{\mathfrak{n}}_{\text{rand}}(\{x,y\},c)], v_i) }
{ \sum\limits_{Q \in \combiwithr{[n], 6}} \Prob(Q) \min\limits_{p \in \mathbb{B}} \frac{1}{6} \sum\limits_{i \in Q} d(p, v_i) }, \\
&\leq \sup_{n \in \mathbb{Z}^+,~P \in \mathbb{B}^n} \max_{Q \in \combiwithr{[n], 6}} \frac{ \!\!\displaystyle \frac{1}{180}\sum\limits_{c \in Q} \sum\limits_{\substack{\{x,y\} \\\in \combi{Q\setminus\{c\}, 2}}} \sum\limits_{i \in Q \setminus \{x,y,c\}}  \mathbb{E}[d(\tilde{\mathfrak{n}}_{\text{rand}}(\{x,y\},c)], v_i) }{ \frac{1}{6} \sum\limits_{i \in Q} d(p, v_i) }, \\
&= EPD(\tilde{\mathfrak{n}}_{\text{rand}}). \qedhere
\end{align*}
%This completes the proof.
\end{proof}
\subsection{Proof of Lemma ~\ref{distortion_Nash_bargaining_randomised}}{\label{sec_proof:distortion_Nash_bargaining_randomised}}
[Restatement of Lemma~\ref{distortion_Nash_bargaining_randomised}]
{\it $\text{EPD}(\tilde{{\mathfrak{n}}}_{\text{rand}}) \leq 1.66.$}
\begin{proof}
The proof is similar to that of Lemma \ref{pd_triadic_nash_relaxed} and uses a construction of a bilinear optimization problem. We give a proof sketch here. Consider a disagreement point $c$ and preferred budgets of bargaining agents, $a$ and $b$. Let $Q$ denote $\{a,b,c\}.$ Suppose $\{\tilde{Z}_{j,P}\}_{j \in [m]}$ is sampled from $\tilde{\mathfrak{n}}_{\text{rand}}(a,b,c).$  Consider the following two cases.
 
 %\textbf{Case 1:} $X_{j,Q}(ab)+X_{j,Q}(bc)+X_{j,Q}(ac)+X_{j,Q}(abc)\leq 1$. 
\ref{Word:case1}:
Recall from \S\ref{exact_soln_constr} that %since we choose projects uniformly at random without replacement, we have 
$\mathbb{E}[r^{a}_j]$ is proportional to  $s^a_j$ which is the same as $ X_{j,Q}(a)$. Using the fact that $\sum_{j=1}^{m} s^a_j = X_Q(a)$ and $\sum_{j=1}^{m} r^a_j = \textsc{Excess}/2$, we have  $\mathbb{E}[r^{a}_j] =\frac{X_{j,Q}(a) \cdot\textsc{excess}/2}{X_{Q}(a)}.$ %\footnote{$r^a_j$ is proportional to $s^a_j = X_{j,Q}(a)$ [\ref{Word:case1} of \S \ref{exact_soln_constr}] and $\sum_{j=1}^{m} s^a_j = X_Q(a)$}
Now, from our construction of  $\tilde{Z}(S)$ using $\alpha_{j,P}(S)$ as described in Case 1 in  \S\ref{sec_alt_barg_soln}, we have the following for any set $\{S ~\vert~ S \subseteq P; a\in S; b,c\not \in S\}$.
\begin{align}
  \mathbb{E}[\tilde{Z}_{j,P}(S)]=\frac{X_{j,Q}(a)  \textsc{excess}/2}{X_{Q}(a)} \frac{X_{j,P}(S)}{X_{j,Q}(a)} = \frac{\textsc{excess} \cdot X_{j,P}(S)}{2 X_Q(a)}   %\label{exp-1}
\end{align}
Taking the sum over all projects $j \in [m]$, we get
\begin{align}
\mathbb{E}[\tilde{Z}_{P}(S)] = \gamma^1_a\cdot X_P(S) \text { for all  }\{S ~\vert~ S \subseteq P; a\in S, b,c\notin S\} \ \text{ for some } 0 \leq \gamma^1_a \leq 1 \label{exp_1}.
\end{align}
A similar result can be shown on the sets $\{S ~\vert~ S \subseteq P, b\in S, a,c\notin S\}$.
\begin{align}
    \mathbb{E}[\tilde{Z}_{P}(S)] = \gamma^1_b\cdot X_P(S) \text { for all  }\{S ~\vert~ S \subseteq P; b\in S, a,c\notin S\} \ \text{ for some } 0 \leq \gamma^1_b \leq 1 \label{exp_2}
\end{align}

%A similar result can be proven on the sets $\{S ~\vert~ S \subseteq P, b\in S, a,c\notin S\}$.

\ref{Word:case2}: %$X_{j,Q}(ab)+X_{j,Q}(bc)+X_{j,Q}(ac)+X_{j,Q}(abc) \geq 1$
Recall from \S\ref{exact_soln_constr} that %since we choose projects uniformly at random without replacement, we have 
$\mathbb{E}[r^{a}_j]$ is proportional to  $t^a_j$ which is the same as $ X_{j,Q}(bc)$. Using the fact that $\sum_{j=1}^{m} t^a_j = X_Q(bc)$ and $\sum_{j=1}^{m} r^a_j = \textsc{Excess}/2$, we have  $\mathbb{E}[r^{a}_j] =\frac{X_{j,Q}(bc)
\cdot\textsc{excess}/2}{X_{Q}(bc)}.$

%$\mathbb{E}[r^{a}_j]=\frac{X_{j,Q}(bc) \cdot \textsc{excess}/2}{X_{Q}(bc)}.$\footnote{$r^a_j$ is proportional to $t^a_j=X_{j,Q}(bc)$ [\ref{Word:case2} of \S\ref{exact_soln_constr}] and $\sum_{j=1}^{m} t^a_j = X_Q(bc).$}
%
From our construction of $\tilde{Z}(S)$ using $\alpha_{j,P}(S)$ as described in Case 2 in \S\ref{sec_alt_barg_soln}, we have that for any sets $\{S ~\vert~ S \subseteq P, S \not\ni a, S\ni b,c\}$:
\begin{align}
 \mathbb{E}[\tilde{Z}_{j,P}(S)] 
 %= X_{j,P}(S)-\frac{X_{j,Q}(bc) \cdot \textsc{excess}/2}{X_{Q}(bc)} \cdot \frac{X_{j,P}(S)}{X_{j,Q}(bc)} 
   =  X_{j,P}(S) \left(1-\frac{\textsc{excess}}{2 \cdot X_Q(bc)}\right).    
\end{align}
   Taking the sum over all projects $j \in [m]$, we get 
   \begin{align}
       \mathbb{E}[\tilde{Z}_{P}(S)] = \gamma^2_a\cdot X_P(S) \text { for all  }\{S ~\vert~ S \subseteq P; a\notin S; b,c\in S\}\text { for some }0 \leq \gamma^2_a \leq 1 \label{exp_3}
   \end{align} 
A similar result can be proven for $\{S ~\vert~ S \subseteq P; b\notin S; a,c\in S\}$ 
   \begin{align}
       \mathbb{E}[\tilde{Z}_{P}(S)] = \gamma^2_b\cdot X_P(S) \text { for all  }\{S ~\vert~ S \subseteq P; a\notin S; b,c\in S\}\text { for some }0 \leq \gamma^2_b \leq 1 \label{exp_4}
   \end{align} 
   \iffalse
\begin{align*}
   \mathbb{E}[\tilde{Z}_{j,P}(S)] = & X_{j,P}(S)-\frac{X_{j,Q}() \cdot \textsc{excess}/2}{X_{Q}(a)} \cdot \frac{X_{j,P}(S)}{X_{j,Q}(a)}\\
   = & X_{j,P}(S) \left(1-\frac{\textsc{excess}}{2 \cdot X_Q(a)}\right).\\
\end{align*}
\fi

Using equations~\eqref{exp_1},~\eqref{exp_2},~\eqref{exp_3}, and~\eqref{exp_4} we formulate our objective function as the numerator of EPD minus 1.66 times its denominator. %~\ref{sec_proof:distortion_Nash_bargaining_randomised}. We solve it using the Gurobi package \cite{gurobi} in Python. 

Recall from the proof of Lemma ~\ref{pd_triadic_nash_relaxed} that we divided our problem into $2^{{{6 \choose 3}}}$ cases with each case presenting a 20 bit length sequence with 0 and 1 at each position defines whether the the triplet $Q$ corresponding to that position belongs in \ref{Word:case1} or \ref{Word:case2}.

%To prove this lemma we may write the maximization problem under the following formulation and we solve the following problem for every $\kappa$ in $\hat{\kappa}$ (denoting the set of all cases). %Recall that $\hat{C}$ denotes the cases which are unique upto permutations amongst voters as defined in proof of \ref{no_assumption_Nash}.

Similar to the proof of Lemma ~\ref{pd_triadic_nash_relaxed}, we first choose a set of 6 budgets given by $P=\{P_1,P_2,\ldots,P_6\}$ and use it as the common budget space. Similar to the proof, we define $X(S)= X_P(S)$ capturing the overlap of budgets in $S\subseteq P$. $v$ is defined to be optimal budgets on this space if 6 budgets and thus $V(S)=X_{P \cup \{v\}}(S \cup \{v\})$ for any subset $S$ of $T$. Also, we define $\tilde{Z}(S)$ from the randomized construction for bargaining scheme $\tilde{\mathfrak{n}}_{\text{rand}}$ as in Section~\ref{sec_alt_barg_soln} and in the optimization problem formulation, we define $\bar{Z}^{\{x,y\},c}(S)=\mathbb{E}[\tilde{Z}^{\{x,y\},c}(S)]$ when budgets $x,y$ bargain with $c$ being the disagreement point. Also note that $\bar{d}$ denotes the expected distance $\tilde{d}$ as defined in Section~\ref{sec_alt_barg_soln}. 

Also, given $Q=\{q_1,q_2,q_3\}$, we define the overlap in the common budget space of three voters as $\mathfrak{X}(S)=X_Q(S)$ and may consider two cases $\mathfrak{X}(q_1q_2q_3)+\mathfrak{X}(q_2q_3)+\mathfrak{X}(q_1q_3)+\mathfrak{X}(q_2q_3)$ being less than 1 or greater than 1 [\ref{Word:case1} or \ref{Word:case2}]. Since there are ${{6 \choose 3}}$ such subsets, we have $2^{20}$ instances, with each instance denoted by a string of length 20 with 0 or 1 at each position would denote whether the set $Q$ corresponding to that position belonging to case 1 or case 2. Let us use $f(.)$ as the bijective function to denote mapping from 3 sized subsets $Q$ to index of a bit in case $\kappa$. Let us denote the collection of all such instances by $\mathbb{K}$ and for every $\kappa \in \mathbb{K}$, we solve the following optimization problem \eqref{opt_bilinear}. 

The conditions \eqref{basic_condn1},\eqref{basic_condn2},\eqref{basic_condn3},\eqref{basic_condn4},\eqref{basic_condn5},\eqref{basic_condn6} follow from the constraints on $\tilde{Z}$ [\eqref{eq1_tilde_Z} and \eqref{eq2_tilde_Z}], $X(.)$[Lemma~\ref{total_allocation_budget_space}]  and Corollaries \ref{sum_proj_budget_z} and \ref{new_budget_z_ineq}. The conditions \eqref{pareto_condn1}, \eqref{pareto_condn2},\eqref{pareto_condn3},\eqref{pareto_condn4} and \eqref{equal_selection_condn1}, \eqref{equal_selection_condn2} follows from the Lemma \ref{Nash_barg_outcome} coupled with the fact that $\tilde{Z}_P$ obeys all of them as constructed in \S \ref{sec_alt_barg_soln} and applying the projection operation from $\tilde{Z}$ to $\tilde{Z}_Q$.

The conditions associated with first case follows from ~\eqref{exp_1},\eqref{exp_2} showing the proportionality of $\mathbb{E}[\tilde{Z}(S)]$ with $X(S)$ for every $S \ni a; S\not \ni b,c$ under the first case with a similar result holding for $S \ni b; S\not \ni a,c$ under the first case. Similarly, the conditions associated with first case follows from ~\eqref{exp_3},\eqref{exp_4} showing the proportionality of $\mathbb{E}[\tilde{Z}(S)]$ with $X(S)$ for every $S \ni a,c; S\not \ni b$ under the second case with a similar result holding for $S \ni b,c; S\not \ni a$ under the second case.

%and for the other case too. Also, for simplicity we often refer to budgets $P_i$ just by the index $i$ often.

%Note that $\bar{d}$ denotes the expected distance $\tilde{d}$ as defined for the hypothetical bargaining scheme $\tilde{\mathcal{N}}_{\text{rand}}$ and note that $p^{\{x,y\},c}$ denotes the solution obtained under randomised hypothetical bargaining scheme $\tilde{\mathcal{N}}_{\text{rand}}(x,y,c)$. Also note that we denote $\bar{Z}^{\{x,y\},c}(S)=\mathbb{E}[{Z}^{\{x,y\},c}(S)]$ which was computed in proof of Lemma~\ref{distortion_Nash_bargaining_randomised}. 

\begin{equation}
    \textit{Maximize} \ \ \ \frac{1}{180} \sum\limits_{c \in [6]} \sum_{\substack{\{x,y\} \in \\ \combi{[6] \backslash \{c\},2}}} \sum_{\substack{i \in [6]\\ \backslash (\{x,y,c\})}} {\bar{d}}(\tilde{Z}^{\{x,y\},c},P_i) - 1.66\frac{1}{6}\sum\limits_{i \in [6]} d(v,P_i) \label{opt_bilinear}
\end{equation}

where $$ \tilde{Z}^{\{x,y\},c}(.) \sim \tilde{\mathfrak{n}}_{\text{rand}}(a,b,c)$$
$${\bar{d}}(\tilde{Z}^{(\{x,y\},c)},P_i) = 2-2\sum\limits_{S \in \mathcal{P}([6]\backslash \{i\})} {\bar{Z}}^{(\{x,y\},\{c\})}(S \cup \{i\}) \label{avg_distance_defn}$$\\
    $$ d(v,P_i) = 2-2\sum\limits_{S \in \mathcal{P}([6]\backslash \{i\})} V(S \cup \{i\}) \label{opt_distance_defn}$$

Note that the variables under this optimization problem are $\{V(S)\}_{S \in \mathcal{P}([6])}; \{X(S)\}_{S \in \mathcal{P}([6])}\newline ; \{\bar{Z}^{(c,(x,y))}(S)\}^{(x,y) \in \combi{[6]\setminus \{c\}, 2}; c \in [6]}_{S \in \mathcal{P}([6])}; \{\alpha^{(\{x,y\},c)}\}_{(x,y) \in \combi{[6]\setminus \{c\}, 2}; c \in [6]}; \{\beta^{(\{x,y\},c)}\}_{(x,y) \in \combi{[6]\setminus \{c\}, 2}; c \in [6]}$ under the following set of conditions.

Also, recall that the problem is indexed by a 20-bit string $\kappa$ (denoting the instance) and $f$ is a bijective mapping from all 3- sized subsets of [6] to all natural numbers from 1 to 20.

\begin{align}
    \sum_{S \in \mathcal{P}([6]\setminus \{i\})} X(S\cup \{i\}) &=1 \ \forall i \in [6]\label{basic_condn1}\\
    X(S) & \geq 0 \ & \forall S\in \mathcal{P}([6])\label{basic_condn2}\\
    V(S) & \leq X(S) \ & \forall S\in \mathcal{P}([6])\label{basic_condn3}\\
    V(S) & \geq 0 \ & \forall S\in \mathcal{P}([6])\label{basic_condn4}\\
    \bar{Z}^{(\{x,y\},c)}(S) \leq X(S) & \forall S\in \mathcal{P}([6]) \text{ } & \ \forall \{x,y\}\in \combi{[6]\setminus \{c\},2} \ \forall c\in [6]\label{basic_condn5}\\
    \bar{Z}^{(\{x,y\},c)}(S) \geq 0 \ & \forall S\in \mathcal{P}([6])\text{ } & \ \forall \{x,y\}\in \combi{[6]\setminus \{c\},2} \ \forall c\in [6]\label{basic_condn6} 
\end{align}

     \begin{align}
     \bar{Z}^{(\{x,y\},c)}(\{c\}\cup U')= \ & 0\ & \forall U' \in \mathcal{P}([6]\backslash \{c,x,y\})\text{ } \ \forall \{x,y\}\in \combi{[6]\setminus \{c\},2} \ \forall c\in [6] \label{pareto_condn1}\\
     \bar{Z}^{(\{x,y\},c)}(U')= \ & 0\ &\forall U' \in \mathcal{P}([6]\backslash \{c,x,y\}) \ \forall \{x,y\}\in \combi{[6]\setminus \{c\},2} \ \forall c\in [6] \label{pareto_condn2}\\
     \bar{Z}^{(\{x,y\},c)}(\{x,y\}\cup U')= \ & X(\{x,y\} \cup U')\ & \forall U' \in \mathcal{P}([6]\backslash \{c,x,y\})\ \forall \{x,y\}\in \combi{[6]\setminus \{c\},2} \ \forall c\in [6]\label{pareto_condn3}\\
     \bar{Z}^{(\{x,y\},c)}(\{x,y,c\} \cup U')= & X(\{x,y,c\} \cup U') \ &\forall U' \in \mathcal{P}([6]\backslash \{c,x,y\}) \ \forall \{x,y\}\in \combi{[6]\setminus \{c\},2} \ \forall c\in [6] \label{pareto_condn4}
     \end{align}

     \begin{align}{}
         & \sum_{\substack{U' \in \mathcal{P}([6]\\\backslash \{c,x,y\})}} \bar{Z}^{(\{x,y\},c)}(\{x,c\} \cup U') + \bar{Z}^{(\{x,y\},c)}(\{x\} \cup U') \nonumber \\ & = \Biggl(1-\sum\limits_{i=2}^{3} \sum_{\substack{R' \in \\ \combi{ \{c,x,y\},i}}} \sum_{\substack{U' \in \mathcal{P}([6]\\\backslash (\{c,x,y\}))}} X(R'\cup U')\Biggr)/2 + \sum_{\substack{U' \in \mathcal{P}([6]\\\backslash (\{c,x,y\}))}} X(\{x,c\}\cup U') 
          \text{ }\forall \{x,y\}\in \combi{[6]\setminus \{c\},2} \forall c\in [6] \label{equal_selection_condn1}\\
         &  \sum_{\substack{U' \in \mathcal{P}([6]\\\backslash \{c,x,y\}))}} \bar{Z}^{(\{x,y\},c)}(\{y,c\} \cup U') + \bar{Z}^{(\{x,y\},c)}(\{y\} \cup U') \nonumber \\ & = \Biggl(1-\sum\limits_{i=2}^{3} \sum_{\substack{R' \in \\ \combi{ \{c,x,y\},i}}} \sum_{\substack{U' \in \mathcal{P}([6]\\\backslash (\{c,x,y\}))}} X(R'\cup U')\Biggr)/2 + \sum_{\substack{U' \in \mathcal{P}([6]\\\backslash (\{c,x,y\}))}} X(\{y,c\}\cup U') 
          \text{ }\forall \{x,y\}\in \combi{[6]\setminus \{c\},2} \forall c\in [6] \label{equal_selection_condn2}
    \end{align}
    
    For every $\{x,y\} \in \combi{[6]\setminus \{c\},2} \text{ } $ and  for all $ c \in [6]$  we have one of the following two cases:  
    \begin{itemize}
        \item If ${\kappa}_{f(\{x,y,c\})} == 0$ (denoting the index corresponding to subset)
    
    $$ \bar{Z}^{(\{x,y\},c)}(\{x\} \cup U') = \bar{Z}^{(\{x,y\},c)}(\{y\} \cup U') = 0 \ \forall U' \in \mathcal{P}([6]\backslash \{c,x,y\})$$
    $$ \mathbf{0\leq \alpha^{(\{x,y\},c)},\beta^{(\{x,y\},c)} \leq 1} $$
    $$ \mathbf{\bar{Z}^{(\{x,y\},c)}(\{x,c\} \cup U') = \alpha^{(\{x,y\},c)}.X(\{x,c\} \cup U') \text{ }\forall U' \in \mathcal{P}([6]\backslash \{c,x,y\})}$$
    $$ \mathbf{\bar{Z}^{(\{x,y\},c)}(\{y,c\} \cup U') = \beta^{(\{x,y\},c)}.X(\{y,c\} \cup U')  \text{ }\forall U' \in \mathcal{P}([6]\backslash \{c,x,y\})}$$
    $$ \sum\limits_{i=2}^{3} \sum_{\substack{R' \in \\ \combi{ \{c,x,y\},i}}} \sum_{\substack{U' \in \mathcal{P}([6]\\\backslash (\{c,x,y\}))}} X(R'\cup U') \geq 1$$
    
    \item  Else if $\kappa_{f(\{x,y,c\})} == 1$ (denoting the index corresponding to subset)
    
    $$ \bar{Z}^{(\{x,y\},c)}(\{x,c\} \cup U') = X(\{x,c\} \cup U') \ \forall U' \in \mathcal{P}([6]\backslash \{c,x,y\})$$
    $$ \bar{Z}^{(\{x,y\},c)}(\{y,c\} \cup U') = X(\{y,c\} \cup U') \ \forall U' \in \mathcal{P}([6]\backslash \{c,x,y\})$$
    $$ \mathbf{0\leq \alpha^{(\{x,y\},c)},\beta^{(\{x,y\},c)} \leq 1 }$$
    $$ \mathbf{\bar{Z}^{(\{x,y\},c)}(\{x\} \cup U') = \alpha^{(\{x,y\},c)}.X(\{x\} \cup U') \text{ }\forall U' \in \mathcal{P}([6]\backslash \{c,x,y\})}$$
    $$ \mathbf{\bar{Z}^{(\{x,y\},c)}(\{y\} \cup U') = \beta^{(\{x,y\},c)}.X(\{y\} \cup U')  \text{ }\forall U' \in \mathcal{P}([6]\backslash \{c,x,y\})}$$
    $$ \sum\limits_{i=2}^{3} \ \ \ \sum_{R' \in \combi{\{c,x,y\},i}} \ \ \ \sum_{U' \in \mathcal{P}([6]\\\backslash (\{c,x,y\}))} X(R'\cup U') \leq 1$$
    \end{itemize}

Note that as shown in Lemma~\ref{toggle_redn}, showing that maximization problem under instance $\kappa$ goes
to 0 is exactly same as showing instance $\hat{\kappa}$ goes to 0. Thus we remove all such cases from $\mathbb{K}$ where the bits representing the case are just toggled. Also, there are multiple cases which can be obtained from the other after a permutation of the voting preferences. 

On removal of all such cases, we find exactly {1244} cases remaining and we use a bilinear gurobi solver\cite{gurobi} to solve these cases and obtain the maximum result to be 0 in all cases, showing the upper bound of 1.66 as desired.

The scripts for the same can be found in \cite{Pythonscripts}.
%We use optimization package by Gurobi to solve every case in $\hat{\kappa}$ except the case which is denoted by all ones.
\end{proof}

\begin{lemma}{\label{toggle_redn}}
    The optimal value of the maximization problem as defined by \eqref{opt_bilinear} is zero for instance $\kappa$ iff the optimal value for the instance $\hat{\kappa}$ (toggling every bit in $\kappa$) equals zero.
\end{lemma}

\begin{proof}
    We prove this problem by constructing a mapping between the two instances.

    Consider any X(.), $V(.)$ and $\bar{Z}^{(\{x,y\},c)}(.)$ which satisfies the constraints given by instance $\kappa$. Thus we have\\

    $ {\sum\limits_{i=2}^{3} \sum\limits_{\substack{R' \in \\ \combi{ \{c,x,y\},i}}} \sum\limits_{\substack{U' \in \mathcal{P}([6]\\\backslash (\{c,x,y\}))}} X(R'\cup U') \geq 1}$ if $\kappa_{f(\{x,y,c\})}=0$ 

    $ {\sum\limits_{i=2}^{3} \sum\limits_{\substack{R' \in \\ \combi{ \{c,x,y\},i}}} \sum\limits_{\substack{U' \in \mathcal{P}([6]\\\backslash (\{c,x,y\}))}} X(R'\cup U') \leq 1}$ otherwise.  

Now, we construct $\hat{X}(.)$, $\hat{V}(.)$ and $\hat{Z}^{(\{x,y\},c)}(.)$ and prove that it satisfies the constraints given by case $\hat{\kappa}$. And we define it as follows for every set $S \in \mathcal{P}([6])$ and $\{x,y\} \in \combi{[6]\setminus \{c\},2}$ and $\{c\} \in [6]$ Note that the set of voters is given by $P=[6]$. 

\begin{align}
    \hat{X}({S})= & \frac{X(P\setminus S)}{\sum\limits_{S \in \mathcal{P}([6])} X(S)-1} \label{X_hat_defn}\\
    \hat{Z}^{(\{x,y\},c)}(S)= & \frac{(X(P\setminus S)-{\bar{Z}}^{(\{x,y\},c)}(P\setminus S))}{\left(\sum\limits_{S \in \mathcal{P}([6])} X(S) - 1\right)} \label{Z_hat_defn}\\
    \hat{V}(S)= & \frac{(X(P\setminus S)-{V}(P\setminus S))}{\left(\sum\limits_{S \in \mathcal{P}([6])} X(S) - 1\right)}\label{V_hat_defn}
\end{align}

Note that each of these terms is positive since the denominator is non-negative due the following points 
\hspace{2 em}
\begin{itemize}[itemsep=.35cm, leftmargin=1cm]
  \item $\sum\limits_{S \in \mathcal{P}([6])} X(S) \geq 1$ follows since $\sum_{S \in \mathcal{P}([6]); S\ni i} X(S)=1 $ from condition \eqref{basic_condn1},\eqref{basic_condn2}.
    \item $X(P\setminus S)\geq {V}(P\setminus S)),{\bar{Z}}^{(\{x,y\},c)}(P\setminus S))$ from condition \eqref{basic_condn5},\eqref{basic_condn6}.
\end{itemize}
%a) $\sum\limits_{S \in \mathcal{P}([6])} X(S) \geq 1$ follows since $\sum_{S \in \mathcal{P}([6]); S\ni i} X(S)=1 $ from condition \eqref{basic_condn1}, \eqref{basic_condn2} and since $X(P\setminus S)\geq {V}(P\setminus S)),{\bar{Z}}^{(\{x,y\},c)}(P\setminus S))$

    %$\tilde{X}({S})= \frac{X(P-S)}{\sum\limits_{S \in \mathcal{P}([6])} X(S)-1}$ where $P$ denotes the set of all voters.

Let us compute 
\begin{align*}
    & \sum\limits_{i=2}^{3} \sum_{\substack{R' \in \\\combi{ \{c,x,y\},i}}} \sum_{\substack{U' \in \mathcal{P}([6]\\\backslash (\{c,x,y\}))}} \hat{X}(R'\cup U')\\
    \overset{(a)}{=} & \frac{1}{\sum\limits_{S \in \mathcal{P}([6])} X(S)-1} \sum\limits_{i=0}^{1} \sum_{\substack{R' \in \\ \combi{ \{c,x,y\},i}}} \sum_{\substack{U' \in \mathcal{P}([6]\\\backslash (\{c,x,y\}))}} X(R' \cup U')\\
    \overset{(b)}{=} & \frac{1}{\sum\limits_{S \in \mathcal{P}([6])} X(S)-1}. \Bigl(\sum_{S \in \mathcal{P}([6])} X(S) - \sum\limits_{i=2}^{3} \sum_{\substack{R' \in \\ \combi{ \{c,x,y\},i}}} \sum_{\substack{U' \in \mathcal{P}([6]\\\backslash (\{c,x,y\}))}} {X}(R'\cup U')\Bigr)\\
    \overset{(c)}{\leq} & 1 \text{ if } \kappa_{f(\{x,y,c\})}=0\\
    \overset{(d)}{\geq} & 1 \text{ otherwise}    
\end{align*}

Note that $(a)$ follows from the definition of $\hat{X}$. $(c),(d)$ follows from the following.
% $$
%     \sum\limits_{i=2}^{3} \sum\limits_{\substack{R' \in \\ \combi{ \{c,x,y\},i}}} \sum\limits_{\substack{U' \in \mathcal{P}([6]\\\backslash (\{c,x,y\}))}} {X}(R'\cup U')  \begin{array}{lr}
%         \geq 1, & \kappa_{f(\{x,y,c\})}=0\\
%         \leq 1 & \text{otherwise}
%     \end{array}
% $$
%\newline
\begin{align*}
    \sum\limits_{i=2}^{3} \sum\limits_{\substack{R' \in \\ \combi{ \{c,x,y\},i}}} \sum\limits_{\substack{U' \in \mathcal{P}([6]\\\backslash (\{c,x,y\}))}} {X}(R'\cup U') & \geq 1 \text{ if }\kappa_{f(\{x,y,c\})}=0 \\
    \sum\limits_{i=2}^{3} \sum\limits_{\substack{R' \in \\ \combi{ \{c,x,y\},i}}} \sum\limits_{\substack{U' \in \mathcal{P}([6]\\\backslash (\{c,x,y\}))}} {X}(R'\cup U') & \leq 1 \text{ if }\kappa_{f(\{x,y,c\})}=1
\end{align*}

Hence $\hat{X}(.)$ satisfies the constraints for the case $\hat{\kappa}$ as $\hat{\kappa}_{f(\{x,y,c\})}$ is the toggled bit of ${\kappa}_{f(\{x,y,c\})}$

%Now, we define $\hat{Z}^{(\{x,y\},c)}(S)= \frac{(X(P-S)-\hat{Z}^{(\{x,y\},c)}(P-S))}{\Bigl(\sum\limits_{S \in \mathcal{P}([6])} X(S) - 1\Bigr)}$ for every set $S \subseteq P$ and $\hat{Z}_v(S)= \frac{(X(P-S)-\hat{Z}_{v}(P-S))}{\Bigl(\sum\limits_{S \in \mathcal{P}([6])} X(S) - 1\Bigr)}$.

Now we show that $\sum\limits_{\substack{S:S \ni i; \\ S \in \mathcal{P}([6])}} \tilde{X}(S)= \frac{\sum_{S \not\ni i} X(S)}{\Bigl(\sum\limits_{S \in \mathcal{P}([6])} X(S) - 1\Bigr)} = 1$ and also $\hat{X}(S) \geq 0$ and \newline
$0 \leq \hat{V}(S), \hat{Z}^{(\{x,y\},c)}(S) \leq \hat{X}(S)$ $\forall S \in \mathcal{P}([6])$, thus satisfying Equations \eqref{basic_condn1},\eqref{basic_condn2},\eqref{basic_condn3},\eqref{basic_condn4},\eqref{basic_condn5},\eqref{basic_condn6}.

For a given $(x,y) \in \combi{[6]\setminus\{c\},2}, c\in [6]$ and a set $U \in \mathcal{P}([6]\setminus \{a,b,c\})$, we define $\hat{U}'$ as follows satisfying $\hat{U}' \in \mathcal{P}([6]\setminus \{a,b,c\})$. 
\begin{equation}{\label{U_prime_defn}}
    \hat{U}' = ([6]\setminus\{x,y,c\})\setminus U'
\end{equation}
$\hat{Z}^{(\{x,y\},c)}(\{c\} \cup U') =  \frac{X(\{x,y\}\cup \hat{U}')-\bar{Z}^{(\{x,y\},c)}(\{x,y\}\cup \hat{U}')}{\Bigl(\sum\limits_{S \in \mathcal{P}([6])} X(S) - 1\Bigr)} = 0$ 
since $\bar{Z}^{(\{x,y\},c)}(\{x,y\} \cup \hat{U}')= X(\{x,y\} \cup \hat{U}')$ and $\hat{U}' \in \mathcal{P}([6]\setminus \{c,x,y\})$ from equation \eqref{pareto_condn3}. \\

%as $P\setminus (\{c,x,y\}\cup U') \in \mathcal{P}([6]/\{c,x,y\})$.

Similarly, we can show  $\hat{Z}^{(\{x,y\},c)}(U') = 0$, $\hat{Z}^{(\{x,y\},c)}(\{x,y\}\cup U')= X(\{x,y\}\cup U')$ and $\hat{Z}^{(\{x,y\},c)}(\{x,y\}\cup \{c\}\cup U') = X(\{x,y\}\cup \{c\}\cup U')$ thus showing $\hat{X}$ and $\hat{Z}$ satifying the conditions in \eqref{pareto_condn1},\eqref{pareto_condn2},\eqref{pareto_condn3} and \eqref{pareto_condn4}.

%Now consider, 

%$\hat{Z}^{(\{x,y\},c)}(\{x,c\}\cup U') = \frac{X(\{y\}\cup \hat{U}') - Z(\{y\}\cup \hat{U}')}{\sum\limits_{S \in \mathcal{P}([6])}X(S)-1} =0 $. This follows as $Z(\{y\}\cup \hat{U})= X(\{y\}\cup \hat{U})$ and $\hat{U}' \in\mathcal{P}([6]/\{c,x,y\})$. % and $X(.), V(.)$ and $Z(.)$ satisfies the constraint corresponding to case $\hat{c}= \{0,0,0,\ldots,0\}$. 

\begin{align*}
    & \sum_{\substack{U' \in \mathcal{P}([6]\\\backslash (\{c,x,y\}))}} \hat{Z}^{(\{x,y\},c)}(\{x,c\}\cup U') + \hat{Z}^{(\{x,y\},c)}(\{x\}\cup U')\\
    & \overset{(a)}{=} \sum_{\substack{\hat{U}' \in \mathcal{P}([6]\\\backslash (\{c,x,y\}))}} \frac{X(\{y\}\cup \hat{U}') - \bar{Z}^{(\{x,y\},c)} (\{y\}\cup \hat{U}') + X(\{y,c\}\cup \hat{U}') - \bar{Z}^{(\{x,y\},c)} (\{y,c\}\cup \hat{U}') }{\sum\limits_{S \in \mathcal{P}([6])}X(S)-1}\\
    & \overset{(b)}{=} \sum_{\substack{{U}' \in \mathcal{P}([6]\\\backslash (\{c,x,y\}))}} \frac{X(\{y\}\cup {U}') - \bar{Z}^{(\{x,y\},c)} (\{y\}\cup {U}') + X(\{y,c\}\cup {U}') - \bar{Z}^{(\{x,y\},c)} (\{y,c\}\cup {U}') }{\sum\limits_{S \in \mathcal{P}([6])}X(S)-1}\\
    & \overset{(c)}{=} \Biggl({- 1/2 + 1/2.\Biggl(\sum\limits_{i=2}^{3} \sum_{\substack{R' \in \\ \combi{ \{c,x,y\},i}}} \sum_{\substack{U' \in \mathcal{P}([6]\\\backslash (\{c,x,y\}))}} X(R'\cup U')\Biggr)} \\
    & + \sum_{\substack{U' \in \mathcal{P}([6]\\\backslash (\{c,x,y\}))}} X(\{y\}\cup {U}') + X(\{y,c\}\cup {U}') - X(\{y,c\}\cup U') \Biggr) \\
    & \times \frac{1}{\sum\limits_{S \in \mathcal{P}([6])}X(S)-1}\\
    &  = \Biggl({- 1/2 + 1/2.\Biggl(\sum\limits_{i=2}^{3} \sum_{\substack{R' \in \\ \combi{ \{c,x,y\},i}}} \sum_{\substack{U' \in \mathcal{P}([6]\\\backslash (\{c,x,y\}))}} X(R'\cup U')\Biggr) + \sum_{\substack{U' \in \mathcal{P}([6]\\\backslash (\{c,x,y\}))}} X(\{y\}\cup {U}')}\Biggr)\times \frac{1}{\sum\limits_{S \in \mathcal{P}([6])}X(S)-1}\\
    & \overset{(d)}{=} \frac{\Biggl({-1/2 + 1/2 \sum\limits_{S \in \mathcal{P}([6])} X(S) -1/2\cdot\Biggl(\sum\limits_{i=0}^{1} \sum\limits_{\substack{R' \in \\ \combi{ \{c,x,y\},i}}} \sum\limits_{\substack{U' \in \mathcal{P}([6]\\\backslash (\{c,x,y\}))}} X(R'\cup U') \Biggr)}\Biggr)}{\sum\limits_{S \in \mathcal{P}([6])}X(S)-1} + \sum_{\substack{U' \in \mathcal{P}([6]\\\backslash (\{c,x,y\}))}} \hat{X}(\{x,c\}\cup {U}') \\
    & \overset{(e)}{=} (1/2) \Biggl(1-\Biggl(\sum\limits_{i=2}^{3} \sum\limits_{\substack{R' \in \\\combi{ \{c,x,y\},i}}} \sum\limits_{\substack{U' \in \mathcal{P}([6]\\\backslash (\{c,x,y\}))}} \hat{X}(R'\cup U') \Biggr)\Biggr) + \sum_{\substack{U' \in \mathcal{P}([6]\\\backslash (\{c,x,y\}))}} \hat{X}(\{x,c\}\cup {U}')  
%    & \overset{(a)}{=} \hat{X}(\{x,c\}\cup U') + \\
\end{align*}

Note that $(a)$ follows from the definition of $\hat{Z}^{(\{x,y\},c)}$ in \eqref{Z_hat_defn} with $\hat{U}'$ [defined in \eqref{U_prime_defn}]. Also note in $(b)$, we replace the variable $\hat{U}'$ by $U'$ as both of them are defined with summation and can take any value in the powerset of $[6]\setminus \{a,b,c\}$. However, $(c)$ follows from the equality constraints satisfied by $\bar{Z}^{(\{x,y\},c)}(\{x,c\} \cup U') + \bar{Z}^{(\{x,y\},c)}(\{x\} \cup U')$ in \eqref{equal_selection_condn1}.  $(d)$ follows from just using the fact that $\sum\limits_{i=0}^{3} \sum\limits_{\substack{R' \in \\ \combi{ \{c,x,y\},i}}} \sum\limits_{\substack{U' \in \mathcal{P}([6]\\\backslash (\{c,x,y\}))}} \hat{X}(R'\cup U') = \sum\limits_{S \in \mathcal{P}([6])} X(S)$ and $(e)$ follows again from the definition of $\hat{X}(.)$. Thus, we show that it satisfies the 
equality constraint in \eqref{equal_selection_condn1} and similarly we can show it satisfies the second equality constraint in \eqref{equal_selection_condn2}. 
desired equality constraints and similarly, we can also show that it satisfies the next equality constraint \eqref{equal_selection_condn2}.

Suppose we have $\kappa_{f(\{x,y,c\})}=0$, for any set $\{x,y,c\} \subseteq [6]$, thus we have 
\begin{align*}
    \hat{Z}^{(\{x,y\},c)}(\{x\}\cup U') & \overset{(e)}{=} \frac{X(\{y,c\} \cup \hat{U}')- \bar{Z}^{(\{x,y\},c)}(\{y,c\}\cup \hat{U}') }{\sum\limits_{S \in \mathcal{P}([6])}X(S)-1}\\
    & \overset{(f)}{=} \hat{X}(\{x\}\cup U') (1-\beta^{(\{x,y\},c)})
\end{align*}
Note that $(e)$ follows from the definition of $\hat{Z}$ and $(f)$ follows since $\hat{X}(\{x\}\cup U') = \frac{X(\{y,c\} \cup \hat{U}')}{\sum_{S \in \mathcal{P}([6])}X(S)-1}$ from the definition of $\hat{X}(.)$

Similarly, we have $\hat{Z}^{(\{x,y\},c)}(\{y\}\cup U')=(1-\alpha^{(\{x,y\},c)})\hat{X}(\{y\}\cup U')$. Hence we may observe that it satisfies all the conditions specific to $\hat{\kappa}_{f(\{x,y,c\})}=1$. This works since $\hat{\kappa}$ is the list of toggled bits from $\kappa$. 

Now suppose $\kappa_{f(\{x,y,c\})}=1$, for any set $\{x,y,c\} \subseteq [6]$, thus we have 
\begin{align*}
    \hat{Z}^{(\{x,y\},c)}(\{x,c\}\cup U') & \overset{(e)}{=} \frac{X(\{y\} \cup \hat{U}')- \bar{Z}^{(\{x,y\},c)}(\{y\}\cup \hat{U}') }{\sum\limits_{S \in \mathcal{P}([6])}X(S)-1}\\
    & \overset{(f)}{=} \hat{X}(\{x,c\}\cup U') (1-\beta^{(\{x,y\},c)})
\end{align*}
Note that $(e)$ follows from the definition of $\hat{Z}$ and $(f)$ follows since $\hat{X}(\{x,c\}\cup U') = \frac{X(\{y\} \cup \hat{U}')}{\sum_{S \in \mathcal{P}([6])}X(S)-1}$ from the definition of $\hat{X}(.)$

Similarly, we have $\hat{Z}^{(\{x,y\},c)}(\{y,c\}\cup U')=(1-\alpha^{(\{x,y\},c)})\hat{X}(\{y,c\}\cup U')$. Hence we may observe that it satisfies all the conditions specific to $\hat{\kappa}_{f(\{x,y,c\})}=0$. This works since $\hat{\kappa}$ is the list of toggled bits from $\kappa$.

%$\hat{Z}^{(\{x,y\},c)}(\{x\}\cup U') = \frac{X(\{y,c\} \cup \hat{U}')- \bar{Z}^{(\{x,y\},c)}(\{y,c\}\cup \hat{U}') }{\sum\limits_{S \in \mathcal{P}([6])}X(S)-1}$ and $\hat{X}(\{x\}\cup U') = \frac{X(\{y,c\} \cup \hat{U}')}{\sum_{S \in \mathcal{P}([6])}X(S)-1}$. 

%Thus, we get $\hat{Z}^{(\{x,y\},c)}({x}\cup U') = (1-\beta^{(\{x,y\},c)}) \hat{X}({x}\cup U')$ and similarly, we get $\hat{Z}^{(\{x,y\},c)}({y}\cup U') = (1-\alpha^{(\{x,y\},c)}) \hat{X}({y}\cup U')$.

Now let us compute the objective function with the defined variables $\hat{X}$, $\hat{V}$ and $\hat{Z}_{.}(.)$. Here we use $\hat{d}$ for the same terms to distinguish from the case $\kappa$.
Now we observe that 

\begin{align*}
     \hat{d}(\tilde{Z}^{(\{x,y\},c)},P_i)
    = & \ 2 -2 \sum_{S \in \mathcal{P}([6]\backslash i)} \hat{Z}^{(\{x,y\},c)}(S\cup \{i\})\\
    \overset{(a)}{=} & \ 2 - 2 \frac{\sum\limits_{S\not \ni i; S \in \mathcal{P}([6])} X(S)-\bar{Z}^{(\{x,y\},c)}(S)}{\sum_{S \in \mathcal{P}([6])}X(S)-1}\\
    \overset{(b)}{=} & \frac{2.\left(1- \sum\limits_{S \in \mathcal{P}([6]\backslash i)} \bar{Z}^{(\{x,y\},c)} (S\cup \{i\})\right) }{\sum_{S \in \mathcal{P}([6])}X(S)-1}\\
    = & \ \frac{\bar{d}(\tilde{Z}^{(\{x,y\},c)},P_i)}{\sum\limits_{S \in \mathcal{P}([6])}X(S)-1}
\end{align*}

Similarly, $\hat{d}(v,P_i)= \frac{d(v,P_i)}{\sum\limits_{S \in \mathcal{P}([6])}X(S)-1}$

Note $(a)$ follows from the from the construction of $\hat{Z}$. %and the summation index $S$ changed from $S$ to $[6]\setminus (S \cup \{i\})$ 

$(b)$ follows from the fact that $\sum\limits_{S\ni i; S \in \mathcal{P}([6])} X(S)=1$ and $\sum\limits_{S \in \mathcal{P}([6])} \tilde{Z}_{j,P}(S)=1$ as defined in the constraints for the original problem $\kappa$

Thus, the optimization problem reduces to 

\begin{align*}
    & \frac{1}{180} \sum\limits_{c \in [6]} \sum_{\substack{\{x,y\} \in \\ \combi{[6] \backslash \{c\},2}}} \sum_{\substack{i \in [6]\\ \backslash (\{x,y,c\})}} \hat{d}(\tilde{Z}^{(\{x,y\},c)},P_i) - 1.66\frac{1}{6}\sum\limits_{i \in [6]} \hat{d}(v,P_i)\\
    & = \left(\frac{1}{180} \sum\limits_{c \in [6]} \sum_{\substack{\{x,y\} \in \\ \combi{[6] \backslash \{c\},2}}} \sum_{\substack{i \in [6]\\ \backslash (\{x,y,c\})}} \bar{d}(\tilde{Z}^{(\{x,y\},c)},P_i) - 1.66\frac{1}{6}\sum\limits_{i \in [6]} {d}(v,P_i)\right)\cdot \frac{1}{\sum\limits_{S \in \mathcal{P}([6])}X(S)-1}
\end{align*}

Now, we thus show that we can have a unique mapping between the two cases and the objective functions are also proportional. Hence the optimal solution of instance $\kappa$ being zero implies the other instance $\hat{\kappa}$ is zero too and vice-versa.
\end{proof}

\end{changemargin}

%\end{document}

\end{document}